\documentclass[12pt]{article}%
\usepackage[nosort]{cite}
\usepackage{graphicx}
\usepackage{multicol}
\usepackage{amsfonts}
\usepackage{amssymb}
\usepackage{amsmath}
\usepackage{dsfont}
\usepackage{heck}
\usepackage{afterpage}
\usepackage{setspace}
\usepackage{verbatim}
\usepackage{color}
\usepackage{longtable}
\usepackage{float}
\usepackage{subcaption}
\usepackage{epsfig}
\usepackage{epstopdf}
\usepackage{adjustbox}
\usepackage{todonotes}
\newcommand{\0}{{(0)}}
\usepackage{tikz}
\usepackage[margin=1in]{geometry}
\usepackage{titletoc}
\usepackage{hyperref}%
\usepackage{mathdots}
\providecommand{\U}[1]{\protect\rule{.1in}{.1in}}
\pdfoutput=1
\newsavebox{\mysavebox}

\usetikzlibrary[shapes.geometric]
\usetikzlibrary{positioning}
\usetikzlibrary{calc,intersections,through,backgrounds}
\tikzset{>=stealth}
\usetikzlibrary{decorations.pathmorphing}
\usetikzlibrary{decorations.markings}
\tikzset{>=stealth}
\hypersetup{colorlinks,citecolor=black,filecolor=black,linkcolor=black,urlcolor=black,pdftex}
\usetikzlibrary{decorations.markings}

\numberwithin{equation}{section}

\newcommand{\ba}{\begin{eqnarray}}
\newcommand{\ea}{\end{eqnarray}}

\newcommand{\Tr}{\, {\rm Tr}}

\newcommand{\be}{\begin{equation}}
\newcommand{\ee}{\end{equation}}

\tikzstyle{startstop} = [rectangle, rounded corners, minimum width=3cm, minimum height=1cm,text centered, draw=black, fill=blue!10]
\tikzstyle{startstop} = [rectangle, rounded corners, minimum width=3cm, minimum height=1cm,text centered, draw=black, fill=blue!10]
\tikzstyle{io} = [trapezium, trapezium left angle=70, trapezium right angle=110, minimum width=3cm, minimum height=1cm, text centered, draw=black, fill=blue!30]
\tikzstyle{process} = [rectangle, minimum width=3cm, minimum height=1cm, text centered, draw=black, fill=orange!30]
\tikzstyle{decision} = [diamond, minimum width=3cm, minimum height=1cm, text centered, draw=black, fill=green!30]
\tikzstyle{arrow} = [thick,->,>=stealth]
\begin{document}

\date{July 2017}

\title{Green-Schwarz Automorphisms and 6D SCFTs}

\institution{UNC}{\centerline{${}^{1}$Department of Physics, University of North Carolina, Chapel Hill, NC 27599, USA}}

\institution{PENN}{\centerline{${}^{2}$Department of Physics and Astronomy, University of Pennsylvania, Philadelphia, PA 19104, USA}}

\institution{HVD}{\centerline{${}^{3}$Jefferson Physical Laboratory, Harvard University,
Cambridge, MA 02138, USA}}

\authors{Fabio Apruzzi\worksat{\UNC , \PENN}\footnote{e-mail: {\tt fabio.apruzzi@unc.edu}},
Jonathan J. Heckman\worksat{\PENN}\footnote{e-mail: {\tt jheckman@sas.upenn.edu}},
and Tom Rudelius\worksat{\HVD}\footnote{e-mail: {\tt rudelius@g.harvard.edu}}}

\abstract{All known interacting 6D superconformal field theories
(SCFTs) have a tensor branch which includes anti-chiral two-forms and
a corresponding lattice of string charges.
Automorphisms of this lattice preserve the
Dirac pairing and specify discrete global and gauge
symmetries of the 6D theory. In this paper we
compute this automorphism group for 6D SCFTs. This
discrete data determines the geometric
structure of the moduli space of vacua. Upon compactification,
these automorphisms generate Seiberg-like dualities, as well
as additional theories in discrete quotients by the 6D global symmetries.
When a perturbative realization is available, these discrete quotients
correspond to including additional orientifold planes in the string construction.}

\maketitle

\setcounter{tocdepth}{2}

\tableofcontents

%\enlargethispage{\baselineskip}

\newpage

\section{Introduction}

Six-dimensional supeconformal field theories (SCFTs)\ provide a
higher-dimensional perspective on many aspects of lower-dimensional quantum field theories.
A canonical example of this phenomenon is the compactification of
theories with $\mathcal{N}=(2,0)$ supersymmetry to four dimensions.
Reduction on a $T^{2}$ yields a geometric perspective on S-duality for
$\mathcal{N}=4$ Super Yang-Mills theory, and compactification on a more general
Riemann surface leads to $\mathcal{N}=2$ generalizations of S-duality
\cite{Vafa:1997mh, Witten:1997sc, Argyres:2007cn, Gaiotto:2009we}.
Similar considerations hold for compactifications
to lower-dimensional systems. The comparatively large number
of 6D theories with $\mathcal{N}=(1,0)$ supersymmetry recently classified in
references \cite{Heckman:2013pva, DelZotto:2014hpa, Heckman:2015bfa}
(see also \cite{Apruzzi:2013yva, Gaiotto:2014lca, Bhardwaj:2015xxa})
provides a vast generalization of this paradigm to lower-dimensional physical theories with
less supersymmetry. For earlier work on the
construction and study of 6D SCFTs, see for example \cite{Witten:1995ex, Witten:1995zh, Strominger:1995ac, Seiberg:1996qx,
WittenSmall, Ganor:1996mu,MorrisonVafaII,Seiberg:1996vs, Bershadsky:1996nu,
Brunner:1997gf, Blum:1997fw, Aspinwall:1997ye, Intriligator:1997dh, Hanany:1997gh}.
Given this, it is important to isolate calculable
quantities for these theories and their compactified lower-dimensional descendants.
For a partial list of references on this topic, see e.g.
\cite{Klemm:1996bj,Evans:1997hk, Landsteiner:1997vd, Brandhuber:1997cc,Landsteiner:1997ei,Kapustin:1998xn,
Landsteiner:1998pb,Argyres:2002xc, Alday:2009aq,
Chacaltana:2010ks,Chacaltana:2012zy,Xie:2012hs,Nekrasov:2012xe,
Beem:2014kka, Ohmori:2014pca, Heckman:2014qba, Ohmori:2014kda, Intriligator:2014eaa, Haghighat:2014vxa, DelZotto:2015isa,
Gaiotto:2015usa, Ohmori:2015pua, Gadde:2015tra, Franco:2015jna, DelZotto:2015rca,
Hanany:2015pfa, Beem:2015aoa, Cordova:2015fha, Aganagic:2015cta, Ohmori:2015pia,
Coman:2015bqq, Hayashi:2015vhy, Cordova:2016xhm,
Morrison:2016nrt, Kim:2016foj, Shimizu:2016lbw, DelZotto:2016pvm, Heckman:2016xdl,
Apruzzi:2016nfr, Razamat:2016dpl, Cordova:2016emh,
DelZotto:2017pti, Haghighat:2017vch, Chang:2017xmr, Mekareeya:2017jgc, Mekareeya:2017sqh}.

One of the robust \textquotedblleft topological\textquotedblright
elements of all $(2,0)$ theories is its Dirac pairing for string charges.
These pairings are classified by the Dynkin diagrams of the
simply laced algebras, a fact which is transparent in the IIB\ realization of
these theories via compactification on $\mathbb{C}^{2}/\Gamma_{ADE}$, with
$\Gamma_{ADE}$ an ADE\ discrete subgroup of $SU(2)$. The resolution of this
orbifold singularity yields a geometric realization of the corresponding
ADE\ root system, and upon compactification on a $T^{2}$ yields an
$\mathcal{N}=4$ Super Yang-Mills theory with ADE gauge group.

The geometry of the root lattice also dictates the structure of the moduli
space. For example, letting $\mathcal{W}$ denote the Weyl group, the $(2,0)$
tensor branch moduli space decomposes into a positive cone $\mathbb{R}%
^{5T}/\mathcal{W}$, where $T$ is the number of $\mathcal{N}=(2,0)$ tensor
multiplets (see e.g. \cite{Seiberg:1997ax}).
This structure persists for lower-dimensional compactifications.
For example, the Coulomb branch of $\mathcal{N}=4$ Super Yang-Mills theory
with gauge group $G$ is $\mathbb{R}^{6T}/\mathcal{W}_{G}$.
As another example, compactification to
two-dimensional theories provides a natural analogue of this in which the
Weyl group defines an orbifold CFT, with twisted sectors given by
its conjugacy classes (see e.g. \cite{Ganor:1996pc, Ganor:1996xg}).

In this paper we determine the analogous structure for all 6D SCFTs realized
via F-theory compactification. More precisely, we shall be interested in the discrete gauge and global
symmetries associated with the lattice of string charges.

The main tool at our disposal is the topological nature of the Green-Schwarz-Sagnotti-West terms
present in the tensor branch deformation of a 6D\ SCFT \cite{Green:1984bx, Sagnotti:1992qw, Sadov:1996zm, Morrison:2012np}.
These couplings take the schematic form:
\begin{equation}
L_{6D}\supset\int\mu_{i,g} B^{(i)}\wedge\text{Tr}(F^{(g)}\wedge F^{(g)}),
\label{muGS}%
\end{equation}
Here, $B^{(i)}$ is an anti-chiral two--form with $i$ an index labelling the
$(1,0)$ tensor multiplets, and $F^{(g)}$ is a two-form field strength with $g$
an index which runs over both dynamical gauge fields as well as background
fields. Such background fields are present when we have a non-trivial background flavor
symmetry, R-symmetry, or spin connection. Anomaly cancellation enforces a
rather rigid structure on the coefficients $\mu_{i,g}$. Even so, there is at
first some apparent freedom in how we specify their values. Indeed, the invariant
quantity which enters the anomaly polynomial eight-form is:%
\begin{equation}
M_{g,h}=\mu_{g,i}^{T}\left(  \frac{1}{A}\right)  ^{ij}\mu_{j,h}, \label{Mmu}%
\end{equation}
where here, $A_{ij}$ is the Dirac pairing for the 6D string charges, which is
interpreted geometrically as the intersection pairing for the base of an
F-theory compactification on an elliptically fibered Calabi-Yau threefold.

This apparent ambiguity would at first seem to suggest more than
one set of Green-Schwarz terms will give a consistent 6D SCFT, but it
is resolved once we impose the further constraint that all effective
strings have positive tension. Geometrically, this is the condition
that each effective divisor of the F-theory base has positive volume.
Different choices of the $\mu$'s correspond to formally continuing some tensions
to negative values. In field theory terms, we are simply performing
a mild version of \textquotedblleft duality\textquotedblright\ in six
dimensions. The reason for the terminology is that in lower dimensions, these
operations often appear as Seiberg-like dualities, in accord with both the brane moves
studied in reference \cite{Hanany:1996ie} as well as the geometric realization of such maneuvers
considered in reference \cite{Cachazo:2001sg}.

As an illustrative example, consider the case of $N$ M5-branes probing the
geometry $\mathbb{R}_{\bot}\times\mathbb{C}^{2}/\Gamma_{ADE}$. Moving onto a
partial tensor branch corresponds to keeping the M5-branes at the
ADE\ singularity, whilst moving them to separate points on the $\mathbb{R}%
_{\bot}$ factor. The relative separation between the M5-branes defines a
chamber of the partial tensor branch. Moving the M5-branes through one another
amounts to formally continuing some vevs to negative values. This leads to a
compensating shift in the Green-Schwarz terms, as dictated by the ambiguity in
specifying the $\mu$'s of line (\ref{Mmu}).

Such ambiguities are captured in terms of linear maps on the lattice of
string charges $\Lambda$ that preserve the intersection pairing
$A_{ij}$. These linear maps are automorphisms of the lattice, or equivalently of
the Dirac / intersection pairing. They form a group, which we denote as
Aut$(\Lambda)$. Our goal in this work will be to determine
Aut$(\Lambda)$ for all 6D\ SCFTs and to explain how it dictates the structure
of lower-dimensional theories obtained from compactification. In some sense
this data is complementary to the defect group $\Lambda^{\ast}/\Lambda$ of a
6D\ SCFT \cite{DelZotto:2015isa}.

To compute Aut$(\Lambda)$, it is convenient to use the F-theory realization of
6D\ SCFTs. The main point is that all of the associated lattices $\Lambda$ are
readily available in this context, and are specified by a configuration of
$\mathbb{P}^{1}$'s which can simultaneously contract to zero size. This is the
condition that the intersection pairing $A$ is positive definite. F-theory
imposes additional conditions on admissible $A$'s, since we must also be able
to define a consistent elliptically fibered Calabi-Yau threefold over a
candidate base. In F-theory, we can also blowdown curves of
self-intersection $-1$ in a descending sequence until we are left with a
configuration of curves, none of which has self-intersection $-1$. The
endpoint configuration of curves also defines a lattice $\Lambda_{\text{end}}%
$. Its automorphism group is related to that of $\Lambda$ as:
\begin{equation}
\text{Aut}(\Lambda)=\text{Aut}(\Lambda_{\text{end}})\times\text{Aut}\left(
\mathfrak{sp}_{Q}\right)  ,
\end{equation}
where Aut$\left(  \mathfrak{sp}_{Q}\right)  $ is the automorphism group for
the root lattice of $\mathfrak{sp}_{Q}$, and $Q$ indicates the number of
blowdowns of $-1$ curves which must be performed to pass from $\Lambda$ to
$\Lambda_{\text{end}}$. An additional property of Aut$(\Lambda_{\text{end}})$
is the existence of a maximal normal subgroup $\mathcal{W}_{\text{end}}$,
which is a close analog of the Weyl group present for $\mathcal{N}=(2,0)$
theories. In terms of this, we find a further refinement:%
\begin{equation}
\text{Aut}(\Lambda_{\text{end}})=\mathcal{O}_{\text{end}}\ltimes
\mathcal{W}_{\text{end}}\text{.}%
\end{equation}

The physical interpretation of this discrete data is as follows: first, we see that the
group $\text{Aut}(\Lambda)$ contains a maximal normal subgroup
$\mathcal{W} = \mathcal{W}_{\text{end}} \times \text{Aut}(\mathfrak{sp}_Q)$, which we identify with
discrete gauge symmetries of the system. Additionally,
$\mathcal{O}_{\text{end}}$ is a candidate global discrete symmetry. This is borne out by the fact that
the base of the F-theory model enjoys a group of discrete isometries
specified by $\mathcal{O}_{\text{end}}$. This is true
both at the conformal fixed point as well as in the resolved geometry. In the full F-theory model, this symmetry
may be broken because there could be a non-trivial elliptic fibration.
Said differently, this global symmetry depends on the Higgs branch moduli.

The structure of the tensor branch moduli space for $\mathcal{N} = (1,0)$ theories
can be far more involved than it is for  $\mathcal{N}=(2,0)$ theories. For example, though there is still a notion of a fundamental domain
of moduli space for a theory with $T$ tensor multiplets, the orbits of this patch under
the automorphism group sometimes do not produce a tessellation of $\mathbb{R}^{T}$,
leading to non-trivial forbidden zones. These different possibilities
are conveniently handled using the F-theory characterization of 6D SCFTs,
where these moduli correspond to resolution parameters of curves.

Though we leave a more complete analysis for future work, we can already point to the ways in which
the data of Aut$(\Lambda)$ shows up in compactifications of 6D SCFTs.
First of all, the discrete gauge symmetries $\mathcal{W}$ lead
to Seiberg-like dualities once we compactify and flow to another four-dimensional theory.
One particularly interesting feature of 6D SCFTs
is the presence on the tensor branch of generalizations of quiver gauge theories
with exceptional gauge groups and ``conformal matter.'' We also consider
the related structures for 2d theories obtained from compactifying on a four-manifold.

The discrete global symmetries of the 6D theory also lead to several novel structures upon further compactification. For
example, adding a chemical potential for the background global symmetries of $\mathcal{O}_{\text{end}}$ yields a new theory which
is the equivalent of introducing \textquotedblleft discrete twists\textquotedblright\
in compactifications of class $\mathcal{S}$ theories (see e.g. \cite{Tachikawa:2009rb, Drukker:2010jp, Tachikawa:2010vg}).
Another (and conceptually distinct) operation of a more stringy flavor is associated with formally quotienting the theory by this symmetry.
An interesting feature of this is that it also provides a
field theoretic characterization of various orientifold planes in compactifications
of F-theory, including the effects of $O7^{+}$-planes (see e.g. \cite{Witten:1997bs, deBoer:2001px, Tachikawa:2015wka}),
along the lines proposed in reference \cite{Bhardwaj:2015oru}.

The rest of this paper is organized as follows: first, in section
\ref{sec:AUTO} we discuss in general terms the automorphism group for a
lattice of strings, and the physical data it captures in compactifications of
a 6D\ SCFT, and compute it for all 6D SCFTs. Section \ref{sec:TENSOR}
studies the structure of the tensor branch moduli space, as dictated by the automorphism group.
Section \ref{sec:COMPACT} presents some examples of how the automorphism group specifies
defining data in compactifications of 6D\ SCFTs, including the realization of
various orientifold planes. We present our conclusions and directions for future research in
section \ref{sec:CONC}. Some additional details on how to calculate the
anomaly polynomial of a 6D\ SCFT via \textquotedblleft analytic
continuation\textquotedblright\ in the rank of the gauge groups present on the
tensor branch are presented in Appendix A.

\section{Green-Schwarz Automorphisms \label{sec:AUTO}}

One of the essential elements in all known interacting 6D\ SCFTs is the
existence of a tensor branch. On this branch, we have a collection of
$\mathcal{N}=(1,0)$ tensor multiplets. Recall that a tensor multiplet contains
a real scalar $t$, an anti-chiral two-form potential $B_{\mu\nu}^{(-)}$ with
anti-self-dual field strength, and corresponding fermionic superpartners. The
appearance of a two-form potential signals the presence of strings, with
tension controlled by the vevs of the real scalars. The conformal fixed point
corresponds to taking all $t$'s to zero simultaneously. In a theory with $T$
tensor multiplets, the Dirac pairing for string charges is a $T\times T$
positive definite symmetric matrix which acts on the lattice of string charges
via a canonical pairing:%
\begin{equation}
A:\Lambda\rightarrow\Lambda.
\end{equation}
This pairing also specifies the metric on moduli space for the $t$'s. Indeed,
indexing the multiplets by the variables $i$ and $j$, the metric on moduli
space takes the form $A_{ij}dt^{i}dt^{j}$, in the obvious notation. In this coordinate system,
the tension of a string is given by:
\begin{equation}
t_{i} = A_{ij} t^{j}.
\end{equation}

An important quantity in all 6D\ SCFTs is the associated anomaly polynomial.
This is a formal eight-form constructed from the background field strengths
for the $SU(2)$ R-symmetry, the curvature of the spin connection, and possible
flavor symmetries. In addition to the one loop contribution from
chiral modes, there are additional ``tree level'' contributions from the
Green-Schwarz terms:
\begin{equation}
L_{6D}\supset\int\mu_{i,g}B^{(i)}\wedge\text{Tr}(F^{(g)}\wedge F^{(g)}),
\label{muGSagain}%
\end{equation}
where here we also include possible couplings to vector multiplets and their
associated gauge field strengths. The contribution to the eight-form anomaly
polynomial takes the form:%
\begin{equation}
I_{8D}\supset\mu_{g,i}^{T}\left(  \frac{1}{A}\right)  ^{ij}\mu_{j,h}\text{
Tr}(F^{(g)})^{2}\text{Tr}(F^{(h)})^{2}.
\end{equation}
The anomaly polynomial determines, for example, the conformal anomalies of the
6D\ SCFT (see e.g. \cite{Cordova:2015fha, Cordova:2015vwa}). Another important
feature of the anomaly polynomial is that when the number of simple gauge
group factors and tensor multiplets is equal, the $\mu$'s are square matrices
and there is then a unique solution to the anomaly cancellation conditions
\cite{Cordova:2015fha, Ohmori:2014kda} (for additional explicit calculations see also
\cite{Heckman:2015ola, Heckman:2016ssk, Mekareeya:2016yal}). This can also be
extended to all 6D\ SCFTs by interpreting \textquotedblleft unpaired
tensors\textquotedblright\ as a generalized type of 6D conformal matter
\cite{DelZotto:2014hpa}. In Appendix A we present another method in which we
formally introduce a possibly trivial gauge group to pair with each such
tensor multiplet.

The only quantity which actually enters into the anomaly polynomial is the
combination:%
\begin{equation}
M_{g,h}=\mu_{g,i}^{T}\left(  \frac{1}{A}\right)  ^{ij}\mu_{j,h}.
\end{equation}
Based on this, it is natural to ask whether there is more than one choice of
$\mu$'s available. The main point is that once we demand all string tensions are positive,
namely $t_{i} > 0$, we have well defined couplings to the
chemical potentials defined by the anti-chiral two-forms. As such, there is a
unique choice for the $\mu_{i,g}$ in this patch of moduli space. Provided it
makes sense, we can ask what happens to the spectrum of strings if we now pass
to formally negative values of some of the $t_{i}$'s. We clearly must seek a
new basis of positive tension objects, and correspondingly the values of the
$\mu_{i,g}$ may change.

To determine the geometry of the tensor branch moduli space, we seek integral
linear transformations $\sigma_{i}^{j}$ of the coefficients $\mu_{i,g}$
which act on the tensor index:%
\begin{equation}
\mu_{i,g}\mapsto\sigma_{i}^{j}\mu_{j,g}\text{,}%
\end{equation}
and preserve the form of the anomaly polynomial, i.e. they preserve the matrix
$M_{g,h}$. Equivalently, we seek all transformations $\sigma$ such that:%
\begin{equation}
\sigma^{T}\cdot\frac{1}{A}\cdot\sigma=\frac{1}{A}. \label{SAS}%
\end{equation}

The collection of all such $\sigma$'s forms a group. First of all, we have the
identity element. Second, if we have two transformations $\sigma$ and
$\sigma^{\prime}$ which both preserve $M_{g,h}$, then their composition will
also preserve $M_{g,h}$. To establish the existence of an inverse element, we
first verify that $\sigma^{-1}$ is an integral transformation. To see this, we
first compute the determinant of equation (\ref{SAS}). This yields the relation:
\begin{equation}
\left(  \det\sigma\right)  ^{2}=1,
\end{equation}
so $\det\sigma=\pm1.$

As such, the inverse of the integral linear map $\sigma$ will also have
integer entries, and so $\sigma^{-1}$ is also an integral transformation.
Next, consider the inverse of equation (\ref{SAS}),%
\begin{equation}
\sigma^{-1}\cdot A\cdot\left(  \sigma^{-1}\right)  ^{T}=A.
\end{equation}
Multiplication by $\sigma$ on the left and $\sigma^{T}$ on the right yields%
\begin{equation}
A=\sigma\cdot A\cdot\sigma^{T}.
\end{equation}
Taking the inverse,%
\begin{equation}
\frac{1}{A}=\left(  \sigma^{-1}\right)  ^{T}\cdot\frac{1}{A}\cdot\sigma^{-1},
\end{equation}
so the inverse is also an integral transformation preserving $M_{g,h}$.

In fact, what we have just established is that the group of $\sigma$'s is also
the automorphism group for the quadratic form defined by $A_{ij}$, the
intersection pairing of the lattice $\Lambda$. This is also known as the
automorphism group of the lattice, and so we shall often write Aut$(\Lambda)$
to reflect this fact.

What then is the physical interpretation of this group action? In the case of
the ADE$\ (2,0)$ theories, there is a further decomposition we can perform:%
\begin{equation}
\text{Aut}(\Lambda)=\mathcal{O}_{ADE}\ltimes\mathcal{W}_{ADE},
\end{equation}
where $\mathcal{W}_{ADE}$ is the Weyl group i.e. the group of inner
automorphisms of the ADE algebra, and $\mathcal{O}_{ADE}$ is the group of
outer automorphisms. This has a clean interpretation upon compactification on torii.

The $(2,0)$ theories yield maximally supersymmetric gauge theories with
ADE\ gauge algebra. In this case, we can identify $\mathcal{W}_{ADE}$ with a
collection of discrete gauge transformations, and $\mathcal{O}_{ADE}$ as
possible ways to twist the theory to reach non-simply laced algebras in lower dimensions.

In the more general case of $(1,0)$ theories, we do not have the luxury of a
lower-dimensional gauge theory when we compactify on torii.\ Instead, we must
make do with the structure already apparent in six dimensions.
Along these lines, we see that on the tensor branch, the discrete gauge transformations
must flip the sign of at least one tensor
branch scalar $t_i$. These are the natural analogs of the Weyl group transformations
in the $(2,0)$ theories. Indeed, they correspond to redundancies in our description of the
tensor multiplets, so we interpret these as discrete gauge symmetries.\footnote{One might ask
whether these are examples of higher-form discrete symmetries in the sense of reference \cite{Gaiotto:2014kfa}.
One way to see that they are ``standard'' discrete symmetries rather than
higher-form symmetries is that on a topologically
trivial background spacetime, such symmetries are abelian, whereas our symmetry group is often non-abelian.}
Other transformations which leave all moduli positive
are the natural analog of the outer automorphisms, and correspond to global symmetries.

From the perspective of F-theory compactification, it is immediate that the data of the global symmetries is indeed
intrinsic to a given SCFT. The reason is that we specify a base as a resolution of
an orbifold of the form $\mathbb{C}^{2} / \Gamma_{U(2)}$ for $\Gamma$ a discrete subgroup of $U(2)$.
The discrete isometries of this geometry are the global symmetries. Indeed, even after resolving the singularity,
these isometries persist, and in the $(2,0)$ case are what we identify with the outer automorphisms of the
corresponding Lie algebra.

More precisely, such isometries of the base are really just \textit{candidate} global symmetries. Indeed,
in a full F-theory compactification we often must specify a non-trivial elliptic fibration to the model.
Geometrically, a discrete isometry of the base need not extend to the full Calabi-Yau threefold. In physical
terms, the elliptic fibration is controlled by the Higgs branch moduli. So, we see that candidate global symmetries
will depend on this data.

Having argued that we can have a discrete global symmetry, it is natural to ask whether we can
gauge it, or whether it is actually anomalous (unless supplemented by additional degrees of freedom).
This sort of gauging operation does not affect the local structure of correlation functions, and instead leads to
global distinctions in the spectrum of extended objects in the theory. It would be interesting to evaluate the corresponding
't Hooft anomalies, but this is beyond the scope of the present paper.

A related though different operation involves quotienting by such a symmetry, as one would do in an orientifold construction.
In such cases, we expect a $(2,0)$ theory to become a $(1,0)$ theory. To illustrate this point, consider the
A-type $(2,0)$ theories. These theories possess a $\mathbb{Z}_2$ outer automorphism which acts by reflection on the nodes of the Dynkin
diagram. In string theory terms, if we attempt to quotient by this symmetry, we need to introduce an orientifold plane. This automatically
breaks half of the supersymmetry, yielding a $(1,0)$ theory instead. Additional branes must also be included to locally satisfy Gauss' law constraints. This is consistent with the fact that there is no way to perform a ``discrete quotient'' of a $(2,0)$ theory which is also
$\mathcal{N} = (2,0)$ supersymmetric. Instead, such a quotient yields a $(1,0)$ theory. We study this question in much greater detail
in section \ref{sec:COMPACT} where we consider discrete quotients of compactified theories.

Our plan in the remainder of this section will be to compute Aut$(\Lambda)$
for all 6D\ SCFTs. As a warmup, we first briefly review the case of the
$(2,0)$ theories, where this data is captured by the automorphisms of the
corresponding ADE root lattice. In the case of the $(1,0)$ SCFTs, there are
analogous results, including a generalization of a Weyl group as well as outer
automorphisms. There are, however, some important differences in the case of
non-generic Higgs branch moduli, a point we return to in section
\ref{ssec:Higgs}.

\subsection{$\mathcal{N}=(2,0)$ Theories}

In this section we consider the automorphism group for the $(2,0)$ theories.
In this case, the Green-Schwarz terms of equation (\ref{muGSagain}) involve
the $\mathfrak{sp}(4)$ R-symmetry background field strength and curvature from
the spin connection. Using the classification of $(2,0)$ theories via discrete
subgroups $\Gamma_{ADE}\subset SU(2)$ and the corresponding IIB backgrounds on
$\mathbb{R}^{5,1}\times\mathbb{C}^{2}/\Gamma_{ADE}$, we also know that the
$(1,0)$ tensor branch is geometrically realized as the resolution of these
orbifold singularities. Indeed, in the resolved geometry, we have a collection
of $\mathbb{P}^{1}$'s which intersect according to the ADE\ Dynkin diagram.
The automorphism group for each intersection form is simply the automorphism
group of the ADE\ root lattice. All of the automorphism groups take the form
of a semi-direct product of the outer automorphisms of the algebra with the
inner automorphisms associated with the Weyl group:%
\begin{equation}
\text{Aut}(\Lambda_{ADE})=\mathcal{O}_{ADE}\ltimes\mathcal{W}_{ADE}.
\end{equation}
In particular, the outer automorphisms for each of the ADE\ root systems are,
for $N>1$ and $M>4$:%
\begin{equation}%
\begin{tabular}
[c]{|c|c|c|c|c|c|c|c|}\hline
& $A_{1}$ & $A_{N}$ & $D_{4}$ & $D_{M}$ & $E_{6}$ & $E_{7}$ & $E_{8}$\\\hline
$\mathcal{O}_{ADE}$ & $1$ & $\mathbb{Z}_{2}$ & $S_{3}$ & $\mathbb{Z}_{2}$ &
$\mathbb{Z}_{2}$ & $1$ & $1$\\\hline
\end{tabular}
\end{equation}
where $S_{3}$ is the symmetric group on three letters.

The moduli space of the tensor branch is given by $\mathbb{R}^{5T}%
/\mathcal{W}_{ADE}$, where the factor of five is due to the fact that we are
dealing with $(2,0)$ rather than $(1,0)$ tensor multiplets. Additionally,
compactification on a manifold can be accompanied by a twist by an outer
automorphism $\mathcal{O}_{ADE}$, possibly composed with an element of
$\mathcal{W}_{ADE}$. Let us note that this operation can be understood field theoretically
as activating a background chemical potential for the discrete flavor symmetry. This is a distinct notion
from the operation of ``discrete quotient'' which we shall encounter in section \ref{sec:COMPACT}.

\subsection{$\mathcal{N}=(1,0)$ Theories}

In this subsection we compute the Green-Schwarz automorphisms of all
6D\ SCFTs. At this point, it is convenient to use the geometric language of
F-theory compactification, though we stress that all of this analysis can be
carried out in purely field theoretic terms.

In the F-theory realization of 6D SCFTs, we introduce a
non-compact K\"{a}hler surface $B$ with some configuration of simultaneously
collapsing $\mathbb{P}^{1}$'s. We obtain a consistent F-theory background when
we can also define an elliptically fibered Calabi-Yau threefold with base $\mathcal{B}$.
The homology lattice of the base determines the lattice of string charges:%
\begin{equation}
H_{2}^{\text{cpct}}(\mathcal{B},%
%TCIMACRO{\U{2124} }%
%BeginExpansion
\mathbb{Z}
%EndExpansion
)=\Lambda,
\end{equation}
and the intersection pairing corresponds to the Dirac pairing. References
\cite{Heckman:2013pva, Heckman:2015bfa} provide a classification of all such
bases, as well as all possible elliptic fibrations over a given base. For our
present purposes, the main point is that each such base $\mathcal{B}$ is generated by
starting with an \textquotedblleft endpoint configuration\textquotedblright%
\ of curves which contains no $-1$ curves, and then performing some prescribed
number of blowups. There is a minimal number of blowups (which may be
zero)\ necessary to define a consistent elliptic fibration, but additional
blowups are sometimes possible.

Although the specific geometry depends on the particular location of each such
blowup, the structure of the lattice of string charges is insensitive to this
data \cite{DelZotto:2015isa, DelZotto:2014fia}. Indeed, given the endpoint
configuration $\mathcal{B}_{\text{end}}$ with lattice $H_{2}^{\text{cpct}}%
(\mathcal{B}_{\text{end}},%
%TCIMACRO{\U{2124} }%
%BeginExpansion
\mathbb{Z}
%EndExpansion
)=\Lambda_{\text{end}}$, blowing up $Q$ times yields the lattice:%
\begin{equation}
\Lambda=\Lambda_{\text{end}}\oplus%
%TCIMACRO{\U{2124} }%
%BeginExpansion
\mathbb{Z}
%EndExpansion
^{\oplus Q}\text{,}%
\end{equation}
This follows from the fact that for a curve $\Sigma$, blowing up the curve
generates a shift in the divisor class as:%
\begin{equation}
\lbrack\Sigma]\mapsto\lbrack\Sigma]-E, \label{shifto}%
\end{equation}
with $E$ the exceptional divisor class. From this perspective, the
automorphism group splits into two pieces: the contribution from
$\Lambda_{\text{end}}$ and the contribution from $%
%TCIMACRO{\U{2124} }%
%BeginExpansion
\mathbb{Z}
%EndExpansion
^{\oplus Q}$. In most cases, there is an upper bound on the value of $Q$
dictated by the choice of endpoint configuration. For example, some
configurations cannot be blown up at all, the $E_{8}$ lattice being one
such case \cite{Heckman:2013pva}.

One important consideration is that this linear change of basis for the
lattice can change the intersection pairing. Along these lines, consider two
lattices $\Lambda$ and $\Lambda^{\prime}$ which are related to each other by a
change of basis, as indicated for example by line (\ref{shifto}):%
\begin{equation}
L:\Lambda^{\prime}\rightarrow\Lambda. \label{LattMap}%
\end{equation}
The intersection pairing of the two lattices are related as%
\begin{equation}
L^{T}\cdot A\cdot L=A^{\prime}.
\end{equation}
Automorphisms of the two lattices are related via the transformation%
\begin{equation}
L^{-1}\cdot\sigma\cdot L=\sigma^{\prime},
\end{equation}
in the obvious notation. As an example, take $\Lambda$ to be the configuration
of two $-1$ curves which do not intersect, and $\Lambda^{\prime}$ to be the
$1,2$ configuration. The lattice transformation $L$ is%
\begin{equation}
L=\left[
\begin{array}
[c]{cc}%
1 & -1\\
0 & 1
\end{array}
\right]  \text{.}%
\end{equation}

Taking into account this structure, we see that the automorphism group is not
sensitive to the locations of the various blowups. Consequently, we learn that
for any choice of blowups of an endpoint configuration, the group
Aut$(\Lambda)$ is given by the product:%
\begin{equation}
\text{Aut}(\Lambda)=\text{Aut}(\Lambda_{\text{end}})\times\text{Aut}(%
%TCIMACRO{\U{2124} }%
%BeginExpansion
\mathbb{Z}
%EndExpansion
^{\oplus Q}).
\end{equation}
For the factor coming from the rank $Q$ E-string theory, we find that the
automorphism group is identical to that for the root system of the Lie algebra
$\mathfrak{sp}(Q)$, where in our notation $\mathfrak{sp}(1)\simeq
\mathfrak{su}(2)$. Said differently, this is just the Weyl group of the root system.

Let us discuss each of these factors in turn. In the case of Aut$(%
%TCIMACRO{\U{2124} }%
%BeginExpansion
\mathbb{Z}
%EndExpansion
^{\oplus Q})$, the group is given by possible flops of each individual curve,
as well as permutations amongst these curves. All told, the automorphism group
for this part is:%
\begin{equation}
\text{Aut}(%
%TCIMACRO{\U{2124} }%
%BeginExpansion
\mathbb{Z}
%EndExpansion
^{\oplus Q})=S_{Q}\ltimes\left(  \mathbb{Z}_{2}\right)  ^{Q}.
\end{equation}
As already remarked, this is also the
Weyl group for the Lie algebra $\mathfrak{sp}(Q)$. Indeed, if we consider a
particular sequence of blowups, we reach the configuration of curves:%
\begin{equation}
\underset{Q}{\underbrace{1,2,...,2}},
\end{equation}
Here and in what follows, the notation $a,b$ denotes a pair of curves of
self-intersection $-a$ and $-b$ that intersect at one point. We note that this
is also the configuration of curves used to realize the rank $Q$ E-string
theory. Upon compactification on a circle, it is well known that the rank $Q$
E-string theory reduces to an $\mathfrak{sp}(Q)$ 5D gauge theory with seven
flavors. At the conformal fixed point, this enhances to an $E_{8}$ flavor symmetry.

Consider next the contribution from the factor Aut$(\Lambda_{\text{end}})$. In
this case, it is convenient to make use of the explicit classification of
endpoint configurations given in reference \cite{Heckman:2013pva}. These take
the form of generalized A- and D-type Dynkin diagrams, while the E-series
still involves only the standard $-2$ curves:
\begin{align}
\text{A-type}  &  \text{: \ \ }n_{1},...,n_{l}\\
\text{D-type}  &  \text{: \ \ }2,\overset{2}{m_{1}}...,n_{l}\\
E_{6}  &  \text{:}\text{ \ \ }2,2,\overset{2}{2},2,2\\
E_{7}  &  \text{:}\text{ \ \ }2,2,\overset{2}{2},2,2,2\\
E_{8}  &  \text{:}\text{ \ \ }2,2,\overset{2}{2},2,2,2,2.
\end{align}
In the last three cases, we simply have the automorphism group of the
corresponding $E$-type root system. We therefore confine our attention to the
A- and D-type endpoint configurations. Most of these computations are a
straightforward application of symmetries manifest in the configuration, and
we have verified this structure using the software package \texttt{MAGMA}.

\subsubsection{A-type Endpoints}

From reference \cite{Heckman:2013pva}, we know that all the A-type end points
can be described by a collection of curves of the following form:
\begin{equation}
M^{1}N^{1}\ldots M^{a}N^{a}\ldots M^{a_{l}}N^{a_{l}},
\end{equation}
where $M^{a}=\{2\,2\,\ldots\,2\}$ is a sequence of $m_{a}$ curves of
self-intersection $-2$, and $N^{a}=\{n_{1}\,n_{2}\,\ldots\,n_{N^{a}}\}$ is a
sequence of curves with $n_{i}>2$ for all $i$. Here, the notation indicates
that we have a collection of curves of self-intersection $-n$, which intersect
pairwise at a single point, as indicated by the ordering of the sequence.
In fact, from the classification results of \cite{Heckman:2013pva}, we know that
$l \leq 3$.

To determine the structure of the automorphism group, and in particular its
action on the tensor branch moduli space, we observe that the automorphism
group will be a subgroup of that for the A-type configuration with just $-2$
curves. In this sense, all we really need to do is track the automorphisms
which survive when we change from a configuration of $-2$ curves
to one where some curves in the endpoint are replaced by $-n$ curves with $n >
2$.

Along these lines, we find that all of the Weyl reflections for the $-2$
curves naturally survive. Additionally, for a $-2$ curve with class $\alpha$
which intersects a $-n$ curve with class $\beta$, the change under Weyl
reflection is again:
\begin{equation}
\alpha\mapsto-\alpha\,\,\,\text{and}\,\,\,\beta\mapsto\beta+\alpha.
\end{equation}
What then becomes of the remaining Weyl reflections which act on the $-n$
curves? The basic point is that we are restricted to a very limited class of reflections
in which all curves simultaneously transform.

Now, by inspection the automorphism group always contains the element $-Id$.
This element is an order two element which acts on the divisor classes as:
\begin{equation}
\alpha_{i} \mapsto - \alpha_{i}.
\end{equation}
The symmetries of the endpoint configuration dictate whether this is an element
of the normal subgroup of $\text{Aut}(\Lambda)$ which generates discrete gauge
symmetries. To see why, we observe that in the case where
the configuration of curves enjoys a reflection symmetry, this element is
a composition of an outer automorphism and the discrete gauge transformation:%
\begin{equation}
\alpha_{i}\mapsto-\alpha_{T+1-i}, \label{longflip}%
\end{equation}
which acts on all of the divisor classes. When there is no such reflection
symmetry, the map $-Id$ is already a discrete gauge transformation. Note that
for both $-Id$ as well as the map of line (\ref{longflip}), the map is an
order two element. Regardless of whether we have a symmetric or asymmetric
endpoint configuration of curves, the analog of the Weyl group for an A-type
endpoint configuration is then given by:
\begin{equation}
\mathbb{Z}_{2}\ltimes\left(  S_{m_{1}+1}\times...S_{m_{l}+1}\right)
\end{equation}
where in the case of an asymmetric endpoint configuration, this is actually a
direct product, and in the case of a symmetric endpoint configuration, the
$\mathbb{Z}_{2}$ group action is dictated by the map of line (\ref{longflip}).
Note that in the latter case, Aut$(\Lambda)$ is a semi-direct product involving this
$\mathbb{Z}_{2}$ reflection symmetry.

It is also helpful to explicitly spell out the various types of groups we
encounter for A-type lattices. The automorphism group of the endpoint configuration
divides into four cases:

\begin{itemize}
\item When the endpoint is a single curve,%
\begin{equation}
\mathrm{Aut}(\Lambda_{\text{end}})=\mathbb{Z}_{2}.
\end{equation}

\item When the endpoint is a sequence of $m$ curves of self-intersection $-2$,
we have%
\begin{equation}
\mathrm{Aut}(\Lambda_{\text{end}})=\mathbb{Z}_{2}\ltimes S_{m+1},
\end{equation}
where $S_{m+1}$ is the symmetric group on $m+1$ letters.

\item When the endpoint configuration is symmetric but contains at least one
curve of self-intersection $-x$ for $x\geq3$, we have:
\begin{equation}
\mathrm{Aut}(\Lambda_{\text{end}})=\mathbb{Z}_{2}\ltimes\left(  \mathbb{Z}%
_{2}\ltimes(S_{m_{1}+1}\times...S_{m_{l}+1}\right)  ).
\end{equation}
Here, the product over symmetric groups just involves the Weyl groups of each
block of $-2$ curves. The middle $\mathbb{Z}_{2}$ factor acts as in line
(\ref{longflip}). Note that in contrast to the case of configurations of all
$-2$ curves, the number of automorphisms is drastically smaller. This is a
consequence of the fact that the notion of \textquotedblleft Weyl
reflection\textquotedblright\ is far more restrictive for curves which do not
have self-intersection $-2$. Finally, the overall semi-direct product by the
leftmost $\mathbb{Z}_{2}$ factor acts on the configuration of curves by
left/right reflection.

\item In the case where the endpoint configuration does not possess such a
$\mathbb{Z}_{2}$ reflection symmetry, we obtain a quite similar answer for the
automorphism group:
\begin{equation}
\mathrm{Aut}(\Lambda_{\text{end}})=\mathbb{Z}_{2}\times(S_{m_{1}+1}%
\times...S_{m_{l}+1}). \label{Autobots}%
\end{equation}

\end{itemize}

Let us note that in all cases, the automorphisms of the endpoint configuration
takes the general form%
\begin{equation}
\mathrm{Aut}(\Lambda_{\text{end}})=\mathcal{O}_{\text{end}}\ltimes
\mathcal{W}_{\text{end}},
\end{equation}
where $\mathcal{O}_{\text{end}}$ are possible automorphisms in the diagram
describing the endpoint configuration of curves, and $\mathcal{W}_{\text{end}%
}$ is a normal subgroup of $\mathrm{Aut}(\Lambda_{\text{end}})$ naturally
generalizing the Weyl group. To illustrate the above notions, let us now turn
to some explicit examples.

\noindent\textbf{Examples}

As a first example, consider the endpoint configuration of curves:%
\begin{equation}
7232222.
\end{equation}
There is no left/right symmetry, and we have $m_{1}=1,m_{2}=4$, so we
get
\begin{equation}
\mathrm{Aut}(\Lambda_{\text{end}})=\mathbb{Z}_{2}\times S_{2}\times S_{5}.
\end{equation}

Next, consider the endpoint configuration:
\begin{equation}
322223,
\end{equation}
which enjoys a $\mathbb{Z}_2$ reflection symmetry. Here, we have $m_{1}=4$ so we get
\begin{equation}
\mathrm{Aut}(\Lambda_{\text{end}})=\mathbb{Z}_{2}\ltimes\left(  \mathbb{Z}%
_{2}\ltimes S_{5}\right)  .
\end{equation}

Finally, consider the endpoint configuration:
\begin{equation}
2232322.
\end{equation}
for which we have
\begin{equation}
\mathrm{Aut}(\Lambda_{\text{end}})=\mathbb{Z}_{2}\ltimes(\mathbb{Z}_{2}\ltimes
S_{3}\times S_{2}\times S_{3}),
\end{equation}
and the $\mathbb{Z}_{2}$ associated with reflection exchanges the two $S_{3}$ factors.

\subsubsection{D-type Endpoints}

Consider next the D-type endpoint configurations. Much as in the case of the
generalized A-type configurations, all of the automorphisms are inherited from
the automorphisms of the D-type configuration with just $-2$ curves.

Recall that in the case of the $D_{N}$ Dynkin diagram with $-2$ curves, the
automorphism group is:
\begin{equation}
\text{Aut}(D_{N}) = \mathcal{O}_{D_{N}} \ltimes\mathcal{W}_{D_{N}},
\end{equation}
where for $N \geq5$, the outer automorphism group is $\mathbb{Z}_{2}$, and for
$N = 4$ it is $S_{3}$. The Weyl group automorphisms are given by:
\begin{equation}
\mathcal{W}_{D_{N}} = S_{N} \ltimes(\mathbb{Z}_{2})^{N} / \mathbb{Z}_{2}.
\end{equation}
here, the overall quotient by $\mathbb{Z}_{2}$ is from the kernel of the map
$(\mathbb{Z}_{2})^{N} \rightarrow\mathbb{Z}_{2}$ given by multiplication of
all factors.

Proceeding now to the more general endpoint configurations, we calculate the
automorphism group by recognizing that all automorphisms are given by
appropriate subgroups of the $\text{Aut}(D_{N})$ series. The key point is that
the diagram breaks up into pieces partitioned by the $-n$ curve(s) with $n >
2$. The only reflection on such curves is given by the long element of the
$D_{N}$ Weyl group, and it acts via a $\mathbb{Z}_{2}$ group action.
Additionally, we see that the rest of the diagram now breaks up into at most
one D-type diagram for $-2$ curves, and an A-type Dynkin Diagram. Generically,
these discrete gauge symmetries decompose as
\begin{equation}
\mathcal{W}_{D-type} = \mathbb{Z}_{2} \ltimes(\mathcal{W}_{D_{N}}
\times\mathcal{W}_{A_{M}})
\end{equation}
where $N$ and $M$ denote the number of $-2$ curves present in the
configuration. For a smaller number of curves, additional possibilities are
present. For example, we can consider the $D_{4}$ Dynkin diagram as well as
the endpoint where the central curve is a $-3$ curve instead.

Again, we observe that in all cases, the automorphisms of the endpoint
configuration take the general form:%
\begin{equation}
\mathrm{Aut}(\Lambda_{\text{end}})=\mathcal{O}_{\text{end}}\ltimes
\mathcal{W}_{\text{end}},
\end{equation}
where $\mathcal{O}_{\text{end}}$ are possible automorphisms in the diagram
describing the endpoint configuration of curves, and $\mathcal{W}_{\text{end}%
}$ is a normal subgroup of $\mathrm{Aut}(\Lambda_{\text{end}})$ which is a
natural generalization of the Weyl group.

\subsection{Higgs Branch Tuning \label{ssec:Higgs}}

Our discussion so far has focused on the structure of the tensor branch moduli
space and the automorphism group for the lattice of string charges. In the
context of physical applications, it is important to understand the interplay
between the tensor branch and Higgs branch moduli. In geometric terms, these
correspond to K\"{a}hler moduli and complex structure moduli, respectively.
More precisely, the complex structure are joined by the intermediate Jacobian
of the Calabi-Yau in determining the structure of the Higgs branch moduli space.

Additional complex structure moduli can appear through suitable tuning of
coefficients in the Weierstrass model. For example, in the case of a
configuration of $-2$ curves realizing an A-type Dynkin diagram, we can
consider various fiber enhancements, leading to a rich structure of possible
6D\ SCFTs \cite{Heckman:2015bfa}:
\begin{equation}
\lbrack\mathfrak{su}_{k_{0}}]\overset{\mathfrak{su}_{k_{1}}}{2}%
,\overset{\mathfrak{su}_{k_{2}}}{2},...,\overset{\mathfrak{su}_{k_{T-1}}%
}{2},\overset{\mathfrak{su}_{k_{T}}}{2}[\mathfrak{su}_{k_{T+1}}],
\label{eq:2squiver}%
\end{equation}
where we have indicated the Lie algebra over each $-2$ curve, as dictated by
the singular elliptic fibrations. To the left and the right, we have also
indicated non-compact flavor branes. Anomaly cancellation requires
$2k_{i}=k_{i-1}+k_{i+1}$ for all of the gauge groups supported on compact $-2$ curves.

Now, depending on the nature of our fiber enhancements, we see that the the
$\mathbb{Z}_{2}$ automorphism corresponding to left/right reflection on the
configuration of $-2$ curves may no longer be a symmetry of the geometry.
Indeed, in the above example we would also need to require $k_{i}=k_{T+1-i}$
for such a reflection symmetry to hold. We take this to mean that some
of the candidate automorphisms originating from the lattice of string charges
may be broken by Higgs branch moduli.

\subsection{RG\ Flows}

It is also natural to study the behavior of the automorphism group under
RG\ flows from one conformal fixed point to another. In a 6D\ SCFT,
supersymmetry preserving flows are limited to deformations triggered by
background operator vevs \cite{Cordova:2016xhm} (see also
\cite{Heckman:2015ola, Heckman:2015axa}). Under a tensor branch flow, we
decompactify some of the curves of the base. Doing so, we see that we pass to
a sublattice:
\begin{equation}
\Lambda_{\text{IR}}\subset\Lambda_{\text{UV}}.
\end{equation}
Even so, we cannot quite say that the automorphisms of one are always
contained in the other. Indeed, we can already see there could be emergent
discrete gauge symmetries in the infrared. To illustrate, consider the case of
the 6D\ SCFT with endpoint $3,3$. After performing one blowup, we reach a
consistent F-theory base, namely $4,1,4$. The automorphism group of this
configuration is:%
\begin{equation}
\text{Aut}(\Lambda_{4,1,4})=\left(  \mathbb{Z}_{2}\ltimes%
%TCIMACRO{\U{2124} }%
%BeginExpansion
\mathbb{Z}
%EndExpansion
_{2}\right)  \times%
%TCIMACRO{\U{2124} }%
%BeginExpansion
\mathbb{Z}
%EndExpansion
_{2}.
\end{equation}
We can also study the automorphism group obtained from decompactifying one of
our curves. Due to the symmetry of the configuration, it is enough to consider
the decompactification of either a $-4$ curve, or a $-1$ curve. In these two
cases, we reach the automorphism groups:%
\begin{align}
\text{Aut}(\Lambda_{4,1})  &  =%
%TCIMACRO{\U{2124} }%
%BeginExpansion
\mathbb{Z}
%EndExpansion
_{2}\times%
%TCIMACRO{\U{2124} }%
%BeginExpansion
\mathbb{Z}
%EndExpansion
_{2}\\
\text{Aut}(\Lambda_{4\oplus4})  &  =\mathbb{Z}_{2}\ltimes\left(
%TCIMACRO{\U{2124} }%
%BeginExpansion
\mathbb{Z}
%EndExpansion
_{2}\times%
%TCIMACRO{\U{2124} }%
%BeginExpansion
\mathbb{Z}
%EndExpansion
_{2}\right)  .
\end{align}
Whereas the group structure of the first case is directly inherited from that
of the original UV theory, the case of two disconnected $-4$ curves is not of
this type. For example, when we have two disconnected $-4$ curves, we can
independently reflect these two curves. This is not possible in the original
$4,1,4$ theory.

More generally, we see that when we decompactify a curve of an endpoint
configuration, there is a strict containment relation for the automorphism groups
of the endpoints:
\begin{equation}
\text{Aut}(\Lambda^{\text{end}}_{\text{IR}})\subset\text{Aut}(\Lambda^{\text{end}}_{\text{UV}}).
\end{equation}
If, however, we decompactify a curve which is only present after blowing up an
endpoint, then we must entertain the possibility of emergent discrete gauge
symmetries in the infrared.

Consider next the case of Higgs branch flows. In these
cases, we see that if we start at a tuned point on the Higgs branch,
then a flow to a generic point will land us on a fixed point which may enjoy
different symmetries. For instance, the theory shown in (\ref{eq:2squiver})
will not generally have a left-right reflection symmetry when the fibers are
tuned to give non-trivial gauge groups. However, a Higgs branch flow to an
infrared theory with trivial Higgs branch yields a $(2,0)$ theory, which does have a
$\mathbb{Z}_{2}$ reflection symmetry.

More generally, we see that in a Higgs branch flow to an infrared theory with trivial Higgs branch,
there is a match between the automorphisms of the base lattice, and the automorphisms of the physical
theory. This is simply because in such situations, the minimal resolution of the endpoint configuration of curves
dictates the resolved geometry of the base, and no tuning of complex structure moduli takes place in this procedure.

\section{Tensor Branch Moduli Space \label{sec:TENSOR}}

Starting from one choice of consistent vevs for the $t^{i}$, it is natural to ask whether there
is a group action akin to what is found for the $(2,0)$ theories. This turns
out to be far more subtle in the case of $(1,0)$ theories, and we will
encounter various generalizations which depend both on the nature of the
blowups and how we identify the Weyl group and the outer automorphisms of the system.

There are various ways in which we can
decompose the automorphism group into a collection of \textquotedblleft outer
automorphisms\textquotedblright\ and a Weyl group action, to write%
\begin{equation}
\text{Aut}(\Lambda)=\mathcal{O}_{H}\ltimes\mathcal{W}_{H}. \label{AutDecomp}%
\end{equation}
Here, we have included a subscript $H$ to remind us that the particular choice
of decomposition we take will be dictated by the ambient values of the complex
structure moduli.

Geometrically, the group action $\mathcal{O}_{H}$ corresponds to isometries in
the base of the F-theory model. Since all such bases are resolutions of
orbifold singularities of the form $\mathbb{C}^{2}/\Gamma_{U(2)}$ for $\Gamma$
a finite subgroup of $U(2)$, the behavior of this group action can be studied
by working in the asymptotic limit far from the actual singularity. The group
action $\mathcal{W}_{H}$ instead parameterizes redundancies in our resolution,
namely they generate discrete gauge transformations.

Let us now turn to the structure of the tensor branch moduli space. Since we
are always considering blowups of a singularity of the form $\mathbb{C}%
^{2}/\Gamma_{U(2)}$ for $\Gamma$ a finite subgroup of $U(2)$, the
divisors $\alpha_{i}$ with intersection pairing $\alpha_i \cap \alpha_j = -A_{ij}$
define generators for the Mori cone of effective divisors,
which we write as:\footnote{We thank A. Grassi and D.R. Morrison for
helpful discussions on this point.}
\begin{equation}
\mathcal{C}_{\text{Mori}} = \{ t^i \alpha_i | t^{i} \geq 0 \}.
\end{equation}

Dual to the Mori cone is the K\"ahler cone:
\begin{equation}
\mathcal{C}_{\text{K\"ahler}} = \{ t_i \omega^i | t_{i} \geq 0 \},
\end{equation}
where we have introduced two-forms $\omega^{i}\in H_{\text{cpct}}^{1,1}(\mathcal{B})$ with compact support which satisfy:
\begin{equation}
\underset{\alpha_j}{\int} \omega^{i} = \delta^{i}_{j}.
\end{equation}
The K\"ahler form for the base $\mathcal{B}$ is:
\begin{equation}
J= t_{i}\omega^{i}.
\end{equation}
Observe that the inverse of the intersection pairing appears via:
\begin{equation}
(A^{-1})^{ij} = - \underset{\mathcal{B}}{\int} \omega^{i} \wedge \omega^{j}.
\end{equation}
Alternatively, we can introduce generators:
\begin{equation}
\omega_{i} = A_{ij} \omega^{j},
\end{equation}
so we can instead present the K\"ahler form as:
\begin{equation}
J= t^{i}\omega_{i}.
\end{equation}
The line element for the metric on the tensor branch moduli space is given by:
\begin{equation}
A_{ij} \delta t^{i} \delta t^{j} = \underset{\mathcal{B}}{\int} \delta J \wedge \delta J.
\end{equation}
The K\"ahler cone is specified by $t_{i}\geq0$, whereas the Mori cone has $t^{i} \geq 0$.

In physical terms, we need to demand that all string tensions are non-negative, namely $t_{i} \geq 0$.
We refer to this as the fundamental domain for the moduli space:
\begin{equation}
\mathbb{D}_{0} = \{ t_{i} \geq 0 \}.
\end{equation}
Observe that our positive definite matrix $A_{ij}$ has inverse $(A^{-1})^{ij}$
with all entries positive. This in turn means that
if $t_{i} \geq 0$ for all $i$, we also have:
\begin{equation}
t^{i} = (A^{-1})^{ij} t_{j} \geq 0,
\end{equation}
so in this sense, the physical moduli space is fully specified by positivity
in the K\"ahler cone. This is somewhat different from the
F-theory construction of supergravity theories,
where both positivity in the K\"ahler cone and Mori cone
must be simultaneously imposed \cite{Kumar:2010ru}.

Consider next the group action of $\text{Aut}(\Lambda)$ on the physical theory. In an ``active frame,''
we interpret the $\omega_{i}$ as elements of the vector space $H^{2}_{\text{cpct}}(\mathcal{B} , \mathbb{R})$
which transform as:
\begin{equation}
\omega_{i} \mapsto \sigma^{j}_{i} \omega_{j}.
\end{equation}
A complementary picture which is most convenient for our present purposes is to instead adopt a ``passive frame''
in which the coordinates themselves transform, namely:
\begin{equation}
t_{j} \mapsto (\sigma^{-1})^{i}_{j} t_{i},
\end{equation}
with the transpose $\sigma^{T}$ acting on the dual coordinates $t^{i}$.
Consequently, the chambers of the physical moduli space
are swept out by the orbits of $\mathbb{D}_{0}$
under the group action by $\text{Aut}(\Lambda)$.

Now in the case of the $(2,0)$ theories, acting by the automorphism group
leads to a tessellation of the extended moduli space. For example, the physical
moduli space of vacua is given by%
\begin{equation}
\mathcal{M}_{\text{(2,0) theory}}=\mathbb{R}^{5T}/\mathcal{W}\text{.}%
\end{equation}
Viewed as a $(1,0)$ theory, we can write the tensor branch moduli space as:
$\mathbb{R}^{T}/\mathcal{W}$. Indeed, starting from $\mathbb{D}_{0}$, we
produce all of the other chambers through the orbits of the Weyl group action.
In this description, the group $\mathcal{O}_{ADE}$ specify discrete isometries
of the chamber. That is, they should be viewed as a discrete global symmetries
of the system.\footnote{One might ask whether it is possible to gauge these
discrete symmetries, thus generating new examples of $(2,0)$ theories. The spectrum of local
operators would be the same, but the spectrum of extended objects would be different.
We do not appear to have this freedom in string constructions, so this symmetry would appear to be anomalous
in the field theory. It would nevertheless be interesting to verify this explicitly.}

Turning next to the $(1,0)$ theories, we can now ask a quite similar question
concerning the orbit of the discrete gauge symmetries $\mathcal{W}_{H}$, as
defined in line (\ref{AutDecomp}). In particular, we would like to know
whether we can expect to tessellate the moduli space, or whether there are
forbidden regions of $\mathbb{R}^{T}$ which appear in no orbit.

To aid us in our analysis of this question, we note that the existence of such
phenomena is fully determined by the automorphism group. Indeed, even though
the actual geometry of the moduli space will depend on the precise
decomposition Aut$(\Lambda)=\mathcal{O}_{H}\ltimes\mathcal{W}_{H}$, the
$\mathcal{O}_{H}$ never flip the signs of the K\"{a}hler moduli; they act as
discrete isometries on the fundamental domain $\mathbb{D}_{0}$. Consequently,
they simply map the chamber back to itself, and we are free to consider the
full group action by Aut$(\Lambda)$ in determining the orbit of $\mathbb{D}%
_{0}$.

Suppose, then, that we have two lattices $\Lambda$ and $\Lambda^{\prime}$
related as in line (\ref{LattMap}):%
\begin{equation}
L:\Lambda^{\prime}\rightarrow\Lambda.
\end{equation}
so that automorphisms of the two lattices are related via the transformation:%
\begin{equation}
L^{-1}\cdot\sigma\cdot L=\sigma^{\prime}.
\end{equation}
Precisely because each automorphism maps to another, we see that the
corresponding group action on $\mathbb{R}^{T}$ will be related by conjugation
by $L$, viewed now as a linear map:%
\begin{equation}
L:\mathbb{R}^{T}\rightarrow\mathbb{R}^{T}.
\end{equation}
By construction, this linear map has trivial kernel i.e. it is invertible.

Our plan in this section will be to analyze the variety of phenomena which we
can expect in the extended K\"{a}hler cone. First, we establish that in
theories where the endpoint is either trivial or given by just a collection of
$-2$ curves, there is a tessellation of $\mathbb{R}^{T}$ via orbits of the
fundamental domain. In all other cases, however, we find that the resulting
structure of moduli space is more intricate. We find that when the endpoint
contains at least one curve of self-intersection $-n$ for $n>2$, that there are
\textquotedblleft forbidden zones\textquotedblright\ in
$\mathbb{R}^{T}$ develop which cannot be reached by any element of
$\mathcal{W}$.

\subsection{Tessellating $\mathbb{R}^{T}$}

In this section we study tensor branches which produce a tessellation of
$\mathbb{R}^{T}$ via orbits of $\mathbb{D}_{0}$ under the group action of
Aut$(\Lambda)$. To begin, we suppose that we have managed to find a lattice
which admits a tessellation of $\mathbb{R}^{T}$. In this situation, the tensor
branch moduli space will be%
\begin{equation}
\mathcal{M}=\mathbb{R}^{T}/\mathcal{W}\text{,}%
\end{equation}
in the obvious notation. Next, suppose that we have another lattice $\Lambda
^{\prime}$ related to this one by a change of basis:
\begin{equation}
L:\Lambda^{\prime}\rightarrow\Lambda.
\end{equation}
Since the generators of the two automorphisms map to one another, we know that:
\begin{equation}
\text{Aut}(\Lambda) \simeq \text{Aut}(\Lambda^{\prime}),
\end{equation}
and moreover, the orbits of the fundamental domains $\mathbb{D}_{0}$ and $\mathbb{D}_{0}^{\prime}$
also map to one another. Consequently, the orbits also match, and a tessellation for one
theory determines a tessellation for the other. Note that in general, however, the resulting orbits could have quite
different structure.

We now show that all theories with trivial endpoint or an endpoint
with just $-2$ curves produce a tessellation of $\mathbb{R}^{T}$. Consider
first the case of a trivial endpoint. After $T$ blowups, we always reach the
same automorphism group:
\begin{equation}
\text{Aut}(%
%TCIMACRO{\U{2124} }%
%BeginExpansion
\mathbb{Z}
%EndExpansion
^{\oplus T})=S_{T}\ltimes\left(  \mathbb{Z}_{2}\right)  ^{T}.
\end{equation}
Now, in the case of independent blowups, namely a collection of $-1$ curves
which do not intersect, the intersection form is proportional to the identity
matrix. In this case, the $S_{T}$ factor acts as an outer automorphism, and is
clearly responsible for permuting the different $-1$ curves. For this theory
of $T$ independent E-strings, the tensor branch moduli space is%
\begin{equation}
\mathcal{M}_{T\text{ E-strings}}=\underset{T}{\underbrace{\mathbb{R}/%
%TCIMACRO{\U{2124} }%
%BeginExpansion
\mathbb{Z}
%EndExpansion
_{2}\times...\times\mathbb{R}/%
%TCIMACRO{\U{2124} }%
%BeginExpansion
\mathbb{Z}
%EndExpansion
_{2}}}\text{,}%
\end{equation}
where the $S_{T}$ acts as a permutation on the different factors. This clearly
yields a tessellation of $\mathbb{R}^{T}$.

Contrast this with the case of the $1,2,...,2$ configuration, in which there
are no outer automorphisms. In this situation, all of the automorphisms are
discrete gauge symmetries, and the Weyl group is just that of the Lie algebra
$\mathfrak{sp}_{T}$. We again get a tessellation of moduli space, but the
structure of the moduli space is quite different:%
\begin{equation}
\mathcal{M}_{\text{Rank }T\text{ E-string}}=\mathbb{R}^{T}/\mathcal{W}%
(\mathfrak{sp}_{T})\text{.}%
\end{equation}

Additional examples include all of the conformal matter theories. For example,
the theories with $G\times G$ global symmetry are given by the configurations
of curves:%
\begin{align}
D_{N}\times D_{N}  &  \text{:}\text{ }1\\
E_{6}\times E_{6}  &  \text{:}\text{ }1,3,1\\
E_{7}\times E_{7}  &  \text{:}\text{ }1,2,3,2,1\\
E_{8}\times E_{8}  &  \text{:}\text{ }1,2,2,3,1,5,1,3,2,2,1.
\end{align}
In all of these cases, we expect the left/right symmetry to actually be a
discrete gauge symmetry of the tensor branch. To see why, it is helpful to
consider other blowup patterns, such as the configurations of curves:
\begin{equation}
1,\overset{1}{4},1\text{ \ \ and \ \ }1,\overset{1}{\underset{1}{5}},1.
\end{equation}
These respectively admit an $S_{3}$ and $S_{4}$ symmetry. However, these are
not really global symmetries, since they can be viewed as permutations present
in the theory of $T-1$ independent E-strings in which the common $E_{8}$
flavor symmetry has been gauged. Indeed, this interpretation is compatible
with the fact that there is no normal subgroup $\mathcal{W}$ of Aut$(\Lambda)$
such that Aut$(\Lambda)/\mathcal{W}$ is given by these would be
\textquotedblleft outer automorphisms.\textquotedblright

Consider next the case of endpoints with just $-2$ curves. If we perform no
blowups, then we simply have the standard ADE\ Weyl group action on
$\mathbb{R}^{T}$. We can also perform blowups, in which case we again get a
tessellation of $\mathbb{R}^{T}$.

\subsection{Forbidden Zones}

For more general endpoint configurations, we find that the group action on the
fundamental domain does not yield a tessellation of $\mathbb{R}^{T}$. It could
happen that there are certain points of $\mathbb{R}^{T}$ which lie in no orbit
of the automorphism group.

We now establish that forbidden zones occur whenever we have at least two
curves in the endpoint configuration, one of which has self-intersection $-n$
with $n>2$. Denote by $\Lambda_{\text{end}}$ the corresponding lattice. To
establish this, we recall that the automorphism group Aut$(\Lambda
_{\text{end}})\varsubsetneq$ Aut$(\Lambda)$ is a proper subgroup of the one we
would obtain by replacing our $-n$ curve by a $-2$ curve. Here, $\Lambda$
denotes the lattice obtained by replacing all curves with self-intersection
less than $-2$ by $-2$ curves, namely an A- or D-type root lattice.

Consequently, $\mathbb{R}^{T}$ can be decomposed into the Weyl chambers
generated by $\mathcal{W\subset}$Aut$(\Lambda)$, the corresponding Weyl group.
Indeed, the Weyl group acts transitively on these Weyl chambers so we know
that there is actually a one to one correspondence between elements of the
Weyl group $\mathcal{W}$ and these chambers.

But precisely because the group action on the $t_{i}$ is the same for elements
of Aut$(\Lambda_{\text{end}})$ and Aut$(\Lambda)$, we see that the orbits
swept out by Aut$(\Lambda_{\text{end}})$ will necessarily be a proper subset
of those swept out by Aut$(\Lambda)$. Consequently, we conclude that we cannot
tessellate $\mathbb{R}^{T}$. In fact, we can also identify the
forbidden zones:\ They are all the images generated by elements $\sigma
\in\Lambda\backslash\Lambda_{\text{end}}$. The full forbidden zone is then
given by:%
\begin{equation}
\mathbb{D}_{\text{forbidden}}=\underset{\sigma\in\Lambda\backslash
\Lambda_{\text{end}}}{%
%TCIMACRO{\dbigcup }%
%BeginExpansion
{\displaystyle\bigcup}
%EndExpansion
}\sigma\left(  \mathbb{D}_{0}\right)  .
\end{equation}
Note that some of these orbits may have common points in the closure other
than the origin. The number of connected components in the moduli space is
simply the order of $\left\vert \mathcal{W}_{\text{end}}\right\vert $:%
\begin{equation}
\left\vert \text{Orbit}_{\Lambda_{\text{end}}}\left(  \mathbb{D}_{0}\right)
\right\vert =\left\vert \mathcal{W}_{\text{end}}\right\vert .
\end{equation}

To illustrate the above considerations, it is helpful to now study a few
examples. Consider, for example, an endpoint configuration such as $3,3$ or $7,7$.
In this case, the automorphism group of the endpoint is $\mathbb{Z}_{2}%
\ltimes\mathbb{Z}_{2}$, and the analog of the Weyl group is $\mathbb{Z}_{2}$.
Labelling the moduli as $t_{1}$ and $t_{2}$, the orbit of the fundamental
domain is:%
\begin{equation}
\text{Orbit}_{\Lambda_{\text{end}}}\left(  \mathbb{D}_{0}\right)
=\{t_{1},t_{2}>0\}\cup\{t_{1},t_{2}<0\}.
\end{equation}
By inspection, the forbidden zone is:%
\begin{equation}
\mathbb{D}_{\text{forbidden}}=\{t_{1}>0\text{ ; }t_{2}<0\}\cup\{t_{1}<0\text{
; }t_{2}>0\}.
\end{equation}

As a somewhat more involved example, consider an endpoint configuration such
as $3,2$. Labelling the modulus of the $-3$ curve by $t_{1}$ and that of the
$-2$ curve by $t_{2}$, we now have that the orbit of the fundamental domain
is:%
\begin{equation}
\text{Orbit}_{\Lambda_{\text{end}}}\left(  \mathbb{D}_{0}\right)
=\{t_{1},t_{2}>0\}\cup\{t_{1},t_{2}<0\}\cup\{t_{1}+t_{2}>0\text{ ; }%
t_{2}<0\}\cup\{t_{1}+t_{2}<0\text{ ; }t_{2}>0\}.
\end{equation}
The forbidden zone is:%
\begin{equation}
\mathbb{D}_{\text{forbidden}}=\{t_{1}+t_{2}<0\text{ ; }t_{1}>0\}\cup
\{t_{1}+t_{2}>0\text{ ; }t_{1}<0\}.
\end{equation}

\section{Compactification \label{sec:COMPACT}}

So far, our analysis has focused on the formal structure of Green-Schwarz
automorphisms, and in particular, their role in dictating the geometry of the
tensor branch moduli space. Much as in the case of the $\mathcal{N}=(2,0)$
theories, it is natural to expect that these automorphisms are also important
in compactifications to lower-dimensional systems.

Now, as we have already remarked, there is a natural sense in which the
automorphisms organize into discrete gauge and global symmetries. In this
sense, we can always introduce a decomposition of the automorphism group as%
\begin{equation}
\text{Aut}(\Lambda)=\mathcal{O}_{H}\ltimes\mathcal{W}_{H}.
\end{equation}
Again, we have introduced the subscript $H$ to indicate that this
decomposition depends on the Higgs branch moduli of the physical theory. There
are then two separate effects we would like to trace in any compactified theory.

First, there is the impact of the discrete gauge symmetries associated with
the factor $\mathcal{W}_{H}$. Roughly speaking, this factor controls the
geometry of the moduli space of vacua. Additionally, in configurations with
non-trivial deformations to $\mathcal{N}=1$ theories, these symmetries play
the role of Seiberg-like duality transformations between IR\ theories.

Second, there is the impact of the global symmetries $\mathcal{O}_{H}$.
Another aim of this section will be to deduce necessary consistency conditions
for ``discrete quotient'' by these symmetries. In the context of
$(1,0)$ theories, this procedure can be carried out in a variety of
dimensions, and is in weakly coupled settings associated with the presence of
various orientifold planes. The local Gauss' law constraint can sometimes also
require additional branes to be present. Let us emphasize that this appears to be a
distinct notion from the case of adding discrete twist lines to a class $\mathcal{S}$ theory, this
being more associated with adding a chemical potential for the discrete symmetry.\footnote{We thank T.T. Dumitrescu for
helpful discussions on this point.}

Our plan in this section is as follows. For specificity we focus on the special case of compactifications of
the class $\mathcal{S}_{\Gamma}$ theories \cite{DelZotto:2014hpa, Gaiotto:2015usa, Heckman:2016xdl}.
The discrete gauge symmetries of the 6D theory lead, for 4D vacua to Seiberg-like dualities, and in 2d
vacua lead to twisted sectors labelled by conjugacy classes of the
discrete gauge symmetries. For the global symmetries, gauging in lower
dimensions also leads to new lower-dimensional theories obtained from
\textquotedblleft discrete quotients\textquotedblright\ of the original construction. To
study consistent ways to perform such quotients, we focus on the geometric
realization afforded by F-theory compactification to consistently track the
effects in both the base and fiber of the model.

\subsection{Descendants of the 6D\ Weyl\ Group}

We now turn to the effects of the analog of the Weyl group in compactifications of 6D SCFTs.
We first consider the case of 4D theories obtained from
compactification on a Riemann surface, and then turn to 2d theories obtained
from compactification on a four-manifold.

\subsubsection{4D\ Theories}

As a first class of examples, we consider the impact of discrete gauge
transformations on the structure of compactified theories. To set the stage,
it is helpful to have in mind the case of $N$ M5-branes probing an A-type
singularity $\mathbb{C}^{2} / \mathbb{Z}_{k}$.
As is by now well-known, this leads, on the tensor branch, to an
F-theory model in which the base is:%
\begin{equation}
\lbrack\mathfrak{su}_{k}]\overset{\mathfrak{su}_{k}}{2},\overset{\mathfrak{su}%
_{k}}{2},...,\overset{\mathfrak{su}_{k}}{2},\overset{\mathfrak{su}_{k}%
}{2}[\mathfrak{su}_{k}],
\end{equation}
a theory with $N-1$ tensor multiplets. In this case, the automorphism group is given by%
\begin{equation}
\text{Aut}(\Lambda)=\mathbb{Z}_{2}\ltimes S_{N},
\end{equation}
with $S_{N}$ the permutation group on $N$ letters which acts via the standard
Weyl reflections. In terms of the M5-brane picture, the automorphisms
correspond to moving the M5-branes past one another.

Compactifying this tensor branch deformation on a $T^{2}$ yields a 4D
$\mathcal{N}=2$ quiver gauge theory. For a theory with $N-1$ simple gauge
group factors, each node has gauge group $SU(k)$. Furthermore, the matter
content of each non-abelian gauge theory factor consists of $F=2k$
hypermultiplets in the fundamental representation. Consequently, we also have
a superconformal field theory in four dimensions. In this case, the motion of
the M5-branes is rather trivial, and leads us back to the same theory. We can,
however, consider various mass deformations as we descend to four dimensions.
Additionally, we can tilt the branes by effectively activating a non-trivial
superpotential deformation for the Coulomb branch scalar of the $\mathcal{N}%
=2$ vector multiplet. This sort of operation, and the resulting duality
cascades \cite{Klebanov:2000hb} were considered in reference
\cite{Cachazo:2001sg} where it was found that the Weyl group transformations
then generate a sequence of Seiberg-like dualities \cite{Seiberg:1994pq} as we
flow from the UV\ to the IR.

Assuming that we have generated an appropriate $\mathcal{N}=1$ theory with general ranks
for the gauge groups, the reason for a Seiberg-like duality is as follows. Consider the 7-branes wrapped
over the $-2$ curves. In this setup, the resulting homology class for all the
7-branes is%
\begin{equation}
\left[  \Sigma\right]  =k_{1}\alpha_{1}+...+k_{N - 1}\alpha_{N-1},
\end{equation}
where the $\alpha_{i}$ denote simple roots and $k_{i}-1$ is the rank of each
$SU$ factor. Upon applying a Weyl group transformation on the $i^{th}$ node
(assuming it is in the middle of the quiver), we have the transformation (see
e.g. \cite{Cachazo:2001sg}):%
\begin{align}
\alpha_{i} &  \rightarrow-\alpha_{i}\\
\alpha_{i+1} &  \rightarrow\alpha_{i+1}+\alpha_{i}\\
\alpha_{i-1} &  \rightarrow\alpha_{i-1}+\alpha_{i},
\end{align}
and the coefficient multiplying $\alpha_{i}$ shifts to $k_{i-1}+k_{i+1}-k_{i}%
$, i.e., $F_{i}-k_{i}$, where $F_{i}$ is the number of flavors in the
fundamental representation.

Given this, it is quite natural to ask whether there is an analogous
Seiberg-like duality for 4D theories obtained from M5-branes probing a D- or
E-type singularity. Again, we observe that with no mass deformations switched
on, permuting the M5-branes simply takes us back to the same 4D $\mathcal{N}%
=2$ SCFT. We can, of course, entertain $\mathcal{N}=1$ deformations, as well
as deformations which break conformal symmetry. We expect that in this broader
context, there is a natural generalization of Seiberg duality now using
compactifications of 6D conformal matter. From the perspective of F-theory
compactification, one complication is that now, the charges of seven-branes
are mutually non-local, so the abelian transformation rule given above must be
modified. We leave the development of this intriguing possibility for future work.

\subsubsection{2d Theories}

Another way in which these discrete gauge symmetries show up is in
compactification to two dimensions. Indeed, as has been appreciated in string
constructions, Seiberg-like dualities, as realized by brane maneuvers can be
extended to a variety of dimensions. Some caution is warranted, however
because the quantum dynamics in the infrared can be quite different depending
on the dimensionality of the resulting theory.

Along these lines, we can also consider the compactification of 6D\ SCFTs on
four-manifolds. In the case of $(1,0)$ theories on a K\"ahler surface, this
yields a class of 2d theories with $\mathcal{N}=(0,2)$ supersymmetry
\cite{Apruzzi:2016nfr, Schafer-Nameki:2016cfr, Apruzzi:2016iac}.
Compactification of the tensor branch leads to a class of theories known as
``DGLSMs'' \cite{Apruzzi:2016nfr}, which are a generalization of a $(0,2)$
gauged linear sigma model in which the gauge couplings are now dynamical
fields. Orbits of the Weyl group on the tensor branch moduli space descend to
non-trivial transformations on the gauge couplings of a DGLSM. Additionally,
we know that these orbits serve to also define additional twisted sector states.

The first point is that in the special case where there is a tessellation of
the extended K\"ahler cone as $\mathbb{R}^{T}$ so that the physical tensor
branch moduli space is $\mathbb{R}^{T}/\mathcal{W}$, we immediately recognize
that the dynamical gauge couplings of the compactified theory generate twisted
sectors labelled by the conjugacy classes of $\mathcal{W}$. This holds for
blowups of the trivial endpoint as well as all admissible blowups of the
ADE\ endpoints composed of $-2$ curves. In those cases where we do not have
such a tessellation, we anticipate additional strongly coupled phenomena to be
present, for example, possible singularities as in the case of
conifold points in a conformal field theory. We have already classified the
6D\ SCFTs where forbidden zones can occur, so we have a clear indication about
when to expect such strongly coupled phenomena.

Finally, we expect that the notion of dualities naturally extends to
trialities \cite{Gadde:2013lxa, Franco:2016nwv, Franco:2016qxh}. Here again,
we expect the discrete gauge symmetries of the 6D tensor branch to
characterize at least part of this structure.

\subsection{Discrete Quotients of a 6D\ SCFT}

So far, we have focused on the discrete gauge symmetries inherited from the
automorphism group. The global symmetries of the 6D also impact the theory,
and its compactifications. In compactifications of the $(2,0)$ theory, adding a background
chemical potential for this symmetry along a one-cycle
is sometimes referred to as introducing a \textquotedblleft discrete
twist,\textquotedblright\ (see e.g. \cite{Vafa:1997mh, Tachikawa:2009rb, Drukker:2010jp, Tachikawa:2010vg}).
There is a conceptually separate notion of quotienting by this symmetry to reach a wholly different theory. In perturbative
string theory terms, this is associated with adding orientifolds. In this section we shall
be interested in this operation.

Our plan in this subsection will be to determine necessary
conditions for quotienting by such discrete symmetries. From the perspective of an
F-theory model, we need to ensure that the symmetry present in the isometries
of the base extends consistently to the fibers of the model. This means that
the full quotient will involve changing both the total number of gauge groups,
as well as the specific gauge groups and matter content present in a given
generalized quiver.

Already in six dimensions, we can see that such quotients can lead to interesting
effects. Now, in the case of the $\mathcal{N}=(2,0)$ theories, we do not
expect to generate any theories other than the ADE\ type ones because all of
these models have a purely geometric realization in terms of IIB\ on an
ADE\ singularity. Once we include various dynamical 7-branes, however, we can
also expect to incorporate both $O7^{-}$-planes and $O7^{+}$-planes. Whereas
in F-theory the $O7^{-}$-planes are fully captured by elliptically fibered
Calabi-Yau threefolds, the case of $O7^{+}$-planes involves \textquotedblleft
frozen\textquotedblright\ singularities \cite{Witten:1997bs, deBoer:2001px,
Tachikawa:2015wka}, and this can generate a small number of additional 6D SCFTs \cite{Hanany:1997gh}.
In reference \cite{Bhardwaj:2015oru} a formal quotienting procedure was proposed to explain such models.
The full set of consistency conditions in this case have yet to be worked out, but we
can already see that this leads to the expected structure in lower-dimensional theories.

In compactifications to lower-dimensional systems, we can extract additional
consistency conditions. For concreteness, we focus on the case of
compactifications to four-dimensional vacua with $\mathcal{N}=2$
supersymmetry. That means we confine our attention to compactification on a
$T^{2}$, with possible quotients also included. For concreteness, we
focus on compactifications of class $\mathcal{S}_{\Gamma}$ theories, since in
such cases the analysis is particularly tractable. Mild deformations of this
case can also be extracted from the general considerations we present.

As we lack a worldsheet construction of F-theory, quotienting by these
discrete symmetries will instead be pieced together through complementary
features. The key idea we shall make use of is that fiber-base duality of the
F-theory geometry can lead to a priori distinct 4D theories which nevertheless
share a common geometric origin \cite{DelZotto:2015rca} (see also \cite{ Bhardwaj:2015oru, Hohenegger:2015btj}).
By construction, we retain all compact two-cycles, so we expect that upon compactification, the
dimensions of the Coulomb branches in theories where
we interchange base and fiber will still match. Indeed, as we already remarked for
theories without a discrete quotient, compactifications of the class
$\mathcal{S}_{\Gamma}$ theories at the conformal point take us to affine
ADE\ type quiver gauge theories, and on the partial tensor branch, take us to
generalized quivers. Each of these theories flows to a 4D\ $\mathcal{N}=2$
SCFT, and due to their common geometric origin, they have the same Coulomb
branch geometry and identical superconformal indices \cite{DelZotto:2014hpa}.
These common features mean that we also expect discrete quotients to
persist on both sides, yielding again a pair of 4D $\mathcal{N}=2$ SCFTs.
Note that this analysis does not require the dimension of the Higgs branches to match,
since in the process of taking appropriate decoupling limits, the number of compact
three-cycles (and thus the number of Higgs branch moduli) could a priori
be different.\footnote{Indeed, for a genuine duality, we ought to
also be able to match the dimensions of the Higgs branches on both sides, since this is
in turn related to the values of the conformal anomalies $a$ and $c$.}

Now, precisely because the Coulomb branches match, we can track the effects of
a discrete quotient in both theories. If we can perform a consistent quotient on
both sides, it is strong evidence that the automorphism of the base actually
extends to the full Calabi-Yau threefold geometry. Whereas the discrete quotients
of the partial tensor branch deformation involves various exceptional group
structures, we see that in the affine quiver theory, we always have $SU$ gauge
groups and the quotient will generate another classical group, namely $SU$, $SO$
or $Sp$. Our first task is therefore to consistently identify which of the
affine quivers can be quotiented, and to then match this data to their
counterparts involving a discrete quotient on the partial tensor branch side.

To guide us in our analysis of quotienting the affine quiver gauge theories, we
note that since this operation does not involve introducing a
mass scale, we expect the quotient to also be a superconformal field theory.
For the quivers obtained using classical gauge groups and matter, this proves
to be quite restrictive, and often leads to a discrete set of possibilities.
In the associated string construction, these constraints are interpreted in
terms of a local Gauss' law constraint for the Ramond-Ramond charge. Consequently, the
quotienting by the discrete symmetry typically involves the presence of both an
orientifold plane and D-branes.

The quotient of an affine quiver gauge theory leads us to a restricted set of 4D
$\mathcal{N}=2$ SCFTs. In particular, based on fiber-base duality of the
associated F-theory geometry, we expect that if it exists, there is \ a
corresponding discrete quotient of the partially tensor branch deformation. In
this case, the string construction involves non-perturbative seven-branes, so
aside from the A-type class $\mathcal{S}_{\Gamma}$ theories, we expect that
the quotient will involve a non-perturbative generalization of orientifold
planes, i.e. another choice of seven-branes. We can, nevertheless, piece
together the structure of the resulting theory by appealing to the form of the
associated affine quiver gauge theory.

\begin{figure}[t!]
\begin{center}
\scalebox{1}[1]{
\includegraphics[trim={0 7cm 0 0},clip,scale=0.5]{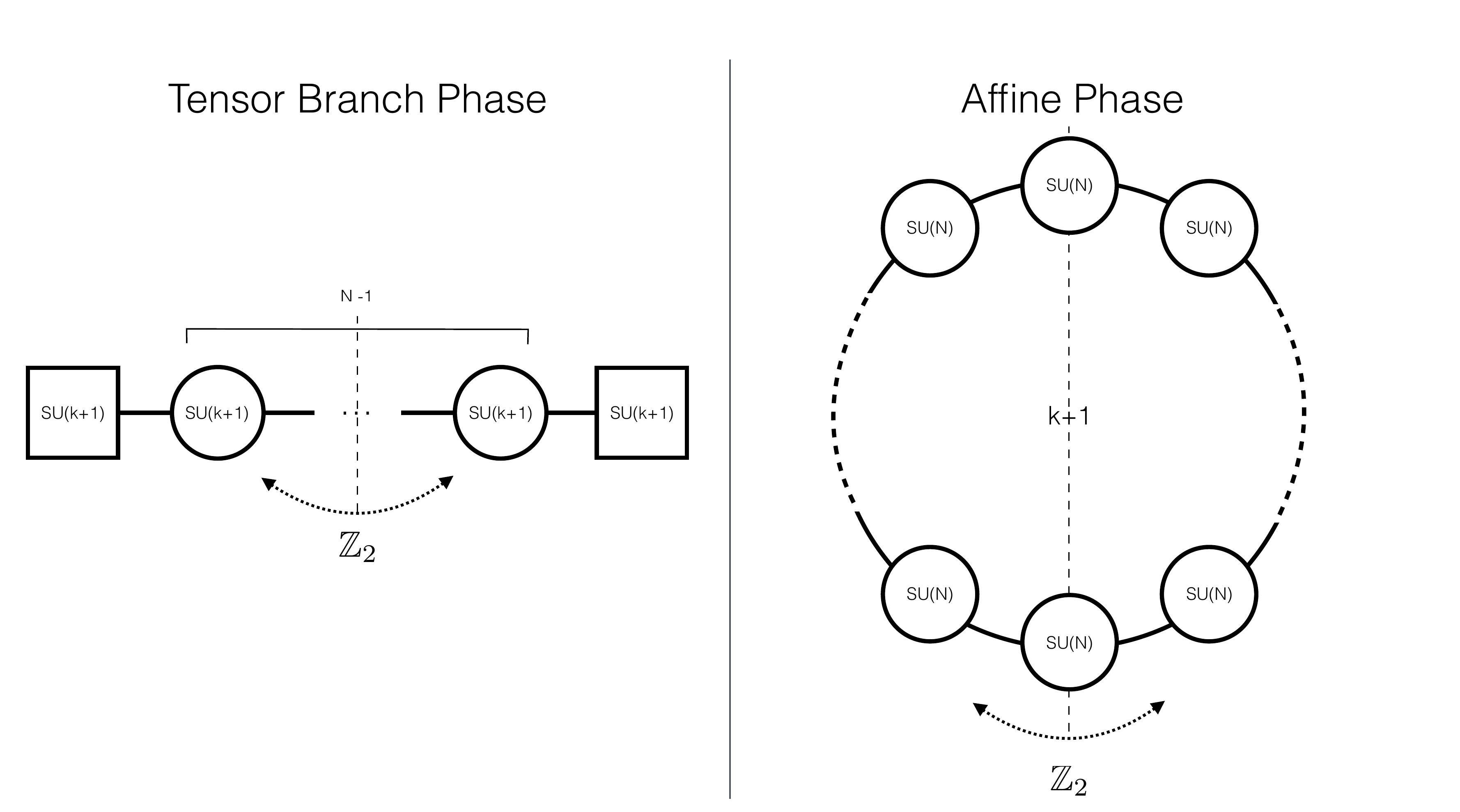}}
\end{center}
\caption{The two quiver phases of A-type class $\mathcal{S}_\Gamma$ theories compactified on a $T^2$. In addition, the effect of quotienting by the global symmetries associated to the outer automorphisms is shown. On the left the reduction on the tensor branch yields $(N-1)$ $SU(k+1)$ gauge nodes and two $SU(k+1)$ flavor symmetry factors. On the right the affine quiver with $(k+1)$ $SU(N)$ gauge nodes. The segments represent standard $\mathcal{N} =2$ hypermultiplets in the bifundamental representation.}
\label{fig:Qs1}
\end{figure}

Since we shall be using the same conditions repeatedly, it is helpful to
collect some general remarks about orientifold projections of the affine
quiver gauge theories in one location. For such theories, a perturbative
string theory analysis is available. For example, the orientifold projection acts on an A-type
symmetric quiver by folding it, as illustrated in figure \ref{fig:Qs1}. We note that when the number
of curves present in the small resolution is odd, then the orientifold action naturally fixes one curve of the geometry,
and an additional node of the corresponding affine quiver is also held fixed. In
the case where the number of curves is even, then all curves of the geometry are interchanged,
but the node present in the affine extension is held fixed. This is all in accord with the partial
tensor branch description, where we have a marked curve in the associated $I_n$ Kodaira-Tate fiber, associated with the
zero section.

Moreover, this orientifold projection maps the $SU$ gauge groups to either
$SO$ or $Sp$ groups, depending on the projection. Since we demand a
construction consistent with an open string construction, large $N$ scaling
already dictates the basic form of these maps:
\begin{align}
&  SU(N)\overset{SO}{\longmapsto}SO(N+\delta_{SO})\\
&  SU(N)\overset{Sp}{\longmapsto}Sp\left(  \frac{N-\delta_{Sp}}{2}\right)  ,
\end{align}
namely the rank decreases by a factor of two, with a possible small shift as
captured by the presence of the $\delta$'s. In the stringy construction, this
is due to the fact that the orientifolds also carry Ramond-Ramond charge, and
this in turn alters the ranks of gauge groups.

Now, our primary focus in this work will be to match various $\mathbb{Z}_{2}$
discrete quotients of the affine quiver phase to corresponding discrete quotients on
the partial tensor branch. For some examples relevant
in the context of 6D SCFTs and their compactification, see for example
\cite{Hayashi:2015vhy, Hanany:1999sj, Bergman:2012kr}.
We also note that there are
various orientifold projections which one can in principle adopt for affine
quivers. As such, we should not expect to find a single orientifold of an
affine quiver, but several possibilities. This is borne out by the fact that
if we only impose the conditions of conformal invariance (by also including
suitable flavors), then we can actually find various sequences of gauge groups
which all lead to conformal fixed points. The fact that there are multiple
choices suggests a richer structure in which we could also attempt to match
various Higgs branch flows. Here, we shall confine our analysis to a few
examples in which we can verify a candidate pair of theories. Indeed, this
suffices for our present purpose, since all we really wish to demonstrate is
that a discrete quotient of the partial tensor branch exists. For this reason, we
shall find it convenient to adopt a \textquotedblleft bottom
up\textquotedblright\ approach where we impose constraints from conformal
invariance, and the dimension of the Coulomb branch, both for the affine
quiver phase, and the partial tensor branch phase.

\subsubsection{4D $\mathcal{N}=2$ SCFTs and Generalized Quivers}

In preparation for our analysis of discrete quotients, in this subsection we collect some
general comments on the types of quiver gauge theories and generalizations we can expect
to encounter. Our guiding principle is to seek out various ways to generate a 4D conformal fixed point
from a 6D theory compactified on a $T^2$. Some such theories have been studied for example in
\cite{Ohmori:2015pua, DelZotto:2015rca, Ohmori:2015pia, Morrison:2016nrt, Bah:2017gph}.

We mainly focus on compactifications of the class
$\mathcal{S}_{\Gamma}$ theories, namely those obtained from $N$ M5-branes
probing an ADE\ singularity. Compactification on a $T^{2}$ yields an SCFT.
There are various orders of limits one can take, depending on whether we
compactify the 6D\ SCFT, or instead a tensor branch
deformation of this theory. We describe each in turn in what follows.

In the case where we remain at the 6D\ conformal fixed point, compactification
on a $T^{2}$ leads us to the theory of $N$ D3-branes probing an
ADE\ singularity. This is described by a quiver gauge theory with gauge groups
$SU(d_{i}N)$, and $d_{i}$ the Dynkin labels of the associated affine Dynkin
diagram of ADE\ type \cite{Douglas:1996sw, Johnson:1996py, Lawrence:1998ja}.
We also have bifundamental hypermultiplets between each node, as dictated by the
structure of the Dynkin diagram.

As is well-known, this yields a class of $\mathcal{N}=2$ SCFTs. Indeed, in our
conventions, the beta function coefficient for an $\mathcal{N}=2$ gauge theory
with classical gauge group with $F$ hypermultiplets in the fundamental
representation is:%
\begin{align}
b(SU(M)) &  =2M-F\\
b(SO(M)) &  =2(M-2)-2F\\
b(Sp(M)) &  =2(2M+2)-2F
\end{align}
where in the above, we have also included the $SO$ and $Sp$ gauge groups as we
shall need them later.\footnote{Recall that in our conventions, $\mathfrak{sp}%
(1)\simeq\mathfrak{su}(2)$.} Observe that for all the affine Dynkin diagram
nodes with gauge group $SU(d_{i}N)$, we have $F=2d_{i}N$, so we indeed realize
a 4D $\mathcal{N}=2$ SCFT. For the ranks and group theory data of the various
perturbative gauge groups see table \ref{tabadabadoo}.

\begin{table}[ptb]
\begin{center}%
\begin{tabular}
[c]{|l|l|l|l|}\hline
$G$ & $SU(k)$ & $Sp(k)$ & $SO(k)$\\\hline
$r_{G}$ & $k-1$ & $k$ & $[k/2]$\\\hline
$\mathrm{Ind}(\mathbf{ Adj})$ & $k$ & $2k+2$ & $k-2$\\\hline
$\mathrm{Ind}(\mathbf{ fund})$ & $\frac{1}{2}$ & $1$ & 1\\\hline
$\mathrm{Ind}(\mathbf{anti})$ & $\frac{k-2}{2}$ & $2k - 2$ & $k - 2$\\\hline
$\mathrm{Ind}(\mathbf{symm})$ & $\frac{k+2}{2}$ & $2k + 2$ & $k + 2$\\\hline
$\mathrm{dim}(\mathbf{fund})$ & $k$ & $2k$ & $k$\\\hline
\end{tabular}
\end{center}
\caption{Relevant group theory data}%
\label{tabadabadoo}
\end{table}

Another way to realize a 4D $\mathcal{N}=2$ SCFT with such gauge groups is to
construct a linear chain of gauge groups. We will find this sort of structure
appearing repeatedly in our analysis of both the original and quotiented theories, so
we collect the relevant points here as well. One such set of examples is
given by taking all of the ranks to be fixed and equal as follows:%
\begin{equation}
\lbrack SU(M)]-SU(M)-...-SU(M)-SU(M)-SU(M)-...-SU(M)-[SU(M)],
\end{equation}
where we have indicated a flavor symmetry in square brackets on the left and
right. Here, each link corresponds to a full hypermultiplet in the
bifundamental representation. There is a natural \textquotedblleft
orientifolded\textquotedblright\ version of this theory given by replacing the
various $SU$ factors by alternating $SO$ and $Sp$ gauge group factors:%
\begin{gather}
\lbrack SO(2a+2)]\overset{1/2}{-}Sp(a)\overset{1/2}{-}...\overset{1/2}{-}%
SO(2a+2)\overset{1/2}{-}Sp(a)\overset{1/2}{-}SO(2a+2)\overset{1/2}{-}%
...\overset{1/2}{-}Sp(a).\overset{1/2}{-}[SO(2a+2)]\\
\lbrack Sp(a)]\overset{1/2}{-}SO(2a+2)\overset{1/2}{-}...\overset{1/2}{-}%
Sp(a)\overset{1/2}{-}SO(2a+2)\overset{1/2}{-}Sp(a)\overset{1/2}{-}%
...\overset{1/2}{-}SO(2a+2).\overset{1/2}{-}[Sp(a)]
\end{gather}
where we have indicated the presence of a half hypermultiplet in the
bifundamental representation by writing $\overset{1/2}{-}$. This is possible
precisely because the bifundamental is in a pseudoreal representation of the
product gauge group. It is also possible to combine all three types of gauge
group factors in a single quiver. An example of this type is:%
\begin{equation}
\lbrack SU(2)]-SU(N)-Sp(N-1)\overset{1/2}{-}SO(2N)\overset{1/2}{-}%
Sp(N-1)\overset{1/2}{-}...,
\end{equation}

There is another way to compactify which yields a class of 4D $\mathcal{N}=2$
SCFTs. This involves first moving onto the partial tensor branch of the 6D
theory, and only then compactifying. Geometrically, we separate the positions
of the M5-branes and then compactify to four dimensions. In doing so, we do
not move onto the tensor branch of the conformal matter sectors. As shown in
references \cite{Ohmori:2015pua, DelZotto:2015rca, Ohmori:2015pia}
the conformal matter descends to a 4D
$\mathcal{N}=2$ SCFT which we view as a generalization of the standard 4D
hypermultiplet. Indeed, it enjoys a flavor symmetry $G_{L}\times G_{R}$. The
contribution to the beta function coefficient of a 4D $\mathcal{N}=2$ gauge
theory with gauge group $G$ of such conformal matter has been computed in
\cite{Ohmori:2015pua, Ohmori:2015pia}, and the result is:\footnote{A note on normalization conventions.
In \cite{Ohmori:2015pua, Ohmori:2015pia}, the contribution to the beta function coefficient from an
$\mathcal{N}=2$ vector multiplet is given as $4h_{G}^{\vee}$, with
$h_{G}^{\vee}$ the dual Coxeter number of the group $G$. For $SU(N)$, this yields $4N$ as
opposed to the \textquotedblleft standard\textquotedblright\ convention of
$2N$ used in the weakly coupled literature.}%
\begin{equation}
b_{G\times G\text{ conf}}(G)=-h_{G}^{\vee},
\end{equation}
where $h_{G}^{\vee}$ is the dual Coxeter number for the gauge group. Now, the
beta function coefficient from the $\mathcal{N}=2$ vector multiplet is:%
\begin{equation}
b_{\text{vec}}(G)=2h_{G}^{\vee},
\end{equation}
so we see that coupling each gauge group to precisely two such conformal
matter sectors leads to a 4D $\mathcal{N}=2$ SCFT \cite{Ohmori:2015pua, Ohmori:2015pia}.

Hence, one can now see that we obtain two a priori different 4D $\mathcal{N}%
=2$ SCFTs from the same class $\mathcal{S}_{\Gamma}$ theory. In the case of
the compactified tensor branch deformation, each node supports the group
$G\times U(1)$, where the $U(1)$ comes from reduction of the tensor multiplet
to four dimensions. Though such $U(1)$ factors flow to weak coupling in the
infrared, the full match of the Coulomb branch dictated by the geometry means
we ought to include these factors in our analysis. Indeed, because of their
common origin in F-theory in which we retain all compact two-cycles, the
resulting 4D theories must have the same Coulomb branch dimension \cite{DelZotto:2014hpa, DelZotto:2015rca}.

The notion of conformal matter can also be generalized beyond ADE\ gauge
groups. For example, given a subgroup $K\subset G$ with imbedding index
$I_{K:G}$, we can also compute the contribution to the beta function of this
subgroup:%
\begin{equation}
b_{G\times G\text{ conf}}(K)=-I_{K:G}\times h_{G}^{\vee}.
\end{equation}

Another generalization along these lines is the compactification of the rank
$Q$ E-string theory on a $T^{2}$. This yields the rank $Q$
Minahan-Nemeschansky theory, which enjoys an $E_{8}$ flavor symmetry. Weakly
gauging this group, the contribution to the beta function is \cite{Aharony:2007dj}:
\begin{equation}
b_{\text{MN}[Q]}(E_{8})=-6Q.
\end{equation}

Finally, we note that compactifying the completely resolved tensor branch deformation of a 6D SCFT
need not generate an SCFT. Instead, we must keep the conformal matter at the origin of its
tensor branch. To illustrate this point, consider the case of a linear quiver with $SO(8)$ gauge groups,
with $D_4 \times D_4$ conformal matter between each gauge group. Geometrically, this conformal matter
is generated by a collapsing $-1$ curve. If we also resolve this $-1$ curve, then compactification of the
fully resolved tensor branch will produce $SO(8)$ gauge groups coupled to no matter fields, and therefore
confines in the infrared. Indeed, to obtain a 4D\ SCFT, we must keep the conformal matter at the origin of the
tensor branch, namely, we only compactify the partial tensor branch deformation.

\subsubsection{A-type $\mathcal{S}_{\Gamma}$ Theories}

Let us begin with discrete quotients of the A-type class $\mathcal{S}_{\Gamma}$
theories. These are realized by $N$ M5-branes probing the transverse geometry
$\mathbb{R}_{\bot}\times\mathbb{C}^{2}/\mathbb{Z}_{k}$, where the total number of tensors is $T=N-1$. Recall that the tensor
branch obtained from separating the M5-branes is given by the 6D F-theory
model:%
\begin{equation}
\lbrack\mathfrak{su}_{k}]\underset{N-1}{\underbrace{\overset{\mathfrak{su}%
_{k}}{2},\overset{\mathfrak{su}_{k}}{2},...,\overset{\mathfrak{su}_{k}%
}{2},\overset{\mathfrak{su}_{k}}{2}}}[\mathfrak{su}_{k}]. \label{Atensor}%
\end{equation}
Compactification on a $T^{2}$ yields a 4D $\mathcal{N}=2$ quiver gauge theory
where each gauge group factor is $SU(k)\times U(1)$. The abelian factors all
flow to weak coupling in the infrared, but the non-abelian factors support a
non-trivial conformal fixed point.

The outer automorphism group of the base is a $\mathbb{Z}_{2}$ symmetry which
amounts to a reflection about the midpoint of this diagram.
We distinguish four different possibilities, depending on whether the axis of
reflection holds fixed a gauge group or a conformal matter link (in this case
a weakly coupled 6D hypermultiplet), and whether the $SU(k)$ gauge groups have
odd or even rank. These choices are captured by $k$ and $N$ even or odd.

To determine the effects of the $\mathbb{Z}_{2}$ quotient of the tensor branch
deformation, we shall now use fiber-base duality to study the closely related
quiver gauge theories obtained from compactification of the 6D conformal fixed
point. Recall that in the absence of a $\mathbb{Z}_{2}$ quotient, the 4D
theory so obtained is given by the circular quiver gauge theory of $SU(N)$
gauge groups:%
\begin{equation}
//\underset{k+1}{\underbrace{SU(N)-...-SU(N)-...-SU(N)}}//,
\end{equation}
where the notation $//$ indicates that we join the left and right by a
hypermultiplet in the bifundamental representation.\ By inspection, we see
that the $\mathbb{Z}_{2}$ symmetry of the tensor branch is also present for
this class of theories. We also see that there is again a natural distinction
between the four distinct choices presented by taking $k$ and $N$ even or odd.

Since in this case the tensor branch admits a weakly coupled
description, we shall find it convenient to also make use of the weakly
coupled IIA\ brane construction of these theories.\ In this case, the sequence
of gauge groups given in line (\ref{Atensor}) are specified by a collection of
NS5-branes for the links with D6-branes suspended between them for the gauge
groups. When we compactify on a $T^{2}$, it is more appropriate to T-dualize
this configuration to that of NS5-branes with D4-branes suspended between each
pair. Applying a $\mathbb{Z}_{2}$ quotient of both sides amounts to
introducing an orientifold plane. For $O6$-planes, there are two general
variants one can consider (see e.g. \cite{deBoer:2001px, Hyakutake:2000mr, Hanany:2000fq, Bergman:2001rp}
given by the $O6^{-}$ or $O6^{+}$-plane. Recall that under a RR $(p+1)$-form in
which the Dp-brane carries $+1$ units of charge, an $Op^{-}$ plane carries
$-2^{p-5}$ units of charge and an $Op^{+}$ plane carries $+2^{p-5}$ units of
charge, other variants being unavailable for $O6$-planes. Now, on the tensor
branch side of the construction, we see that our quotient can therefore only locally satisfy Gauss' law
if we introduce an $O6^{-}$-plane and two D6-branes.
We shall indeed find that this is compatible with the \textquotedblleft bottom
up\textquotedblright\ condition of conformal invariance.

\begin{figure}[t!]
\begin{center}
\scalebox{1}[1]{
\includegraphics[scale=0.45]{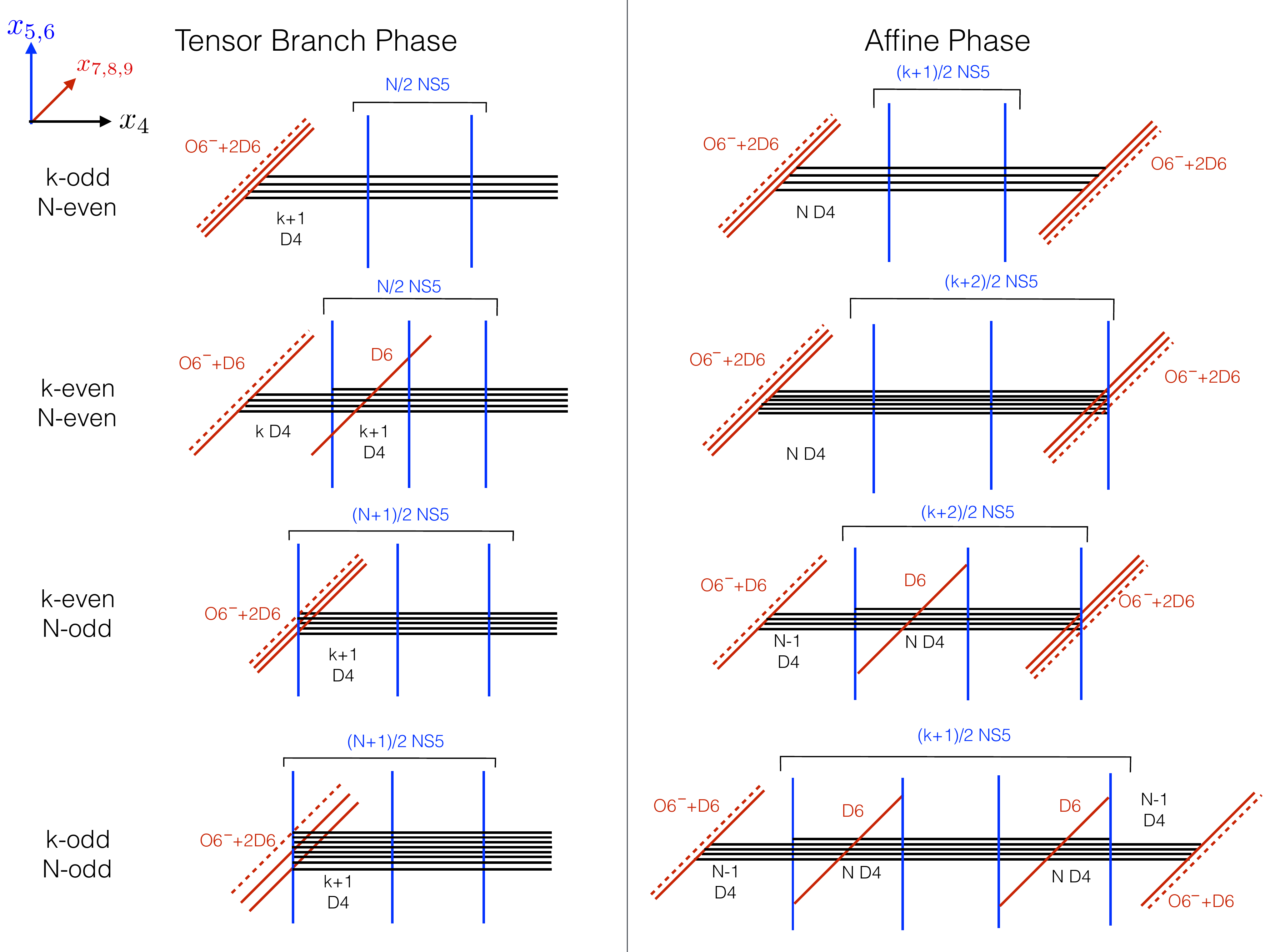}}
\end{center}
\caption{Suspended brane configurations for discrete quotients of the A-type class $\mathcal{S}_\Gamma$ theories on
the tensor branch. We depict linear symmetric suspended branes configurations of D4 (in black)
filling $x^{0},\ldots, x^{3}$ and extended along $x_{6}$, NS5 branes (in blue)
filling $x^{0},\ldots, x^{3}$, wrapping $x^{4},x^{5}$ (i.e. the two torus
directions), and probing $x^{6}$, O6$^{-}$ $+$ 2D6 (in red) filling
$x^{0},\ldots, x^{3}$ and extended along $x_{7},x_{8}, x_{9}$. The other
configurations are given by moving the O6$^{-}$ $+$ $2$D6 on top of an NS5
brane, or by moving one NS5 brane inside the O6$^{-}$ $+$ $2$D6 system
depending on whether $k$ and $N$ are even or odd. In this figure, only the
physical branes have been illustrated. However in the presence of orientifold
$O6^{-}$ there is an equivalent mirror image of the brane system, which in the
affine cases makes the quiver circular, and the $x_{6}$ direction
compact.}%
\label{fig:Bs1}%
\end{figure}

The reduction of the tensor branch quiver to 4D can be described via the
suspended configuration of D4-NS5 branes, where the D4 are extended along
$x^{6}$, with a system of O6$^{-}$ $+$ $2$D6 sitting at $x^{6}=0$, as shown in
figure \ref{fig:Bs1}. The total brane charge of O6$^{-}$ $+$ $2$D6 is zero,
and hence the Gauss' law constraint associated with the charges of these
objects is locally satisfied. There are four distinct possibilities to
consider, depending on whether $k$ and $N$ are respectively even or odd. We
shall therefore step through each possibility in what follows. The basic idea
will be to first consider the D6-branes and $O6^{-}$-plane all on top of each
other, and passing through either the D4-brane (in the case of $N$ even) or
the NS5-brane (in the case of $N$ odd). Moving the D6-branes away from this
fixed locus will then take us to the other possible theories, i.e. the cases
of $k$ even and odd. In figure \ref{fig:Bs1}, we display also other cases
depending on whether $k$ and $N$ are even or odd. For these cases consistency
requires moving the O6$^{-}$ $+$ $2$D6 stack on top of an NS5, or to move an
NS5 brane inside the O6$^{-}$ $+$ $2$D6 system. The affine quiver theory can
be described by the same suspended brane configuration, where now $x_{6}$ is a
compactified $S^{1}$ direction with two O6$^{-}$ $+$ $2$D6 systems at the
opposite ends of the circle. For each case in the tensor branch phase we have
a corresponding affine brane system. It is important to notice that in figure
\ref{fig:Bs1} only the physical branes are shown (namely no images under the
orientifold are included).

\begin{figure}[t!]
\begin{center}
\scalebox{1}[1]{
\includegraphics[scale=0.47]{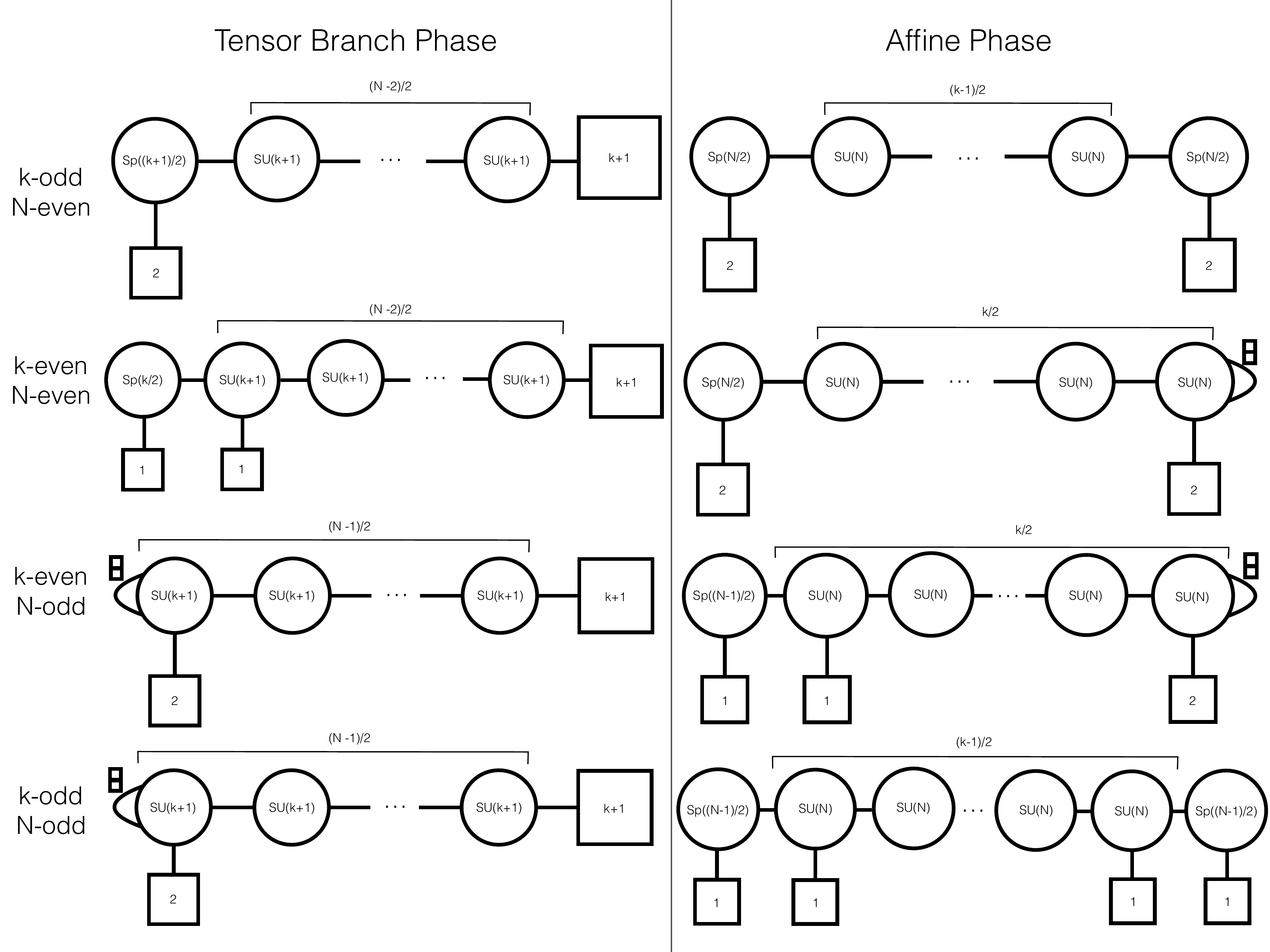}}
\end{center}
\caption{Resulting quiver theories for A-type class $\mathcal{S}_\Gamma$ theories
associated with the brane systems in
\ref{fig:Bs1}. On the left the reduction on the tensor branch to 4D, on the
right the reduction at the fixed point. The possible cases are listed
depending on $k$ and $N$ even or odd. The double box labeling some matter on
the right or left gauge nodes stands for a full antisymmetric hypermultiplet.}%
\label{fig:Qs2}%
\end{figure}

The theories associated with the brane system of figure \ref{fig:Bs1} are
given in figure \ref{fig:Qs2}. Conformal invariance dictates that only $Sp$
groups are allowed. For instance, If we replace the $Sp$ with $SO$ groups,
either the beta function for $SO$ or the one for the close $SU$ group would
have a negative value, and hence the quiver would not be conformal. Note that
this is compatible with the fact that we have $O6^{-}$-planes rather than
$O6^{+}$-planes.

Moreover, having two D6's on top of the O6$^{-}$ plane is consistent with the
flavor symmetries being $SU(2)$ or $U(1)\times U(1)$. Finally, the dimensions
of the Coulomb branches are summarized in table \ref{tab:2}.

\begin{table}[ptb]
\begin{center}%
\begin{tabular}
[c]{|l|l|l|}\hline
& $N$ Even & $N$ Odd\\\hline
$k$ Even & $\frac{kN-k+N}{2}$ & $\frac{kN-k+N-1}{2}$\\\hline
$k$ Odd & $\frac{kN-k+N+1}{2}$ & $\frac{kN-k+N-1}{2}$\\\hline
\end{tabular}
\end{center}
\caption{Dimensions of the Coulomb branch of the various cases}%
\label{tab:2}%
\end{table}

\subsubsection{D-type $\mathcal{S}_{\Gamma}$ Theories}

We now turn to discrete quotients of the D-type class $S_{\Gamma}$ theories. Let
us begin by reviewing the general structure of the D-type theories before
quotientng. These are realized by $N$ M5-branes probing a D-type singularity. Here the total number of tensors is $T=2N-1$. We
shall denote this singularity as $D_{k}$ where we label according to the
associated Dynkin diagram with $k\geq4$ nodes obtained from a small resolution
of the singularity. We move onto the partial tensor branch by separating all
the M5-branes.\ In this phase, each M5-brane defines a 6D conformal matter
sector in the sense of references \cite{DelZotto:2014hpa, Heckman:2014qba}.
The partial tensor branch is described in the F-theory geometry by:%
\begin{equation}
\lbrack\mathfrak{so}_{2k}]\underset{N-1}{\underbrace{\overset{\mathfrak{so}%
_{2k}}{2},\overset{\mathfrak{so}_{2k}}{2},...,\overset{\mathfrak{so}_{2k}%
}{2},\overset{\mathfrak{so}_{2k}}{2}}}[\mathfrak{so}_{2k}].
\end{equation}
By inspection, there is again a $\mathbb{Z}_{2}$ reflection symmetry by which
we can quotient the theory. Compactification on a $T^{2}$ takes us to a 4D
$\mathcal{N}=2$ SCFT. As explained in reference \cite{Ohmori:2015pua, Ohmori:2015pia},
the 6D conformal matter contributes the requisite amount to maintain conformal invariance of
the generalized quiver gauge theory. Indeed, in our normalization conventions,
we have two $D\times D$ conformal matter sectors, and a pair of such sectors
precisely cancels the contribution to the $SO$ beta functions from the
$\mathcal{N}=2$ vector multiplets. This results in a 4D $\mathcal{N}=2$ SCFT
using generalized quivers. Now, the Coulomb branch of such theories is
calculated by including the contributions from both the vector multiplets as
well as the conformal matter sectors. Observe that in contrast to the A-type
case, the matter sectors now contribute non-trivially to the 4D Coulomb
branch, since in the F-theory construction they involve a $-1$ curve with an
$\mathfrak{sp}_{k-4}$ gauge algebra over each such curve. If we instead
compactify the 6D fixed point, we obtained a classical quiver gauge theory
with gauge groups:%
\begin{equation}
SU(N)-\underset{k-3}{\underbrace{\overset{%
\begin{array}
[c]{c}%
SU(N)\\
|
\end{array}
}{SU(2N)}-SU(2N)-...-SU(2N)-\overset{%
\begin{array}
[c]{c}%
SU(N)\\
|
\end{array}
}{SU(2N)}}}-SU(N),\label{Daffine}%
\end{equation}
which again realizes a 4D $\mathcal{N}=2$ SCFT.

By inspection, both phases enjoy a discrete $\mathbb{Z}_{2}$ symmetry, so we
do expect to be able to consistently gauge this symmetry. One complication is
that now on the tensor branch side, we must specify this quotient on either
the $\mathfrak{so}_{2k}$ factor or the conformal matter link. The former case
occurs when $N$ is even while the latter occurs when $N$ is odd.

To facilitate our understanding of these cases, we first study the proposed
group action on the affine quiver gauge theories. Here, we see that the
central spine has a fixed plane, so we expect an $SU(N)$ factor on the left
and right of the quotient of line (\ref{Daffine}), with the fixed spine
composed of an alternating sequence of $SO$ and $Sp$ gauge group factors.
Again, conformal invariance of the entire configuration severely limits the
available possibilities. For example, in the interior of the quotiented spine
of line (\ref{Daffine}), the gauge groups must alternate as
\begin{equation}
...\overset{1/2}{-}Sp(M-1)\overset{1/2}{-}SO(2M)\overset{1/2}{-}%
Sp(M-1)\overset{1/2}{-}SO(2M)\overset{1/2}{-}...
\end{equation}
Additionally, the leftmost and rightmost $SU(2N)$ factors must both be $Sp$
gauge group factors. If they are $SO$ gauge groups instead, we find that the
$SU(N)$ gauge group factors have too much matter to support a conformal fixed
point. This in turn limits us to the special case of $k$ even. For $k$ odd, we
do not find a consistent $\mathbb{Z}_{2}$ quotient. The resulting quotient
leads us to the following quiver gauge theory:%
\begin{equation}
\lbrack SU(2)]-SU(N)-\underset{k-3}{\underbrace{Sp(N-1)\overset{1/2}{-}%
SO(2N)\overset{1/2}{-}...\overset{1/2}{-}SO(2N)\overset{1/2}{-}Sp(N-1)}%
}-SU(N)-[SU(2)].\label{myguy}%
\end{equation}
Let us note that there is another phase compatible with conformal invariance
in which we alternate $Sp(N)$ and $SO(2N+2)$ gauge groups. In this case, the
flavors move to the leftmost and rightmost $Sp$ factors. Anticipating what we
shall find on the partial tensor branch and the dimension of its Coulomb
branch, we shall focus on the case of line (\ref{myguy}) in what follows. In
this case, the complex dimension of the Coulomb branch is:%
\begin{equation}
\dim_{\mathbb{C}}(\text{Coulomb})=Nk-N-\frac{k}{2}-1\text{.}%
\label{dimCoulDtype}%
\end{equation}

Consider next the quotient of the partial tensor branch. Based on our analysis
of the affine quivers, we confine our attention to $k$ even, but with $N$
arbitrary. There are thus two cases to analyze, depending on whether a gauge
group or conformal matter sector is held fixed under the group action.

To start, suppose that an $SO(2k)$ gauge group is held fixed under the
group action. We would like to determine the resulting gauge group from the
quotient $SO(2k)/\mathbb{Z}_{2}$. Now, in an F-theory compactification, there
is a well-known effect known as \textquotedblleft monodromy\textquotedblright%
\ which amounts to applying a quotient on the quotiented affine Dynkin diagram.
For example, this quotient realizes an $Sp(N)$ gauge group from an $SU(2N)$
gauge theory, and produces an $SO(2k-1)$ gauge group from an $SO(2k)$ gauge theory.

The general lesson we would like to extract from our analysis of the A-type
theories, as well as the D-type affine quiver phase is that the $\mathbb{Z}%
_{2}$ quotient has two general effects on the partial tensor branch. First, it
appears to introduce a low rank flavor symmetry, that is, one which does not
scale with the values $N$ and $k$. Additionally, the rank of a gauge group
fixed by the quotient reduces by roughly $1/2$ rather than a constant shift.
We must perform such a reduction in the rank of the gauge group in order to
match the dimension of the Coulomb branch to that present in the affine quiver
phase. This is all compatible with the perturbative type IIA\ construction of
such gauge theories, where we would introduce an additional $O6^{-}$-plane.
The point is that in general, this quotient is a distinct notion from
monodromy of the fiber present in an F-theory construction.

Taking this into account, we conjecture that the quotient of the partial tensor
branch, is, for the case of a fixed $SO(2k)$ gauge group, given by:%
\begin{equation}
\lbrack SO(2)]\overset{1/2}{-}Sp\left(  \frac{k-2}{2}\right)  \underset{%
%TCIMACRO{\QATOP{|}{[Sp(1)\times Sp(1)]}}%
%BeginExpansion
\genfrac{}{}{0pt}{}{|}{[Sp(1)\times Sp(1)]}%
%EndExpansion
}{\overset{CM}{-}}\underset{(N-2)/2}{\underbrace{SO(2k)\overset{CM}{-}%
SO(2k)...\overset{CM}{-}SO(2k)}}\overset{CM}{-}[SO(2k)],\label{Dtensquot}%
\end{equation}
where we have introduced the notation $\overset{CM}{-}$ to denote the
dimensional reduction on a $T^{2}$ of six-dimensional $SO(2k)\times SO(2k)$
conformal matter. This also includes the special case of conformal matter
between the $Sp$ and $SO$ factor at the left. Here, we have gauged the
$Sp\left(  (k-2)/2\right)  $ subgroup of $SO(2k)$, with the commutant
remaining as a global symmetry. This is necessary to have the correct amount
of matter contribute to both the $Sp$ and $SO$ gauge coupling beta functions.
The relevant embedding of subalgebras is:%
\begin{equation}
\mathfrak{so}(2k)\supset\mathfrak{sp}\left(  \frac{k}{2}\right)
\times\mathfrak{sp}(1)\supset\mathfrak{sp}\left(  \frac{k-2}{2}\right)
\times\mathfrak{sp}(1)\times\mathfrak{sp}(1).
\end{equation}
Using this, we can calculate the contribution to the $Sp((k-2)/2)$ beta
function coefficient from the corresponding one for $D_{k}\times D_{k}$
conformal matter:%
\begin{equation}
b_{D_{k}\times D_{k}\text{ conf}}\left(  Sp\left(  \frac{k-2}{2}\right)
\right)  =-h_{SO(2k)}^{\vee}=-(2k-2).
\end{equation}
The total $Sp\left(  \frac{k-2}{2}\right)  $ beta function coefficient is then
given by adding up the contributions from the conformal matter sector, the
vector multiplet, and $F$ hypermultiplets, which we take to be in the
fundamental representation (in accord with the string construction). The
result is:%
\begin{equation}
b_{\text{total}}\left(  Sp\left(  \frac{k-2}{2}\right)  \right)
=2h_{Sp((k-2)/2)}^{\vee}-2h_{SO(2k)}^{\vee}-2F=2k-(2k-2)-2F=2-2F,
\end{equation}
so we must add a half hypermultiplet in the fundamental representation of an $SO(2)$ flavor
symmetry, as indicated in line (\ref{Dtensquot}). As a further piece of
evidence in favor of our conjecture for the quotient, we can also calculate the
dimension of the Coulomb branch, and it indeed matches that of line
(\ref{dimCoulDtype}); we have $N/2$ $D_{k}\times D_{k}$ conformal matter
sectors, each with Coulomb branch dimension $k-3$, $(N-2)/2$ $SO(2k)\times U(1)$ gauge
groups, and one $Sp\left(  \frac{k-2}{2}\right)  \times U(1)$ gauge group. The
total Coulomb branch dimension is therefore:%
\begin{equation}
\dim_{\mathbb{C}}(\text{Coulomb})=\frac{N}{2}(k-3)+\frac{N-2}{2}(k+1)+\frac
{k}{2}=Nk-N-\frac{k}{2}-1\text{,}%
\end{equation}
which precisely matches the dimension of the Coulomb branch in the affine
quiver phase! Note that if we had not reduced the rank of the leftmost $Sp$
gauge group by a factor of $1/2$, the dimension of the quotiented tensor branch
phase would have have been greater than that of the affine quiver. A related
point is that if we had instead attempted to use an $SO(k)$ gauge group rather
than $Sp((k-2)/2)$, we would have encountered an $SO(k)$ flavor symmetry,
contradicting the requirement that the flavor symmetry remains independent of
$k$ and $N$. As a final additional comment, we note that for this construction
to be valid, we require $k$ to be even, a point we already encountered in the
affine quiver phase.

Consider next the case of $N$ even, namely where the $\mathbb{Z}_{2}$ quotient
holds fixed a conformal matter link. Returning to the affine quiver phase, we
see no distinction between the cases $N$ even and odd. This strongly indicates
that the quotient should also make sense in the discrete quotient of the partial
tensor branch. In this case, however, the fixed locus of the quotient will be
a conformal matter sector. Now, for A-type conformal matter, namely weakly
coupled hypermultiplets, we can see that a bifundamental is instead replaced
by a two-index anti-symmetric or symmetric representation (depending on the
type of orientifold plane) of a single $SU$ gauge group.

We shall now conjecture a generalization of such matter fields, as dictated by
consistency with the requirements that we have a 4D\ SCFT, that the dimension
of the Coulomb branch matches that of the affine quiver phase, and that any
flavor symmetries visible in the UV do not scale with the parameters $k$ and
$N$.

\begin{figure}[t!]
\begin{center}
\scalebox{1}[1]{
\includegraphics[scale=0.47]{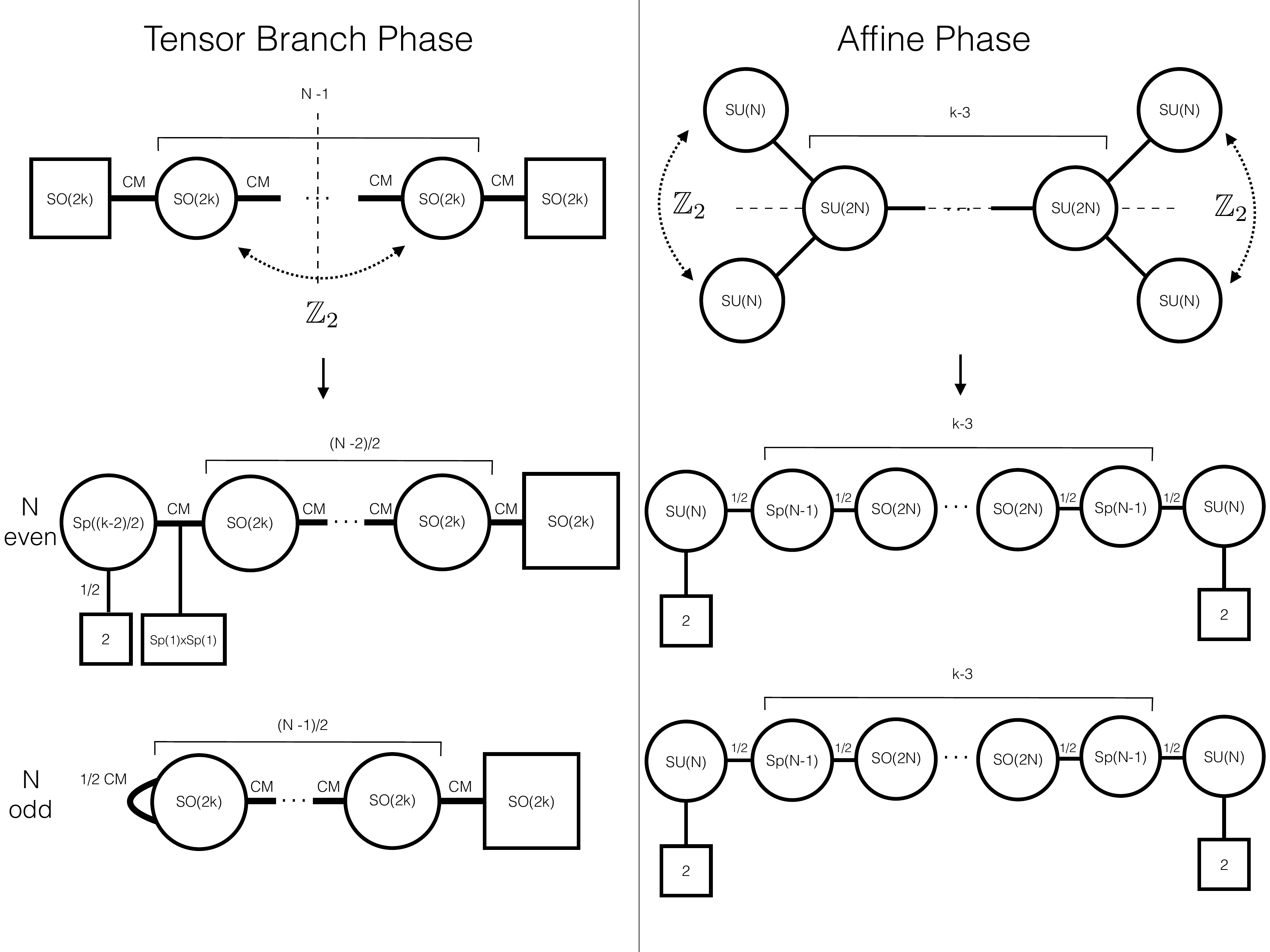}}
\end{center}
\caption{D-type quiver theories. On top the unquotiented theories in the tensor branch phase (left) and affine phase (right). On the bottom the two cases depending on $N$ being even or odd for the two corresponding phases. The segments represents standard $\mathcal N=2$ hypermultiplets, when the $1/2$ appears on a link it stands for half hypermultiplet. The label \textquotedblleft CM\textquotedblright means that it is not just an hypermultiplet, but generalized matter coming from compactification of the conformal matter given by the two connected gauge nodes. }
\label{fig:Qs3}
\end{figure}

To determine the related structure for D-type conformal matter, we first
consider the case of $D_{k}\times D_{k}$ conformal matter compactified on a
$T^{2}$, in which the two $SO(2k)$ factors are flavor symmetries. If, instead
of gauging both such factors, we only gauge the diagonal, we obtain in four
dimensions, an $SO(2k)$ gauge theory where the 4D conformal matter contributes
double that of a $D_{k}\times D_{k}$ conformal matter sector. The reason is
simply that the embedding index is two for the diagonal subgroup. We denote
this system as:%
\begin{equation}
\overset{CM}{\subset}SO(2k)\overset{CM}{-}...,
\end{equation}
in the obvious notation. Observe that we now have effectively three 4D
conformal matter sectors contributing to the $SO(2k)$ beta function, so this
theory will not flow to a conformal fixed point. If, however, we also assume
the existence of a $\mathbb{Z}_{2}$ quotient which acts on the
$\overset{CM}{\subset}$ factor, we again have the requisite amount of matter
to ensure a fixed point for this $SO(2k)$ factor.\footnote{Here we allow for
the definition of a $\frac{1}{2}CM$ sector to have some number of weakly
coupled hypermultiplets in order to ensure conformal invariance of the
$SO(2k)$ sector.} We denote the resulting generalized quiver as:%
\begin{equation}
\overset{\frac{1}{2}CM}{\subset}%
\underset{(N-1)/2}{\underbrace{SO(2k)\overset{CM}{-}...\overset{CM}{-}SO(2k)}%
}\overset{CM}{-}[SO(2k)].
\end{equation}
We should also ask what the resulting contribution is to the dimension of the
Coulomb branch. To address this question, it is helpful to briefly pass to the
tensor branch of the 6D $D_{k}\times D_{k}$ conformal matter. Recall that this
is given by a $-1$ curve which supports an $\mathfrak{sp}(k-4)$ gauge theory
coupled to $4k$ half hypermultiplets in the fundamental representation. We
expect that just as in the other D-type generalized quiver, the $\mathbb{Z}%
_{2}$ quotient decreases the rank of this gauge group factor by roughly $1/2$.

To track in more detail what happens, it is helpful to consider the
dimensional reduction of an isolated $-1$ curve by itself. In six dimensions,
collapsing this curve to zero size generates the rank one E-string 6D SCFT.
Upon reduction on a circle, we obtain an $Sp(1)$ gauge theory with $N_{f}=7$
flavors, which at strong coupling enjoys an $E_{8}$ flavor symmetry
\cite{Morrison:1996xf}. We emphasize that\ geometrically, it is appropriate to view
this as an $Sp(1)$ rather than $SU(2)$ gauge group in the sense that we have
\textquotedblleft already quotiented\textquotedblright\ by the outer
automorphism of $SU(2)$. Now, acting once again by a $\mathbb{Z}_{2}$ quotient
ought to reduce the rank by one half again. This takes us below a rank one
theory, so for this reason we expect this contribution to the Coulomb branch
to vanish for this system. Note, however, that there is still a non-trivial
contribution to the beta function of the weakly gauged flavor symmetry coming
from the matter fields which transformed in the fundamental representation of
$Sp(1)$. Putting this all together, we conclude that the $\frac{1}{2}CM$
sector of type $D_{k}$ will have a Coulomb branch of dimension $(k-4)/2$.

Though it would of course be desirable to have a systematic derivation of
our conjecture, we can already see that it passes a few non-trivial checks.
First, we see that we again retain a conformal fixed point. Additionally, the
dimension of the Coulomb branch receives contributions from the $(N-1)/2$
$SO(2k)\times U(1)$ factors, the $(N-1)/2$ $D_{k}\times D_{k}$ 4D conformal
matter sectors, and the single $\frac{1}{2}CM$ sector. Taking all this into
account, we obtain a Coulomb branch dimension:%
\begin{equation}
\dim_{\mathbb{C}}(\text{Coulomb})=\frac{N-1}{2}(k+1)+\frac{N-1}{2}%
(k-3)+\frac{k-4}{2}=Nk-N-\frac{k}{2}-1\text{,}%
\end{equation}
which matches the dimension found for the affine quiver phase!

There are, however, a few curious features of this theory. Observe
that for $k=4$, the $\frac{1}{2}CM$ sector has no Coulomb branch, but still
contributes to the running of the $SO(2k)$ gauge coupling. It would be most
instructive to better understand this generalized matter sector, either using
methods from IIA\ suspended brane configurations, F-theory, or related
compactifications of class $\mathcal{S}$ theories with discrete quotients.

The theories in the tensor branch and affine phases are summarized in figure \ref{fig:Qs3}.

\subsubsection{E-type $\mathcal{S}_{\Gamma}$ Theories}

Let us now turn to the E-type class $\mathcal{S}_{\Gamma}$ theories and their
discrete quotients. On the partial tensor branch obtained by separating $N$
M5-branes along the $\mathbb{R}_{\bot}$ factor of the transverse geometry
\begin{equation}
\mathbb{R}_{\bot}\times\mathbb{C}^{2}/\Gamma,
\end{equation}
we have the F-theory realization given by:%
\begin{equation}
\lbrack\mathfrak{e}_{k}]\underset{N-1}{\underbrace{\overset{\mathfrak{e}%
_{k}}{2},\overset{\mathfrak{e}_{k}}{2},...,\overset{\mathfrak{e}_{k}%
}{2},\overset{\mathfrak{e}_{k}}{2}}}[\mathfrak{e}_{k}],
\end{equation}
for $k=6,7,8$. Again, there is a $\mathbb{Z}_{2}$ outer automorphism of the
configuration, so we can again ask whether this can be consistently gauged
upon compactification to four dimensions. To address this issue, we again pass
to the affine quiver phase. Now, for the $E_{6}$ case, we see that both the
original Dynkin diagram and its affine extension enjoy a $\mathbb{Z}_{2}$
reflection symmetry. In the $E_{7}$ case, we see that only the affine
extension of the Dynkin diagram enjoys this reflection symmetry. This is
problematic in the context of F-theory constructions, because we typically
select an elliptic fibration with a section. The existence of this section
selects out the affine node, and breaks the symmetry. For this reason, we do
not consider the $E_{7}$ or for that matter, $E_{8}$ cases in what follows.

Let us now study in greater detail the $E_{6}$ case. To begin, we consider the
quiver gauge theory of $N$ D3-branes probing an $E_{6}$ singularity (see e.g.
\cite{Johnson:1996py, Lawrence:1998ja}):
\begin{equation}
SU(N)-SU(2N)-\overset{%
\begin{array}
[c]{c}%
SU(N)\\
|\\
SU(2N)\\
|
\end{array}
}{SU(3N)}-SU(2N)-SU(N).
\end{equation}
Performing a $\mathbb{Z}_{2}$ quotient, we find two possible configurations of
gauge groups:%
\begin{equation}
SU(N)-SU(2N)-\overset{%
\begin{array}
[c]{c}%
Sp(M_{1})\\
\frac{1}{2}\\
SO(M_{2})\\
\frac{1}{2}%
\end{array}
}{Sp(M_{3})},\text{ \ \ or \ \ }SU(N)-SU(2N)-\overset{%
\begin{array}
[c]{c}%
SO(P_{1})\\
\frac{1}{2}\\
Sp(P_{2})\\
\frac{1}{2}%
\end{array}
}{SO(P_{3})},\label{twochoices}%
\end{equation}
where we have omitted possible flavor symmetry factors.

Let us now demonstrate that the latter case in line (\ref{twochoices}) cannot
produce a conformal fixed point. To this end, we note that the condition of
conformal invariance requires all beta function coefficients to vanish.
Writing out the non-trivial constraints, we have the conditions:%
\begin{align}
b(SU(2N))  &  =4N-N-P_{3}-F_{4}\\
b(SO(P_{3}))  &  =2(P_{3}-2)-4N-2P_{2}-2F_{3}\\
b(Sp(P_{2}))  &  =2(2P_{2}+2)-P_{3}-P_{1}-2F_{2}\\
b(SO(P_{1}))  &  =2(P_{1}-2)-2P_{2}-2F_{1},
\end{align}
where we have referred to the number of flavors of the $SU(2N)$ gauge theory
as $F_{4}$. Consider, however, the weighted sum:%
\begin{equation}
b_{\text{sum}}=4b(SU(2N))+3b(SO(P_{3}))+2b(Sp(P_{2}))+b(SO(P_{1})).
\end{equation}
This sum is negative for any non-negative number of flavors, and is given by:%
\begin{equation}
b_{\text{sum}}=-8-4F_{4}-6F_{3}-4F_{2}-2F_{1}<0,
\end{equation}
so we conclude that this sequence of gauge groups will not produce a conformal
fixed point.

Focusing, then, on the remaining case of two $Sp$ factors and a single $SO$
factor, we seek a choice of ranks and flavors which maintains conformal
invariance, and which we can compare with a discrete quotients of a partial tensor branch.

\begin{figure}[t!]
\begin{center}
\scalebox{1}[1]{
\includegraphics[scale=0.47]{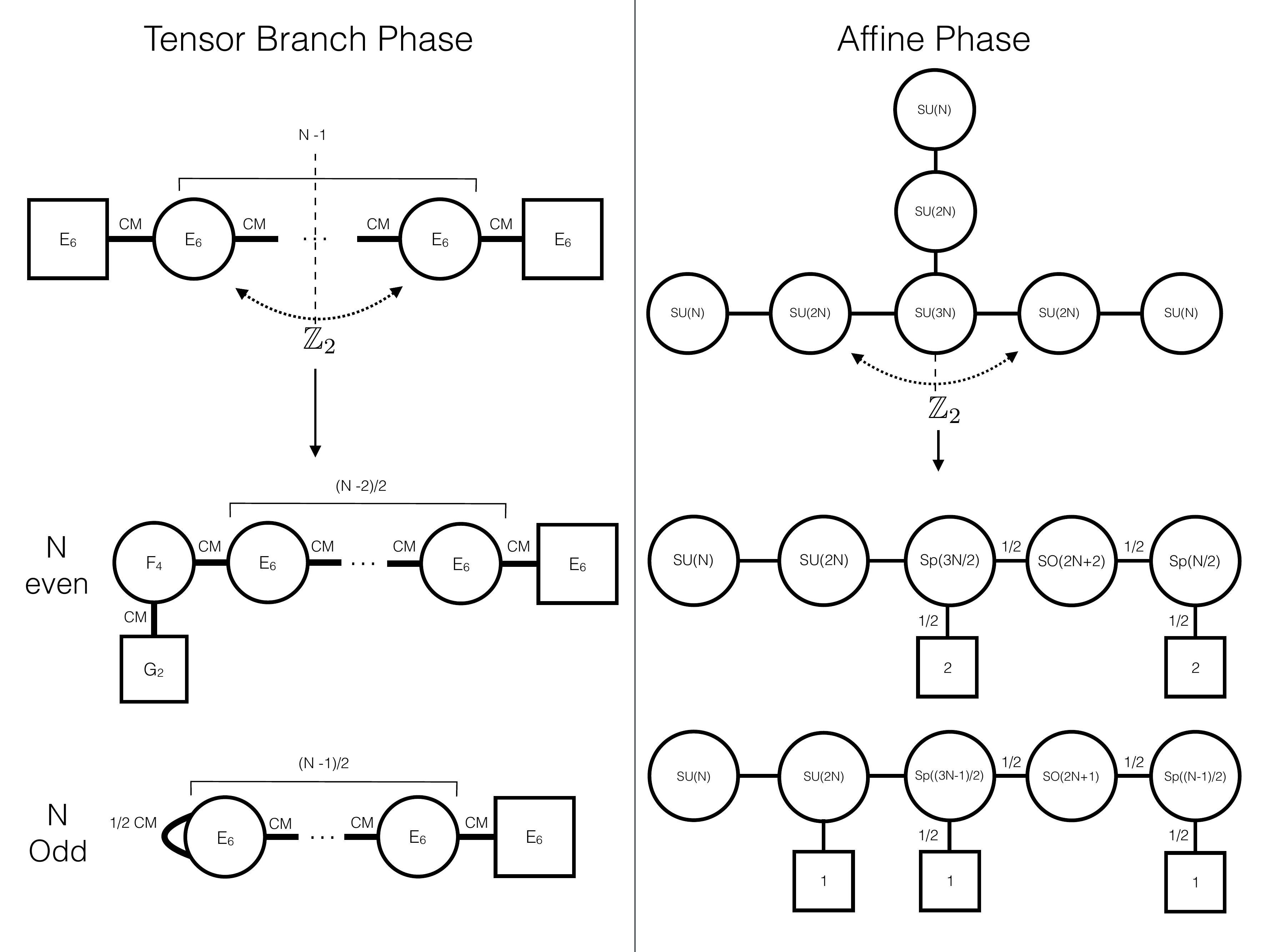}}
\end{center}
\caption{Quiver theories obtained from a discrete quotient of $E_6$-type class $\mathcal{S}_\Gamma$ theories reduced on a $T^2$ with a discrete quotient. On top the unquotiented theories in the tensor branch phase (left) and affine phase (right). On the bottom the two cases depending on $N$ being even or odd for the two corresponding phases. The segments represents standard $\mathcal N=2$ hypermultiplets, when the $1/2$ appears on a link it stands for half hypermultiplet. The label \textquotedblleft CM\textquotedblright denotes
generalized matter coming from compactification of the 6D conformal
matter. }
\label{fig:Qs4}
\end{figure}

As we have already remarked, the possible presence of flavor branes
means there are multiple ways to obtain a conformal fixed
point. Much as in the A- and D-type theories, this also depends on whether $N$
is even or odd. We present some consistent choices in figure \ref{fig:Qs4}. The
particular choice we make anticipates the condition that we can match the
dimension of the Coulomb branch to one obtained from a discrete quotient of the partial tensor branch.
This also depends on whether $N$ is even or odd, and we have in the two cases:
\begin{align}
\dim_{\mathbb{C}}(\text{Coulomb}_{N\text{ even}})  &  =6N-1 \\
\dim_{\mathbb{C}}(\text{Coulomb}_{N\text{ odd}})  &  =6N-3.
\end{align}

We remark that there are additional consistent choices which also produce a
conformal fixed point. The examples presented here are based on the
\textquotedblleft bottom up\textquotedblright\ condition that we match to a
candidate discrete quotient of the partial tensor branch. Much as in the case of
the A- and D-type cases, we can distinguish between whether $N$ is even or
odd, namely whether the discrete quotient holds fixed the gauge group, or the
conformal matter sector.

In the case where $N$ is even, our conjecture for the discrete quotient of the
partial tensor branch is:%
\begin{equation}
\lbrack G_{2}]\overset{CM}{-}F_{4}\overset{CM}{-}%
\underset{(N-2)/2}{\underbrace{E_{6}\overset{CM}{-}...\overset{CM}{-}E_{6}}%
}\overset{CM}{-}[E_{6}].
\end{equation}
In the above, we have identified the $\mathbb{Z}_{2}$ quotient of $E_{6}$ with
the gauge group $F_{4}$. Additionally, the conformal matter between the
$F_{4}$ and $E_{6}$ groups is the standard $E_{6}\times E_{6}$ conformal
matter in which we gauge the $F_{4}$ subgroup (with embedding index one).
We also introduced a rank one Minahan-Nemeschansky theory \cite{Minahan:1996cj}
with $E_{8}$ flavor symmetry in which we gauge the $F_{4}$, retaining a $G_2$ flavor symmetry.
We can calculate the contribution to the $F_{4}$ beta function coefficient from these sectors:%
\begin{equation}
b\left(  F_{4}\right)  =2h_{F_{4}}^{\vee}-h_{E_{6}}^{\vee}-6=0,
\end{equation}
where the contribution from $2h_{F_{4}}^{\vee}$ comes from the $F_4$ vector multiplet,
the contribution $-h_{E_{6}}^{\vee}$ comes from an $(E_6 , E_6)$ conformal matter sector,
and the $-6$ comes from the rank one Minahan-Nemeschansky theory with a gauged $F_4 \subset E_8$. Based
on this, we conclude that all gauge group factors are indeed conformal. Observe also
that the dimension of the Coulomb branch in this case matches that of the
affine quiver phase:%
\begin{equation}
\dim_{\mathbb{C}}(\text{Coulomb}_{N\text{ even}})=6N-1.
\end{equation}

Let us now turn to the case of $N$ odd. Here, a conformal matter sector will
be fixed by the discrete quotient. Now, in the $D_{k}\times D_{k}$ conformal
matter case, we argued that the system%
\begin{equation}
\overset{\frac{1}{2}CM}{\subset}SO(2k)\overset{CM}{-}%
\end{equation}
would retain conformal invariance. We can provide a similar argument in the $E_{6}$ case for the
system:%
\begin{equation}
\overset{\frac{1}{2}CM}{\subset}E_{6}\overset{CM}{-}%
\end{equation}
Here we allow for the definition of a $\frac{1}{2}CM$ sector to have some
number of weakly coupled hypermultiplets in order to ensure conformal
invariance of the $E_{6}$ sector.

We can also calculate the dimension of the Coulomb branch. For $E_{6}\times
E_{6}$ conformal matter, recall that in the resolved phase, we have the tensor
branch description:%
\begin{equation}\label{E6E6confmatt}
1,\overset{\mathfrak{su}_{3}}{3},1,
\end{equation}
which reduces in the 4D theory to a Coulomb branch of dimension five. In anticipation of our discussion of the $\mathbb{Z}_2$
quotient, let us now discuss how to count the Coulomb branch for this theory. First of all, each $-1$ curve contributes a
single tensor multiplet. In the ambient Calabi-Yau threefold, each $-1$ curve is associated with a del Pezzo $9$ surface.
Additionally, the resolved geometry of an isolated $-3$ curve consists of
an affine $\widehat{A}_{2}$ Dynkin diagram of three intersecting Hirzebruch
$\mathbb{F}_{1}$ surfaces \cite{DelZotto:2017pti, Hayashi:2017jze}. We thus see five independent
divisors, and these translate to a count of $h^{1,1}_{\text{cpct}} = 5$ in the local Calabi-Yau geometry.

We now study the $\mathbb{Z}_{2}$ quotient of this theory. By inspection, we see that the two $-1$ curves and their associated surfaces will be identified. Additionally, in the resolution of the $-3$ curve theory with three Hirzebruch surfaces, the $\mathbb{Z}_2$ symmetry of the
affine $\widehat{A}_2$ configuration identifies two of the Hirzebruch surfaces, and leaves the other one invariant (but not pointwise invariant).
So, out of the original five divisors, only three will be independent in the $\mathbb{Z}_2$ quotient. We conclude that the
dimension of the Coulomb branch for the $\mathbb{Z}_2$ quotient of line (\ref{E6E6confmatt}) is:
\begin{equation}
\dim_{\mathbb{C}}(\text{Coulomb[} (1,\overset{\mathfrak{su}_{3}}{3},1)/%
%TCIMACRO{\U{2124} }%
%BeginExpansion
\mathbb{Z}
%EndExpansion
_{2}\text{]})=3.
\end{equation}
Observe that in this case, the $-3$ curve modulus has not been \textquotedblleft
projected out,\textquotedblright\ whereas in the D-type conformal matter case
we did project out the $-1$ curve modulus. Roughly speaking, this is due to
the fact that $3/2>1$, whereas $1/2<1$.

Putting all of this together, we conjecture the following generalized quiver in
the case $N$ odd:%
\begin{equation}
\overset{\frac{1}{2}CM}{\subset}\underset{(N-1)/2}{\underbrace{E_{6}%
\overset{CM}{-}...\overset{CM}{-}E_{6}}}\overset{CM}{-}[E_{6}],
\end{equation}
which has Coulomb branch dimension:%
\begin{equation}
\dim_{\mathbb{C}}(\text{Coulomb}_{N\text{ odd}})=6N-3.
\end{equation}
As a final comment, we note that for the affine quiver phases, the lowest
dimension for the Coulomb branch we could find is $6N-3$, again suggesting
that the dimension of the Coulomb branch for the quotiented conformal matter is
three rather than two.

Based on this, we conclude that a discrete quotient is possible both for $N$ even
and $N$ odd. Again, we stress that we have pieced our analysis together from
various complementary points of view. It would be interesting to study related
examples and extend the correspondence to all possible choices of orientifolds
/ discrete quotients.

The theories in the tensor branch and affine phases are summarized in figure \ref{fig:Qs4}.

\section{Conclusions \label{sec:CONC}}

In this paper we have calculated the discrete gauge
and global symmetries of a 6D\ SCFT generated by automorphisms of the lattice
of strings present on the tensor branch.
These automorphisms capture an important ambiguity in specifying
the structure of Green-Schwarz terms. The ambiguity is resolved
by specifying a chamber of moduli space. We have also elaborated on the similarities and
differences with the case of the $(2,0)$ theories. For example, in the $(1,0)$
case, the resulting moduli space for $T$ tensor multiplets only admits a
tessellation of $\mathbb{R}^{T}$ in rather special circumstances. More
generally, the orbit of the fundamental domain of moduli space may lead to \textquotedblleft forbidden
zones.\textquotedblright\
Finally, we have also taken some preliminary steps in identifying the role of
the automorphism group in compactifications. In the remainder of this section,
we discuss some potential avenues for future investigation.

One of the features which would be quite interesting to understand better is Seiberg-like dualities
for $\mathcal{N} = 1$ gauge theories with exceptional gauge groups and conformal matter.
As we have already noted, the primary complication here is that the transformation rules will need
to involve a non-abelian generalization of the familiar rule $SU(N) \mapsto SU(F - N)$. We anticipate that
the geometric realization in F-theory will provide a route to understand this structure.

We have also seen that the structure of the tensor branch moduli space can be
quite different from the $\mathcal{N}=(2,0)$ case, including the possibility
of forbidden zones. In compactifications to two
dimensions, studying such walls suggests the presence of additional light
states and / or strong coupling effects. It would be most instructive to
understand this issue in explicit examples.

We have also studied some aspects of discrete quotients upon compactification (as well as
sometimes in six dimensions!). In the comparatively simpler setting of
compactifications of class $\mathcal{S}$ theories on a Riemann surface, ``discrete twists,'' namely adding background chemical potentials
for flavor symmetries along one-cycles leads to a broad class of $\mathcal{N} = 2$ theories. There is also an important interplay
between punctures \cite{Gaiotto:2009we} and twists \cite{Drukker:2010jp, Tachikawa:2010vg, Chacaltana:2012ch}.
Extending this to the case of punctures for $(1,0)$ theories such as the class $\mathcal{S}_{\Gamma}$ theories (see e.g.
\cite{Gaiotto:2015usa, Heckman:2016xdl, Razamat:2016dpl}) would seem worthwhile
to investigate further.

\section*{Acknowledgements}

We thank M. Del Zotto, T.T. Dumitrescu, A. Grassi, K. Intriligator, N. Mekareeya, D.R. Morrison, D.S. Park, L. Tizzano, A. Tomasiello,
A. Trimm, and K. Yonekura for helpful discussions. We also thank M. Del Zotto, D.R. Morrison, A. Tomasiello, and A. Trimm
for helpful comments on an earlier draft. JJH thanks the Aspen Center for Physics
Winter Conference on Superconformal Field Theories in $d \geq 4$, NSF grant PHY-1066293,
for hospitality during part of this work. The work of FA and JJH is supported by NSF CAREER grant
PHY-1452037. FA\ also acknowledges support from NSF\ grant PHY-1620311. The
work of TR is supported by NSF grant PHY-1067976 and by the NSF GRF under DGE-1144152.

\appendix

\section{Anomaly Polynomial and Green-Schwarz Redundancy}

In this Appendix we discuss some further details on the structure of the anomaly polynomial,
and redundancy in the Green-Schwarz terms. The main notion
we wish to explore here is the additional structure which results when
the number of simple gauge groups is the same as the number of tensor multiplets. In fact, we will aim
to show how a general form for the anomaly polynomial can be obtained by ``analytic continuation'' in the ranks of the gauge
groups. Along these lines, we shall find it convenient to label the Green-Schwarz terms as:
\begin{equation}
L_{6D}\supset\int\mu_{i}^{j}B^{(i)}\wedge\text{Tr}(F_{(j)}\wedge F_{(j)}),
\end{equation}
where now $\mu$ is a square matrix, and we only include couplings to dynamical gauge fields, i.e.
we drop all couplings to background flavor symmetry field strengths. In this special case, we observe
that the $\mu$'s are square matrices. Moreover, due to the placement of indices, they are
related to our previous presentation of couplings as:
\begin{equation}
\mu_{i}^{k} = \mu_{i,j} (A^{-1})^{jk}.
\end{equation}
In the fundamental chamber where $\mu_{i,j} = A_{ij}$, we have $\mu_{i}^{k} = \delta_{i}^{k}$, the identity matrix.
Indeed, since we can label the chambers of moduli space by elements of the automorphism group, we see that the $\mu$'s can all be
identified with group elements of Aut$(\Lambda)$. Note also that in this basis, both $\mu$ and $\mu^{-1}$ will be integral matrices.

Our plan in this Appendix will be to first review some elements of the anomaly polynomial in general terms,
and then explain how to compute the $\mu$'s in this case via analytic continuation in the ranks of gauge groups.

The anomaly polynomial eight-form of the theory splits into a 1-loop piece and a Green-Schwarz piece.  At the 1-loop level, there is a contribution from the tensor multiplets, the vector multiplets, and the hypermultiplets.  Each tensor multiplet contributes
\begin{equation}
 I_{\rm tens}=\frac{1}{24}\left(c_2(R)^2+\frac{1}{2}c_2(R)p_1(T)+ \frac{1}{240}\left(23 p_1(T)^2 -116p_2(T)\right)  \right),
\label{eq:Itensor}
\end{equation}
where $R$ denotes the $SU(2)$ R-symmetry bundle and $T$ the formal tangent bundle.
From a gauge group $G_i$ with field strength $F_i$, the vector multiplets contribute
\begin{align} \label{eq:apvec}
 I_{{\rm vec}}=&-\frac{{\rm tr}(F_i^4)+ 6 c_2 {\rm tr}(F_i^2)+ d_{G_i} c_2(R)^2 }{24}- p_1(T) \left(\frac{{\rm tr}(F_i^2) + d_{G_i} c_2(R)}{48} \right)-\nonumber \\
 &d_{G_i} \left(\frac{7 p_1(T)^2-4p_2(T)}{5760} \right),
\end{align}
where the trace, ${\rm tr}$ is taken over the adjoint representation of the gauge group $G_i$. Finally, a hypermultiplet in representation $\tilde \rho_i$ of a symmetry with field strength $F_i$ contributes
\begin{align} \label{eq:aphyp0}
 I_{{\rm hyp}}=& \tilde r_i \frac{\left(7p_1(T)^2-4 p_2(T)\right)}{5760}  +\frac{ \text{tr}_{\tilde \rho_i}(F^2_{i} )p_1(T)}{48}  +\frac{\text{tr}_{\tilde \rho_i}(F^4_{i})}{24},
\end{align}
where $\tilde r_i$ is the dimension of the representation $\tilde \rho_i$. Sometimes we can have also half-hypermultiplets.
This will mean that the associated anomaly polynomial contribution is divided by a factor of 2.
There can also be hypermultiplets in mixed representations $\rho_i, \rho_{i+1}$, which contribute
\begin{align} \label{eq:aphyp}
 I_{{\rm hyp-mix}}=& \frac{1}{5760} r_i r_{i+1} \left(7p_1(T)^2-4 p_2(T)\right)+\frac{1}{48} p_1(T) \left(r_i\text{tr}_{\rho_{i+1}}(F^2_{i+1})+r_{i+1}\text{tr}_{\rho_i}(F^2_{i})\right) \nonumber\\
&+\frac{1}{24}\left(r_i\text{tr}_{\rho_{i+1}}(F^4_{i+1})+r_{i+1}\text{tr}_{\rho_i}(F^4_{i})+6 \text{tr}_{\rho_{i+1}}(F^2_{i+1})\text{tr}_{\rho_i}(F^2_{i})\right),
\end{align}
where the trace is taken over the representations $\rho_i$, which is usually the fundamental, and $r_i$ are the dimension of the representations $\rho_i$. The same formula applies if one of the symmetry groups $G_{i}$ is a global symmetry, with the associated $F_i$ now a background field strength.  In IIB language, these flavor symmetries arise from 7-branes wrapping a non-compact component of the discriminant locus.

Following the discussion in Appendix A of reference \cite{Ohmori:2014pca},
we can express the traces of the gauge field strength monomials on some representation as follows
\begin{equation}
{\rm tr}_{\rho_i}(F_i^4) =\alpha_{\rho_{i}} {\rm Tr}(F^4_i) + \frac{3}{4} c_{\rho_{i}} \left({\rm Tr}(F^2_i)\right)^2 \qquad  {\rm tr}_{\rho_i}(F_i^2)= {\rm Ind}(\rho_{i})  {\rm Tr}(F_i^2)
\end{equation}
where Ind is the index of the representation. For the adjoint and the fundamental, these equations become
\begin{align}
&{\rm tr}(F_i^4) =t_{G_{i}} {\rm Tr}(F^4_i) + \frac{3}{4} u_{G_{i}} \left({\rm Tr}(F^2_i)\right)^2 &\qquad & {\rm tr}(F_i^2)= h_{G_i}^\vee {\rm Tr}(F_i^2)\\
&\text{tr}_{fund}(F^4_{i})=\text{Tr}(F^4_{i})&\qquad & \text{tr}_{fund}(F^2_{i}) = s_{G_i} \text{Tr}(F^2_{i}) \ ,
\end{align}
where $\{t_{G_{i}}, u_{G_{i}},  h^{\vee}_{G_i},  s_{G_i} \}$ are the group theory data defined in Appendix A of \cite{Ohmori:2014kda}. In particular, $h_{G_i}^{\vee}$ is the dual Coxeter number of the group $G_i$.  For gauge groups $G_i$ without an independent quartic Casimir, we have $\alpha_{\rho_i} = 0$ for every representation $\rho_i$, and $\text{tr}_{fund}(F^4_{i})=  \frac{3}{4} c_{fund} \left({\rm Tr}(F^2_i)\right)^2 $.

The one-loop contribution of the anomaly polynomial is given by the sum of all these terms
\begin{equation}
I^{1-loop}=I_{\rm tens}+ I_{\rm vec}+ I_{\rm hyp}+ I_{\rm hyp-mix}.
\label{eq:1loop}
\end{equation}
This differs from the prescription of \cite{Ohmori:2014kda} slightly in that we are not including contributions from empty $-1$ curves that are not paired with gauge groups, also known as ``E-strings."  In the following section, we will show that these E-string contributions can be treated via an analytic continuation, so there is no need to place them on a different footing from the rest of the contributions.

In order for the theory to be consistent, we need to cancel all the terms in $I^{1-loop}$ that involve field strengths of gauge groups. This is done by a Green-Schwarz mechanism.  For this, we introduce the intersection pairing on the curves $\Sigma_i$ in $B_2$,
\begin{equation}
A_{ij} = -\Sigma_i \cap \Sigma_j.
\end{equation}
In what follows, we assume that the number of gauge groups is the same as the number of tensor multiplets.  This allows us to conflate gauge group indices with tensor indices, and it also means that the matrix $\mu$ will be square (and in fact invertible). In the following subsection, we will see that by a suitable analytic continuation to $\mathfrak{sp}(0)$ or $\mathfrak{su}(1)$, we can think of all tensor multiplets as being paired with a gauge group, justifying this assumption.

The Green-Schwarz term contribution to the anomaly eight-form reads:
\begin{equation}
I_{GS}= \frac{1}{2} A_{ij}I^i I^j, \qquad i,j=1,\ldots, N_T,
\end{equation}
where the $I^i$ are defined as follows:
\begin{equation} \label{eq:Ii2}
I^i=\left( \sum_{j=1}^{N_T}(\mu^{-1})_{j}^i \frac{{\rm Tr}(F_j^2)}{4}+y_i c_2(R)+K_i\frac{p_1(T)}{12} +z_{i}\frac{{\rm Tr}(F_{(fl)\, i}^2)}{4} \right).
\end{equation}
By charge quantization, the matrix $\mu^{-1}$ must be integral. Gauge anomaly cancellation requires that the combination $I^{1-loop}+I_{GS}$ should be independent of any gauge field strengths ${\rm Tr}(F_j^2)$.  Note that the coefficient in front of $\text{Tr}(F^4_{i})$ cannot be canceled by a GS type mechanism, so this must vanish at 1-loop.  In the final analysis, the anomaly cancellation conditions for each gauge group $G_i$ become \cite{Bershadsky:1997sb}:
\begin{subequations}
\begin{align}
t_{G_i}-\sum_{\rho_i}{\alpha_{\rho_i}n_{\rho_i}}  &  = 0 \label{eq:t} \\
u_{G_i} -\sum_{\rho_i}{c_{\rho_i}n_{\rho_i}}  &  = A_{ii} \label{eq:u} \\
h_{G_i}^\vee -\sum_{\rho_i} {Ind({\rho_{i}}) n_{\rho_{i}}}  &  = -6+3 A_{ii}  \label{eq:h} \\
\sum_{\rho_{i},\rho_{j}}Ind(\rho_{i})Ind(\rho_{j})n_{\rho_{i},\rho_{j}}  &  = -\frac{1}{4} A_{ij}.
\end{align}
\end{subequations}
Here, $n_{\rho_i}$ is the number of hypermultiplets in the $\rho_i$th representation of the gauge group $G_i$, and $n_{\rho_{i},\rho_{j}}$ is the number of hypermultiplets in the mixed representation $(\rho_i, \rho_j)$ of $G_i \times G_j$.  Note that some of the hypermultiplets in the sums in \eqref{eq:t}-\eqref{eq:h} might also be charged under an additional flavor or symmetry group i.e. they might be in mixed representations.

\subsection{All Green-Schwarz Terms via Analytic Continuation} \label{ssec:ancont}

For $-1$ or $-2$ curves without gauge algebras used as matter in a 6D SCFT, one can compute the anomaly polynomial by analytically continuing $Sp(k)$ to $k=0$ and $SU(k)$ to $k=1$, respectively. Indeed, through this analytic continuation, we may compute the GS terms in any phase of any 6D SCFT. We begin with the SCFT quiver as well as the matrix $\mu$ defined in (\ref{muGSagain}).

Next, for a given curve $\Sigma_i$ with $\Sigma_i \cap \Sigma_i = - n$ carrying gauge group $G_i$ with field strength $F_i$, define $I_i^\0$ by the following:
\be
I_i^\0 = h_G^{\vee} c_2(R) + \frac{(n-2)}{4} p_1(T) + \frac{n}{4} \Tr F_i^2  - \sum_{j \in \textrm{nn}} \frac{1}{4} \Tr F_j^2.
\label{eq:I0}
\ee
Here, $h_G^{\vee}$ is the dual coxeter number of $G$, and the sum runs over ``nn,'' the ``nearest neighbors'' of the curve $\Sigma_i$, which are simply the curves $\Sigma_j$ that intersect it at a point.  Note that the coefficient of the gauge field strength ${\rm Tr} F_j^2$ in $I_i^\0$ is given simply by $\frac{1}{4} A_{ij}$.

For an empty $-1$ curve, also known as the rank 1 E-string theory, we use the analytic continuation of $Sp(k)$ to $k=0$ and set $F_i = 0$, yielding
\be
I_{\rm E-string}^\0 = c_2(R) - \frac{1}{4} p_1(T) - \sum_{j \in \textrm{nn}} \frac{1}{4} \Tr F_j^2.
\label{eq:Estring}
\ee
For an empty $-2$ curve, corresponding to the $A_1$ 6D SCFT we use the analytic continuation from $SU(k)$ to $SU(1)$, set $F_i=0$, and associate a global $SU(2)_L$ symmetry with the curve.  This last stipulation amounts to replacing $\sum_{j \in \textrm{nn}} \frac{1}{4}  \Tr F_j^2$ with $c_2(L)$,
\be
I_{A_1}^\0 = c_2(R) - c_2(L).
\label{eq:A1}
\ee
So far, the vectors $I_i^{(0)}$ specify the anomalies for the trivial phase of the geometry, with $\mu= \sigma =Id$.  If we now transform to a different phase by an automorphism $\sigma$, we find a new matrix $\mu = \sigma$.  In this phase, we define the Green-Schwarz vectors $I_i$ by
\be
I_i = \sum_j \mu_{i}^j I_j^\0.
\label{eq:I}
\ee
The Green-Schwarz contribution to the anomaly polynomial is then given by
\begin{equation}
I_{GS} = \frac{1}{2} (A^{-1})^{ij} I_i I_j.
\end{equation}
The full anomaly polynomial is then given by a sum of $I_{GS}$ and $I^{1-loop}$, which is computed via (\ref{eq:1loop}).  For a paired tensor, the rules given in Appendix A of \cite{Ohmori:2014kda} may be used to compute $I^{1-loop}$.  For an unpaired $-1$ tensor, we simply get a contribution from the tensor multiplet, as shown in (\ref{eq:Itensor}).  For an unpaired $-2$ tensor, we add the contribution from the tensor multiplet as well as the contribution from a single free hypermultiplet charged as a half-doublet under the $c_2(L)$ symmetry,
\begin{equation}
I_{free} = \frac{1}{24} c_2(L)^2 + \frac{1}{48} c_2(L) p_1(T) + \frac{7 p_1(T)^2 - 4 p_2(T)}{5760}.
\label{eq:Ifree}
\end{equation}
If the unpaired $-2$ tensor is adjacent to a tensor carrying $SU(2)_G$ gauge symmetry, the $SU(2)_L$ is gauged, and we replace $c_2(L)$ in (\ref{eq:A1}) and (\ref{eq:Ifree}) with $ \frac{1}{4} \Tr (F_{SU(2)_G})^2$.

One might wonder whether this analytic continuation truly gives the full set of allowed Green-Schwarz couplings.  When the number of tensor multiplets is equal to the number of gauge groups (i.e. there are no unpaired tensors), one verifies using line (\ref{eq:Ii2}) that each choice of $\mu$ gives rise to a unique choice for the Green-Schwarz couplings.

In physical theories where there are unpaired tensors, the matrix of couplings for the $\mu$'s are no longer square. By analytic continuation on the groups, however, we can always extend this to a square matrix. This analytic always appears to yield a unique answer. Indeed, the perspective of
F-theory compactification, we note that the geometric phases of the base are completely characterized by the automorphism group of the lattice.  This means that the choices of $\mu$ for a theory with $-1$ or $-2$ curves should be independent of whether these curves are paired with gauge groups or not.  This in turn fixes the Green-Schwarz couplings $I_i$ for unpaired tensors to the values specified by the analytic continuations in (\ref{eq:Estring}) and (\ref{eq:A1}).  Thus, we expect from F-theory that these analytic continuations give the unique Green-Schwarz couplings.

In \cite{Bhardwaj:2015oru}, the possibility of ``outlier" 6D SCFTs that do not admit known F-theory constructions was discussed, and several such theories were proposed.  In particular, one may consider a theory with base
\begin{equation}
4 || 2,2,2,...,2
\end{equation}
where $||$ indicates that the $-4$ curve is tangent to the adjacent $-2$ curve.  In terms of the Dirac pairing, such a tangency implies $A_{12} = A_{21} = -2$.  All of the anomaly cancellation conditions are satisfied for this theory, and (\ref{eq:I0}) is modified straightforwardly by adding a factor of 2 in the sum over nearest neighbors for the tangency, so that the coefficient of the gauge field strength ${\rm Tr} F_j^2$ in $I_i^\0$ is still given by $\frac{1}{4} A_{ij}$.

However, one may also consider an outlier theory with $SU(N)$ gauge group for $N \geq 8$ on a $-1$ tensor with $N_F=N-8$ fundamental hypermultiplets and a symmetric hypermultiplet.  This theory violates \eqref{eq:h} and therefore has non-vanishing gauge-gravitational anomalies at 1-loop, but a suitable Green-Schwarz term can cancel this term.  In particular, all gauge anomalies will cancel if we take
\be
I_1^\0 = N c_2(R) + \frac{1}{4} p_1(T) + \frac{1}{4} \Tr F_{SU(N)}^2  - \frac{1}{4} \Tr F_{N_F}^2.
\label{eq:I0prime}
\ee
Note that the coefficient of $p_1(T)$ is $+1/4$ rather than the $-1/4$ expected from (\ref{eq:I0}).  It appears that the above analytic continuation does \emph{not} apply to such ``outlier" theories. It would be interesting to see whether there is a further refinement in this analysis
by interpreting these outlier theories in terms of a discrete quotient, along the lines presented in this paper.

In the following subsection, we will show how the above formulae work in a handful of simple examples.

\subsection{Examples}

\subsubsection{Rank $Q$ E-string Theory}

Let us consider the rank $Q$ E-string theory. The anomaly polynomial of this theory was computed in \cite{Ohmori:2014pca}:
\begin{equation} \label{eq:apestr}
I^{{\rm E-string}}_{{\rm rank Q}}= \frac{Q^2}{6}\left(c_2(L)- c_2(R) \right)^2 + \frac{Q}{2}\left(c_2(L)- c_2(R) \right) I^2_4 +Q \left(\frac{1}{4}I_4 - I_8 \right) \ .
\end{equation}
The contributions $I_4$ and $I_8$ are given by
\begin{subequations}
\begin{align}
& I_4= \frac{1}{4} \left( {\rm Tr}(F_{E_8}^2) + p_1(T) - 2c_2(L)-2c_2(R)\right) \ , \\
& I_8= \frac{1}{48}\left(c_2(L)-c_2(R))^2+p_2(T)-\frac{1}{4}(2c_2(L)+ 2c_2(R)+p_1(T) \right) \ , \label{eq:apM5}
\end{align}
\end{subequations}
where $F_{E_8}$ is the field strength of the global $E_8$ symmetry for the theory.

We begin with the special case $Q=1$, namely the rank $1$ E-string theory. This can be viewed as the theory of an M5-brane on the M9-wall of Heterotic M-theory.  Here, $c_2(L)$ is the second Chern class of the $SU(2)_L$ bundle associated to the transverse space of the M5 brane, which together with the $SU(2)_R$ makes up the normal bundle associated to the $SO(4)$ global symmetry of the theory.  The $-1$ curve usually appears as generalized matter in the 6D $(1,0)$ theories, and as in \cite{Ohmori:2014kda}, we need to subtract the free hypermultiplet contribution given by
\begin{equation}
 I_{free-hyp}=\frac{1}{24}\left(c_2(L)^2+\frac{1}{2}c_2(L)p_1(T)+ \frac{1}{240}\left(7 p_1(T)^2 -4p_2(T)\right)  \right)\ ,
\end{equation}

Using our formulae from above, we have
\begin{equation} \label{eq:IiEstr}
I^i_{\rm E-string}= c_2(R)-\frac{1}{4}p_1(T)- \frac{1}{4}{\rm Tr}(F_{E_8}^2).
\end{equation}
This gives
\begin{align}
I_{GS}&=\frac{1}{2}(I^i_{\rm E-string})^2 \nonumber \\
&= \frac{1}{2} [ c_2(R)^2 - \frac{1}{2} c_2(R) p_1(T) - \frac{1}{2} c_2(R) {\rm Tr}(F_{E_8}^2) + \frac{1}{16} p_1(T)^2
\nonumber \\
&+ \frac{1}{8} p_1(T) {\rm Tr}(F_{i-1}^2) + \frac{1}{8} p_1(T) {\rm Tr}(F_{i+1}^2) + \frac{1}{16} {\rm Tr}(F_{E_8}^2)^2  ].
\end{align}
To this, we add the $1$-loop piece associated with a tensor multiplet to get
\begin{align}
I_{tot} &= \frac{13}{24} c_2(R)^2  - \frac{11}{48} c_2(R) p_1(T) + \frac{203}{5760} p_1(T)^2 - \frac{29}{1440} p_2(T) -
 \frac{1}{4} c_2(R) \Tr(F_{E_8}^2) \nonumber \\
&+ \frac{1}{16} p_1(T) \Tr(F_{E_8}^2) + \frac{1}{32} \Tr(F_{E_8}^2)^2.
\end{align}
This is precisely the anomaly polynomial of the rank 1 E-string with $I_{free-hyp}$ subtracted off.  Next, we consider the rank $Q$ E-string.  We have
\begin{equation}
A_{ij} = \left[ \begin{array}{cccc} 1 & -1 & 0 &... \\ -1 & 2 &-1 &... \\ \vdots & & & \vdots \\ ... & 0 & -1 & 2   \end{array} \right].
\end{equation}
Using (\ref{eq:Estring}) and (\ref{eq:A1}), we have
\begin{align}
I_1 &= c_2(R) - \frac{1}{4} p_1(T) - \frac{1}{4} \Tr (F_{E_8})^2 \nonumber \\
I_k &= c_2(R) - c_2(L),~~~~k =2,...,Q.
\end{align}
The Green-Schwarz term is then given by $I_{GS} = \frac{1}{2} I_i (A^{-1})^{ij} I_j$.  To this, we add the 1-loop contribution associated with $Q$ tensor multiplets (obtained by taking $N_T$ = Q copies of line (\ref{eq:Itensor}))
and $Q-1$ free hypermultiplets ($Q-1$ times $I_{free}$ of (\ref{eq:Ifree})).  We arrive at
\begin{align}
I_{tot} &= -\frac{1}{24} c_2(L)^2 - \frac{1}{48} c_2(L) p_1(T) - \frac{7}{5760}  p_1(T)^2 + \frac{1}{1440} p_2(T) + \frac{Q}{8}
 c_2(L)^2  \nonumber \\
&+ \frac{Q}{3} c_2(L) c_2(R)  + \frac{Q}{8} c_2(R)^2  - \frac{5 Q}{48}  c_2(L) p_1(T)  -
 \frac{5 Q}{48} c_2(R) p_1(T)  + \frac{7 Q}{192} p_1(T)^2 \nonumber \\
& - \frac{Q}{48} p_2(T)  - \frac{Q^2}{4} c_2(L)^2  + \frac{Q^2}{4}
 c_2(R)^2  + \frac{Q^2}{8} c_2(L) p_1(T)  - \frac{Q^2}{8} c_2(R) p_1(T)  \nonumber \\
& + \frac{Q^3}{6} c_2(L)^2  -
 \frac{Q^3}{3} c_2(L) c_2(R)  + \frac{Q^3}{6} c_2(R)^2  - \frac{Q}{8} c_2(L) \Tr (F_{E_8}^2) -
\frac{Q}{8} c_2(R)  \Tr (F_{E_8}^2) \nonumber \\
& + \frac{Q}{16} p_1(T)  \Tr (F_{E_8}^2) +
\frac{Q^2}{8} c_2(L)  \Tr (F_{E_8}^2) - \frac{Q^2}{8} c_2(R)  \Tr (F_{E_8}^2) +
\frac{Q}{32} \Tr (F_{E_8}^2)^2.
\end{align}
This is precisely (\ref{eq:IiEstr}) with the contribution of a free hypermultiplet subtracted off.

\subsubsection{$SO(10)$-$Sp(1)$}

We next consider the $SO(10)$-$Sp(1)$ gauge theory with Dirac pairing:
\be
A_{ij} = \left[ \begin{array}{cc} 4 & -1 \\ -1 & 1  \end{array} \right].
\ee
We know that there are four phases, given by:
\be
\mu =\sigma= \pm I\,,~~~~ \mu =\sigma= \pm \left[ \begin{array}{cc} 1 & 2 \\ 0 & -1  \end{array} \right].
\ee
For simplicity, we consider only the phase where both $t_1 , t_2> 0 $.
For $\mu = I$, we have $(\mu^{-1})^T = I$, so $I_1 = I_1^\0$, $I_2 = I_2^\0$.  Using (\ref{eq:I0}),
\begin{align}
I_1 &= 8 c_2(R) + \frac{1}{2} p_1(T) + \Tr F_1^2 - \frac{1}{4} \Tr F_L^2 - \frac{1}{4} \Tr F_2^2, \nonumber \\
I_2 &= 2 c_2(R) - \frac{1}{4} p_1(T) + \frac{1}{4} \Tr F_2^2 - \frac{1}{4} \Tr F_R^2 - \frac{1}{4} \Tr F_1^2.
\end{align}
It is easily checked that this produces the correct Green-Schwarz term $I_{\rm GS}= \frac{1}{2} A^{ij} I_i I_j$.

The other solution is just slightly more complicated.  We have
\be
\mu = \left[ \begin{array}{cc} 1 & 2 \\ 0 & -1  \end{array} \right].
\ee
(\ref{eq:I}) yields
\begin{align}
I_1 &= 12 c_2(R) +\frac{1}{2} \Tr F_1^2 + \frac{1}{4} \Tr F_2^2 - \frac{1}{4} \Tr F_L^2- \frac{1}{2} \Tr F_R^2, \nonumber \\
I_2 &= -2 c_2(R) +\frac{1}{4} \Tr F_1^2 - \frac{1}{4} \Tr F_2^2 + \frac{1}{4}  \Tr F_R^2 + \frac{1}{4} p_1(T)^2.
\end{align}
It can also be checked that this reproduces the correct GS term $I_{\rm GS}$. Raising indices with the metric, we find
\begin{align}
I^1 &\supset  \frac{1}{4} (\Tr F_1^2) \nonumber \\
I^2 &\supset \frac{1}{4} (2 \Tr F_1^2 - \Tr F_2^2).
\end{align}
This agrees with charge quantization.  Note also that
\be
\left[ \begin{array}{c} I^1  \\ I^2  \end{array} \right] \supset \frac{1}{4} \left[ \begin{array}{cc} 1 & 0 \\ 2 & -1  \end{array} \right] \left[ \begin{array}{c} \Tr F_1^2  \\ \Tr F_2^2  \end{array} \right] = \frac{1}{4} (\mu^{-1})^T \left[ \begin{array}{c} \Tr F_1^2  \\ \Tr F_2^2  \end{array} \right],
\ee
in accord with line (\ref{eq:Ii2}).

\subsubsection{$SU(N_1)$-$SU(N_2)$ }

We consider now a quiver consisting of two $-2$ curves carrying gauge groups $SU(N_1)$ and $SU(N_2)$, respectively.
In this case, we have:
\be
A_{ij} = \left[ \begin{array}{cc} 2 & -1 \\ -1 & 2  \end{array} \right],
\ee
and
\begin{align}
I_1^\0 &= N_1 c_2(R)  + \frac{1}{2} \Tr F_1^2 - \frac{1}{4} \Tr F_L^2 - \frac{1}{4} \Tr F_2^2, \nonumber \\
I_2^\0 &= N_2 c_2(R)  +  \frac{1}{2} \Tr F_2^2 - \frac{1}{4} \Tr F_R^2 - \frac{1}{4} \Tr F_1^2.
\end{align}
There are a number of different phases.  We will consider simply the one with
\be
\mu = \left[ \begin{array}{cc} 0 & 1 \\ -1 & -1  \end{array} \right].
\ee
So,
\begin{align}
I_1 &= N_2 c_2(R)  +  \frac{1}{2} \Tr F_2^2 - \frac{1}{4} \Tr F_R^2 - \frac{1}{4} \Tr F_1^2, \nonumber \\
I_2 &= -(N_1+N_2) c_2(R)  -  \frac{1}{4} \Tr F_1^2  - \frac{1}{4} \Tr F_2^2+ \frac{1}{4} \Tr F_L^2 + \frac{1}{4} \Tr F_R^2.
\end{align}
This gives the correct GS term, $\frac{1}{2} (A^{-1})^{ij} I_i I_j$, and we have
\begin{align}
I^1 &\supset  \frac{1}{4} (- \Tr F_1^2+ \Tr F_2^2) \nonumber \\
I^2 &\supset \frac{1}{4} (- \Tr F_1^2).
\end{align}
This agrees with charge quantization, and
\be
\left[ \begin{array}{c} I^1  \\ I^2  \end{array} \right] \supset \frac{1}{4} \left[ \begin{array}{cc} -1 & 1 \\ -1 & 0  \end{array} \right] \left[ \begin{array}{c} \Tr F_1^2  \\ \Tr F_2^2  \end{array} \right] = \frac{1}{4} (\mu^{-1})^T \left[ \begin{array}{c} \Tr F_1^2  \\ \Tr F_2^2  \end{array} \right]
\ee

\subsubsection{$Sp(1)$-$SO(10)$-$Sp(1)$}

Finally, we consider a quiver with three simple gauge group factors, namely $Sp(1) \times SO(10) \times Sp(1)$.
We have
\be
A_{ij} = \left[ \begin{array}{ccc} 1 & -1& 0 \\ -1 & 4 & -1 \\ 0 & -1 & 1  \end{array} \right],
\ee
and
\begin{align}
I_1^\0 &= 2 c_2(R)  - \frac{1}{4} p_1(T) + \frac{1}{4} \Tr F_1^2 - \frac{1}{4} \Tr F_L^2 - \frac{1}{4} \Tr F_2^2, \nonumber \\
I_2^\0 &= 8 c_2(R) + \frac{1}{2} p_1(T)  +  \Tr F_2^2 - \frac{1}{4} \Tr F_3^2 - \frac{1}{4} \Tr F_1^2,  \nonumber \\
I_3^\0 &= 2 c_2(R)  - \frac{1}{4} p_1(T)+  \frac{1}{4} \Tr F_3^2 - \frac{1}{4} \Tr F_R^2 - \frac{1}{4} \Tr F_2^2.
\end{align}
There are many choices of $\mu$ that cancel gauge anomalies and satisfy charge quantization.  One choice is
\be
 \mu =  \left[ \begin{array}{ccc} -1 & 0 & 0 \\ 0 & -1 & -2  \\ 0 & 0 & 1 \end{array} \right].
\ee
Using (\ref{eq:I}), we have
\begin{align}
I_1 &= -2 c_2(R)  + \frac{1}{4} p_1(T) - \frac{1}{4} \Tr F_1^2 + \frac{1}{4} \Tr F_L^2 + \frac{1}{4} \Tr F_2^2, \nonumber \\
I_2 &= -12 c_2(R) -  \frac{1}{2} \Tr F_2^2 - \frac{1}{4} \Tr F_3^2 + \frac{1}{4} \Tr F_1^2 + \frac{1}{2} \Tr F_R^2,  \nonumber \\
I_3 &= 2 c_2(R)  - \frac{1}{4} p_1(T)+  \frac{1}{4} \Tr F_3^2 - \frac{1}{4} \Tr F_R^2 - \frac{1}{4} \Tr F_2^2.
\end{align}
This gives the correct $I_{\rm GS}$ and satisfies
\be
\left[ \begin{array}{c} I^1  \\ I^2  \\ I^3 \end{array} \right] \supset \frac{1}{4} \left[ \begin{array}{ccc} -1 & 0 & 0 \\ 0 & -1 & 0  \\ 0 & -2 & 1 \end{array} \right] \left[ \begin{array}{c} \Tr F_1^2  \\ \Tr F_2^2 \\ \Tr F_3^2   \end{array} \right] = \frac{1}{4} (\mu^{-1})^T \left[ \begin{array}{c} \Tr F_1^2  \\ \Tr F_2^2 \\ \Tr F_3^2  \end{array} \right].
\ee

\newpage

\bibliographystyle{utphys}
\bibliography{Auto6DSCFT}

\providecommand{\href}[2]{#2}\begingroup\raggedright\begin{thebibliography}{100}

\bibitem{Vafa:1997mh}
C.~Vafa, ``{Geometric Origin of Montonen-Olive Duality},'' {\em Adv. Theor.
  Math. Phys.} {\bfseries 1} (1998) 158--166,
\href{http://arxiv.org/abs/hep-th/9707131}{{\ttfamily arXiv:hep-th/9707131}}.
%%CITATION = HEP-TH/9707131;%%.

\bibitem{Witten:1997sc}
E.~Witten, ``{Solutions of Four-Dimensional Field Theories via M-theory},''
  \href{http://dx.doi.org/10.1016/S0550-3213(97)00416-1}{{\em Nucl. Phys.}
  {\bfseries B500} (1997) 3--42},
\href{http://arxiv.org/abs/hep-th/9703166}{{\ttfamily arXiv:hep-th/9703166}}.
%%CITATION = HEP-TH/9703166;%%.

\bibitem{Argyres:2007cn}
P.~C. Argyres and N.~Seiberg, ``{S-duality in N=2 supersymmetric gauge
  theories},'' \href{http://dx.doi.org/10.1088/1126-6708/2007/12/088}{{\em
  JHEP} {\bfseries 12} (2007) 088},
\href{http://arxiv.org/abs/0711.0054}{{\ttfamily arXiv:0711.0054 [hep-th]}}.
%%CITATION = ARXIV:0711.0054;%%.

\bibitem{Gaiotto:2009we}
D.~Gaiotto, ``{$\mathcal{N} = 2$ Dualities},''
  \href{http://dx.doi.org/10.1007/JHEP08(2012)034}{{\em JHEP} {\bfseries 08}
  (2012) 034},
\href{http://arxiv.org/abs/0904.2715}{{\ttfamily arXiv:0904.2715 [hep-th]}}.
%%CITATION = ARXIV:0904.2715;%%.

\bibitem{Heckman:2013pva}
J.~J. Heckman, D.~R. Morrison, and C.~Vafa, ``{On the Classification of 6D
  SCFTs and Generalized ADE Orbifolds},''
  \href{http://dx.doi.org/10.1007/JHEP06(2015)017,
  10.1007/JHEP05(2014)028}{{\em JHEP} {\bfseries 05} (2014) 028},
  \href{http://arxiv.org/abs/1312.5746}{{\ttfamily arXiv:1312.5746 [hep-th]}}.
[Erratum: {\textit{JHEP}} {$\mathbf{06}$} (2015) 017].
%%CITATION = ARXIV:1312.5746;%%.

\bibitem{DelZotto:2014hpa}
M.~Del~Zotto, J.~J. Heckman, A.~Tomasiello, and C.~Vafa, ``{6d Conformal
  Matter},'' \href{http://dx.doi.org/10.1007/JHEP02(2015)054}{{\em JHEP}
  {\bfseries 02} (2015) 054},
\href{http://arxiv.org/abs/1407.6359}{{\ttfamily arXiv:1407.6359 [hep-th]}}.
%%CITATION = ARXIV:1407.6359;%%.

\bibitem{Heckman:2015bfa}
J.~J. Heckman, D.~R. Morrison, T.~Rudelius, and C.~Vafa, ``{Atomic
  Classification of 6D SCFTs},''
  \href{http://dx.doi.org/10.1002/prop.201500024}{{\em Fortsch. Phys.}
  {\bfseries 63} (2015) 468--530},
\href{http://arxiv.org/abs/1502.05405}{{\ttfamily arXiv:1502.05405 [hep-th]}}.
%%CITATION = ARXIV:1502.05405;%%.

\bibitem{Apruzzi:2013yva}
F.~Apruzzi, M.~Fazzi, D.~Rosa, and A.~Tomasiello, ``{All $AdS_7$ solutions of
  type II supergravity},''
  \href{http://dx.doi.org/10.1007/JHEP04(2014)064}{{\em JHEP} {\bfseries 04}
  (2014) 064},
\href{http://arxiv.org/abs/1309.2949}{{\ttfamily arXiv:1309.2949 [hep-th]}}.
%%CITATION = ARXIV:1309.2949;%%.

\bibitem{Gaiotto:2014lca}
D.~Gaiotto and A.~Tomasiello, ``{Holography for (1,0) theories in six
  dimensions},'' \href{http://dx.doi.org/10.1007/JHEP12(2014)003}{{\em JHEP}
  {\bfseries 12} (2014) 003},
\href{http://arxiv.org/abs/1404.0711}{{\ttfamily arXiv:1404.0711 [hep-th]}}.
%%CITATION = ARXIV:1404.0711;%%.

\bibitem{Bhardwaj:2015xxa}
L.~Bhardwaj, ``{Classification of 6d $ \mathcal{N}=\left(1,0\right) $ gauge
  theories},'' \href{http://dx.doi.org/10.1007/JHEP11(2015)002}{{\em JHEP}
  {\bfseries 11} (2015) 002},
\href{http://arxiv.org/abs/1502.06594}{{\ttfamily arXiv:1502.06594 [hep-th]}}.
%%CITATION = ARXIV:1502.06594;%%.

\bibitem{Witten:1995ex}
E.~Witten, ``{String theory dynamics in various dimensions},''
  \href{http://dx.doi.org/10.1016/0550-3213(95)00158-O}{{\em Nucl. Phys.}
  {\bfseries B443} (1995) 85--126},
\href{http://arxiv.org/abs/hep-th/9503124}{{\ttfamily arXiv:hep-th/9503124}}.
%%CITATION = HEP-TH/9503124;%%.

\bibitem{Witten:1995zh}
E.~Witten, ``{Some comments on string dynamics},'' in {\em {Future perspectives
  in string theory. Proceedings, Conference, Strings'95, Los Angeles, USA,
  March 13-18, 1995}}.
\newblock 1995.
\newblock
\href{http://arxiv.org/abs/hep-th/9507121}{{\ttfamily arXiv:hep-th/9507121}}.
\newblock
%%CITATION = HEP-TH/9507121;%%.

\bibitem{Strominger:1995ac}
A.~Strominger, ``{Open P-Branes},''
  \href{http://dx.doi.org/10.1016/0370-2693(96)00712-5}{{\em Phys. Lett.}
  {\bfseries B383} (1996) 44--47},
\href{http://arxiv.org/abs/hep-th/9512059}{{\ttfamily arXiv:hep-th/9512059}}.
%%CITATION = HEP-TH/9512059;%%.

\bibitem{Seiberg:1996qx}
N.~Seiberg, ``{Nontrivial fixed points of the renormalization group in
  six-dimensions},''
  \href{http://dx.doi.org/10.1016/S0370-2693(96)01424-4}{{\em Phys. Lett.}
  {\bfseries B390} (1997) 169--171},
\href{http://arxiv.org/abs/hep-th/9609161}{{\ttfamily arXiv:hep-th/9609161}}.
%%CITATION = HEP-TH/9609161;%%.

\bibitem{WittenSmall}
E.~Witten, ``{Small Instantons in String Theory},''
  \href{http://dx.doi.org/10.1016/0550-3213(95)00625-7}{{\em Nucl. Phys.}
  {\bfseries B460} (1996) 541--559},
\href{http://arxiv.org/abs/hep-th/9511030}{{\ttfamily arXiv:hep-th/9511030}}.
%%CITATION = HEP-TH/9511030;%%.

\bibitem{Ganor:1996mu}
O.~J. Ganor and A.~Hanany, ``{Small {$E_8$} instantons and tensionless
  noncritical strings},''
  \href{http://dx.doi.org/10.1016/0550-3213(96)00243-X}{{\em Nucl. Phys.}
  {\bfseries B474} (1996) 122--140},
\href{http://arxiv.org/abs/hep-th/9602120}{{\ttfamily arXiv:hep-th/9602120}}.
%%CITATION = HEP-TH/9602120;%%.

\bibitem{MorrisonVafaII}
D.~R. Morrison and C.~Vafa, ``{Compactifications of F-Theory on Calabi--Yau
  Threefolds -- II},''
  \href{http://dx.doi.org/10.1016/0550-3213(96)00369-0}{{\em Nucl. Phys.}
  {\bfseries B476} (1996) 437--469},
\href{http://arxiv.org/abs/hep-th/9603161}{{\ttfamily arXiv:hep-th/9603161}}.
%%CITATION = HEP-TH/9603161;%%.

\bibitem{Seiberg:1996vs}
N.~Seiberg and E.~Witten, ``{Comments on string dynamics in six-dimensions},''
  \href{http://dx.doi.org/10.1016/0550-3213(96)00189-7}{{\em Nucl. Phys.}
  {\bfseries B471} (1996) 121--134},
\href{http://arxiv.org/abs/hep-th/9603003}{{\ttfamily arXiv:hep-th/9603003}}.
%%CITATION = HEP-TH/9603003;%%.

\bibitem{Bershadsky:1996nu}
M.~Bershadsky and A.~Johansen, ``{Colliding Singularities in F-theory and Phase
  Transitions},'' \href{http://dx.doi.org/10.1016/S0550-3213(97)00027-8}{{\em
  Nucl. Phys.} {\bfseries B489} (1997) 122--138},
\href{http://arxiv.org/abs/hep-th/9610111}{{\ttfamily arXiv:hep-th/9610111}}.
%%CITATION = HEP-TH/9610111;%%.

\bibitem{Brunner:1997gf}
I.~Brunner and A.~Karch, ``{Branes at orbifolds versus Hanany Witten in
  six-dimensions},''
  \href{http://dx.doi.org/10.1088/1126-6708/1998/03/003}{{\em JHEP} {\bfseries
  03} (1998) 003},
\href{http://arxiv.org/abs/hep-th/9712143}{{\ttfamily arXiv:hep-th/9712143}}.
%%CITATION = HEP-TH/9712143;%%.

\bibitem{Blum:1997fw}
J.~D. Blum and K.~A. Intriligator, ``Consistency conditions for branes at
  orbifold singularities,'' {\em Nucl. Phys. B} {\bfseries 506} (1997)
  223--235,
\href{http://arxiv.org/abs/arXiv:hep-th/9705030}{{\ttfamily
  arXiv:hep-th/9705030}}.
%%CITATION = HEP-TH/9705030;%%.

\bibitem{Aspinwall:1997ye}
P.~S. Aspinwall and D.~R. Morrison, ``{Point - like instantons on K3
  orbifolds},'' \href{http://dx.doi.org/10.1016/S0550-3213(97)00516-6}{{\em
  Nucl. Phys.} {\bfseries B503} (1997) 533--564},
\href{http://arxiv.org/abs/hep-th/9705104}{{\ttfamily arXiv:hep-th/9705104}}.
%%CITATION = HEP-TH/9705104;%%.

\bibitem{Intriligator:1997dh}
K.~A. Intriligator, ``{New string theories in six-dimensions via branes at
  orbifold singularities},'' {\em Adv. Theor. Math. Phys.} {\bfseries 1} (1998)
  271--282,
\href{http://arxiv.org/abs/hep-th/9708117}{{\ttfamily arXiv:hep-th/9708117}}.
%%CITATION = HEP-TH/9708117;%%.

\bibitem{Hanany:1997gh}
A.~Hanany and A.~Zaffaroni, ``{Branes and six-dimensional supersymmetric
  theories},'' \href{http://dx.doi.org/10.1016/S0550-3213(98)00355-1}{{\em
  Nucl. Phys.} {\bfseries B529} (1998) 180--206},
\href{http://arxiv.org/abs/hep-th/9712145}{{\ttfamily arXiv:hep-th/9712145}}.
%%CITATION = HEP-TH/9712145;%%.

\bibitem{Klemm:1996bj}
A.~Klemm, W.~Lerche, P.~Mayr, C.~Vafa, and N.~P. Warner, ``{Selfdual Strings
  and $\mathcal{N} = 2$ Supersymmetric Field Theory},''
  \href{http://dx.doi.org/10.1016/0550-3213(96)00353-7}{{\em Nucl. Phys.}
  {\bfseries B477} (1996) 746--766},
\href{http://arxiv.org/abs/hep-th/9604034}{{\ttfamily arXiv:hep-th/9604034}}.
%%CITATION = HEP-TH/9604034;%%.

\bibitem{Evans:1997hk}
N.~J. Evans, C.~V. Johnson, and A.~D. Shapere, ``{Orientifolds, Branes, and
  Duality of 4D Gauge Theories},''
  \href{http://dx.doi.org/10.1016/S0550-3213(97)00384-2}{{\em Nucl. Phys.}
  {\bfseries B505} (1997) 251--271},
\href{http://arxiv.org/abs/hep-th/9703210}{{\ttfamily arXiv:hep-th/9703210}}.
%%CITATION = HEP-TH/9703210;%%.

\bibitem{Landsteiner:1997vd}
K.~Landsteiner, E.~Lopez, and D.~A. Lowe, ``{$\mathcal{N} = 2$ Supersymmetric
  Gauge Theories, Branes and Orientifolds},''
  \href{http://dx.doi.org/10.1016/S0550-3213(97)00559-2}{{\em Nucl. Phys.}
  {\bfseries B507} (1997) 197--226},
\href{http://arxiv.org/abs/hep-th/9705199}{{\ttfamily arXiv:hep-th/9705199}}.
%%CITATION = HEP-TH/9705199;%%.

\bibitem{Brandhuber:1997cc}
A.~Brandhuber, J.~Sonnenschein, S.~Theisen, and S.~Yankielowicz, ``{M-theory
  and Seiberg-Witten Curves: Orthogonal and Symplectic Groups},''
  \href{http://dx.doi.org/10.1016/S0550-3213(97)00531-2}{{\em Nucl. Phys.}
  {\bfseries B504} (1997) 175--188},
\href{http://arxiv.org/abs/hep-th/9705232}{{\ttfamily arXiv:hep-th/9705232}}.
%%CITATION = HEP-TH/9705232;%%.

\bibitem{Landsteiner:1997ei}
K.~Landsteiner and E.~Lopez, ``{New Curves from Branes},''
  \href{http://dx.doi.org/10.1016/S0550-3213(98)00022-4}{{\em Nucl. Phys.}
  {\bfseries B516} (1998) 273--296},
\href{http://arxiv.org/abs/hep-th/9708118}{{\ttfamily arXiv:hep-th/9708118}}.
%%CITATION = HEP-TH/9708118;%%.

\bibitem{Kapustin:1998xn}
A.~Kapustin, ``{Solution of $\mathcal{N} = 2$ Gauge Theories via
  Compactification to Three Dimensions},''
  \href{http://dx.doi.org/10.1016/S0550-3213(98)00520-3}{{\em Nucl. Phys.}
  {\bfseries B534} (1998) 531--545},
\href{http://arxiv.org/abs/hep-th/9804069}{{\ttfamily arXiv:hep-th/9804069}}.
%%CITATION = HEP-TH/9804069;%%.

\bibitem{Landsteiner:1998pb}
K.~Landsteiner, E.~Lopez, and D.~A. Lowe, ``{Supersymmetric Gauge Theories from
  Branes and Orientifold Six Planes},''
  \href{http://dx.doi.org/10.1088/1126-6708/1998/07/011}{{\em JHEP} {\bfseries
  07} (1998) 011},
\href{http://arxiv.org/abs/hep-th/9805158}{{\ttfamily arXiv:hep-th/9805158}}.
%%CITATION = HEP-TH/9805158;%%.

\bibitem{Argyres:2002xc}
P.~C. Argyres, R.~Maimon, and S.~Pelland, ``{The M-theory Lift of Two O6-planes
  and Four D6-branes},''
  \href{http://dx.doi.org/10.1088/1126-6708/2002/05/008}{{\em JHEP} {\bfseries
  05} (2002) 008},
\href{http://arxiv.org/abs/hep-th/0204127}{{\ttfamily arXiv:hep-th/0204127}}.
%%CITATION = HEP-TH/0204127;%%.

\bibitem{Alday:2009aq}
L.~F. Alday, D.~Gaiotto, and Y.~Tachikawa, ``{Liouville Correlation Functions
  from Four-dimensional Gauge Theories},''
  \href{http://dx.doi.org/10.1007/s11005-010-0369-5}{{\em Lett. Math. Phys.}
  {\bfseries 91} (2010) 167--197},
\href{http://arxiv.org/abs/0906.3219}{{\ttfamily arXiv:0906.3219 [hep-th]}}.
%%CITATION = ARXIV:0906.3219;%%.

\bibitem{Chacaltana:2010ks}
O.~Chacaltana and J.~Distler, ``{Tinkertoys for Gaiotto Duality},''
  \href{http://dx.doi.org/10.1007/JHEP11(2010)099}{{\em JHEP} {\bfseries 11}
  (2010) 099},
\href{http://arxiv.org/abs/1008.5203}{{\ttfamily arXiv:1008.5203 [hep-th]}}.
%%CITATION = ARXIV:1008.5203;%%.

\bibitem{Chacaltana:2012zy}
O.~Chacaltana, J.~Distler, and Y.~Tachikawa, ``{Nilpotent orbits and
  codimension-two defects of 6d N=(2,0) theories},''
  \href{http://dx.doi.org/10.1142/S0217751X1340006X}{{\em Int. J. Mod. Phys.}
  {\bfseries A28} (2013) 1340006},
\href{http://arxiv.org/abs/1203.2930}{{\ttfamily arXiv:1203.2930 [hep-th]}}.
%%CITATION = ARXIV:1203.2930;%%.

\bibitem{Xie:2012hs}
D.~Xie, ``{General Argyres-Douglas Theory},''
  \href{http://dx.doi.org/10.1007/JHEP01(2013)100}{{\em JHEP} {\bfseries 01}
  (2013) 100},
\href{http://arxiv.org/abs/1204.2270}{{\ttfamily arXiv:1204.2270 [hep-th]}}.
%%CITATION = ARXIV:1204.2270;%%.

\bibitem{Nekrasov:2012xe}
N.~Nekrasov and V.~Pestun, ``{Seiberg-Witten geometry of four dimensional N=2
  quiver gauge theories},''
\href{http://arxiv.org/abs/1211.2240}{{\ttfamily arXiv:1211.2240 [hep-th]}}.
%%CITATION = ARXIV:1211.2240;%%.

\bibitem{Beem:2014kka}
C.~Beem, L.~Rastelli, and B.~C. van Rees, ``{$ \mathcal{W} $ symmetry in six
  dimensions},'' \href{http://dx.doi.org/10.1007/JHEP05(2015)017}{{\em JHEP}
  {\bfseries 05} (2015) 017},
\href{http://arxiv.org/abs/1404.1079}{{\ttfamily arXiv:1404.1079 [hep-th]}}.
%%CITATION = ARXIV:1404.1079;%%.

\bibitem{Ohmori:2014pca}
K.~Ohmori, H.~Shimizu, and Y.~Tachikawa, ``{Anomaly Polynomial of E-string
  Theories},'' \href{http://dx.doi.org/10.1007/JHEP08(2014)002}{{\em JHEP}
  {\bfseries 08} (2014) 002},
\href{http://arxiv.org/abs/1404.3887}{{\ttfamily arXiv:1404.3887 [hep-th]}}.
%%CITATION = ARXIV:1404.3887;%%.

\bibitem{Heckman:2014qba}
J.~J. Heckman, ``{More on the Matter of 6D SCFTs},''
  \href{http://dx.doi.org/10.1016/j.physletb.2015.05.046}{{\em Phys. Lett.}
  {\bfseries B747} (2015) 73--75},
\href{http://arxiv.org/abs/1408.0006}{{\ttfamily arXiv:1408.0006 [hep-th]}}.
%%CITATION = ARXIV:1408.0006;%%.

\bibitem{Ohmori:2014kda}
K.~Ohmori, H.~Shimizu, Y.~Tachikawa, and K.~Yonekura, ``{Anomaly Polynomial of
  General 6d SCFTs},'' \href{http://dx.doi.org/10.1093/ptep/ptu140}{{\em PTEP}
  {\bfseries 2014} no.~10, (2014) 103B07},
\href{http://arxiv.org/abs/1408.5572}{{\ttfamily arXiv:1408.5572 [hep-th]}}.
%%CITATION = ARXIV:1408.5572;%%.

\bibitem{Intriligator:2014eaa}
K.~Intriligator, ``{6d, $ \mathcal{N}=\left(1,\;0\right) $ Coulomb branch
  anomaly matching},'' \href{http://dx.doi.org/10.1007/JHEP10(2014)162}{{\em
  JHEP} {\bfseries 10} (2014) 162},
\href{http://arxiv.org/abs/1408.6745}{{\ttfamily arXiv:1408.6745 [hep-th]}}.
%%CITATION = ARXIV:1408.6745;%%.

\bibitem{Haghighat:2014vxa}
B.~Haghighat, A.~Klemm, G.~Lockhart, and C.~Vafa, ``{Strings of Minimal 6d
  SCFTs},'' \href{http://dx.doi.org/10.1002/prop.201500014}{{\em Fortsch.
  Phys.} {\bfseries 63} (2015) 294--322},
\href{http://arxiv.org/abs/1412.3152}{{\ttfamily arXiv:1412.3152 [hep-th]}}.
%%CITATION = ARXIV:1412.3152;%%.

\bibitem{DelZotto:2015isa}
M.~Del~Zotto, J.~J. Heckman, D.~S. Park, and T.~Rudelius, ``{On the Defect
  Group of a 6D SCFT},''
  \href{http://dx.doi.org/10.1007/s11005-016-0839-5}{{\em Lett. Math. Phys.}
  {\bfseries 106} no.~6, (2016) 765--786},
\href{http://arxiv.org/abs/1503.04806}{{\ttfamily arXiv:1503.04806 [hep-th]}}.
%%CITATION = ARXIV:1503.04806;%%.

\bibitem{Gaiotto:2015usa}
D.~Gaiotto and S.~S. Razamat, ``{$ \mathcal{N}=1 $ theories of class $
  {\mathcal{S}}_k $},'' \href{http://dx.doi.org/10.1007/JHEP07(2015)073}{{\em
  JHEP} {\bfseries 07} (2015) 073},
\href{http://arxiv.org/abs/1503.05159}{{\ttfamily arXiv:1503.05159 [hep-th]}}.
%%CITATION = ARXIV:1503.05159;%%.

\bibitem{Ohmori:2015pua}
K.~Ohmori, H.~Shimizu, Y.~Tachikawa, and K.~Yonekura, ``{6d $\mathcal{N}=(1,0)$
  theories on $T^2$ and class S theories: Part I},''
  \href{http://dx.doi.org/10.1007/JHEP07(2015)014}{{\em JHEP} {\bfseries 07}
  (2015) 014},
\href{http://arxiv.org/abs/1503.06217}{{\ttfamily arXiv:1503.06217 [hep-th]}}.
%%CITATION = ARXIV:1503.06217;%%.

\bibitem{Gadde:2015tra}
A.~Gadde, B.~Haghighat, J.~Kim, S.~Kim, G.~Lockhart, and C.~Vafa, ``{6d String
  Chains},''
\href{http://arxiv.org/abs/1504.04614}{{\ttfamily arXiv:1504.04614 [hep-th]}}.
%%CITATION = ARXIV:1504.04614;%%.

\bibitem{Franco:2015jna}
S.~Franco, H.~Hayashi, and A.~Uranga, ``{Charting Class $\mathcal S_k$
  Territory},'' \href{http://dx.doi.org/10.1103/PhysRevD.92.045004}{{\em Phys.
  Rev.} {\bfseries D92} no.~4, (2015) 045004},
\href{http://arxiv.org/abs/1504.05988}{{\ttfamily arXiv:1504.05988 [hep-th]}}.
%%CITATION = ARXIV:1504.05988;%%.

\bibitem{DelZotto:2015rca}
M.~Del~Zotto, C.~Vafa, and D.~Xie, ``{Geometric engineering, mirror symmetry
  and $ 6{\mathrm{d}}_{\left(1,0\right)}\to
  4{\mathrm{d}}_{\left(\mathcal{N}=2\right)} $},''
  \href{http://dx.doi.org/10.1007/JHEP11(2015)123}{{\em JHEP} {\bfseries 11}
  (2015) 123},
\href{http://arxiv.org/abs/1504.08348}{{\ttfamily arXiv:1504.08348 [hep-th]}}.
%%CITATION = ARXIV:1504.08348;%%.

\bibitem{Hanany:2015pfa}
A.~Hanany and K.~Maruyoshi, ``{Chiral Theories of Class $ \mathcal{S} $},''
  \href{http://dx.doi.org/10.1007/JHEP12(2015)080}{{\em JHEP} {\bfseries 12}
  (2015) 080},
\href{http://arxiv.org/abs/1505.05053}{{\ttfamily arXiv:1505.05053 [hep-th]}}.
%%CITATION = ARXIV:1505.05053;%%.

\bibitem{Beem:2015aoa}
C.~Beem, M.~Lemos, L.~Rastelli, and B.~C. van Rees, ``{The (2, 0)
  superconformal bootstrap},''
  \href{http://dx.doi.org/10.1103/PhysRevD.93.025016}{{\em Phys. Rev.}
  {\bfseries D93} no.~2, (2016) 025016},
\href{http://arxiv.org/abs/1507.05637}{{\ttfamily arXiv:1507.05637 [hep-th]}}.
%%CITATION = ARXIV:1507.05637;%%.

\bibitem{Cordova:2015fha}
C.~Cordova, T.~T. Dumitrescu, and K.~Intriligator, ``{Anomalies,
  renormalization group flows, and the a-theorem in six-dimensional (1, 0)
  theories},'' \href{http://dx.doi.org/10.1007/JHEP10(2016)080}{{\em JHEP}
  {\bfseries 10} (2016) 080},
\href{http://arxiv.org/abs/1506.03807}{{\ttfamily arXiv:1506.03807 [hep-th]}}.
%%CITATION = ARXIV:1506.03807;%%.

\bibitem{Aganagic:2015cta}
M.~Aganagic and N.~Haouzi, ``{ADE Little String Theory on a Riemann Surface
  (and Triality)},''
\href{http://arxiv.org/abs/1506.04183}{{\ttfamily arXiv:1506.04183 [hep-th]}}.
%%CITATION = ARXIV:1506.04183;%%.

\bibitem{Ohmori:2015pia}
K.~Ohmori, H.~Shimizu, Y.~Tachikawa, and K.~Yonekura, ``{6d
  $\mathcal{N}=\left(1,\;0\right) $ theories on S$^{1}$ /T$^{2}$ and class S
  theories: Part II},'' \href{http://dx.doi.org/10.1007/JHEP12(2015)131}{{\em
  JHEP} {\bfseries 12} (2015) 131},
\href{http://arxiv.org/abs/1508.00915}{{\ttfamily arXiv:1508.00915 [hep-th]}}.
%%CITATION = ARXIV:1508.00915;%%.

\bibitem{Coman:2015bqq}
I.~Coman, E.~Pomoni, M.~Taki, and F.~Yagi, ``{Spectral curves of
  $\mathcal{N}=1$ theories of class $\mathcal{S}_k$},''
\href{http://arxiv.org/abs/1512.06079}{{\ttfamily arXiv:1512.06079 [hep-th]}}.
%%CITATION = ARXIV:1512.06079;%%.

\bibitem{Hayashi:2015vhy}
H.~Hayashi, S.-S. Kim, K.~Lee, M.~Taki, and F.~Yagi, ``{More on 5d descriptions
  of 6d SCFTs},'' \href{http://dx.doi.org/10.1007/JHEP10(2016)126}{{\em JHEP}
  {\bfseries 10} (2016) 126},
\href{http://arxiv.org/abs/1512.08239}{{\ttfamily arXiv:1512.08239 [hep-th]}}.
%%CITATION = ARXIV:1512.08239;%%.

\bibitem{Cordova:2016xhm}
C.~Cordova, T.~T. Dumitrescu, and K.~Intriligator, ``{Deformations of
  Superconformal Theories},''
  \href{http://dx.doi.org/10.1007/JHEP11(2016)135}{{\em JHEP} {\bfseries 11}
  (2016) 135},
\href{http://arxiv.org/abs/1602.01217}{{\ttfamily arXiv:1602.01217 [hep-th]}}.
%%CITATION = ARXIV:1602.01217;%%.

\bibitem{Morrison:2016nrt}
D.~R. Morrison and C.~Vafa, ``{F-theory and $ \mathcal{N} $ = 1 SCFTs in four
  dimensions},'' \href{http://dx.doi.org/10.1007/JHEP08(2016)070}{{\em JHEP}
  {\bfseries 08} (2016) 070},
\href{http://arxiv.org/abs/1604.03560}{{\ttfamily arXiv:1604.03560 [hep-th]}}.
%%CITATION = ARXIV:1604.03560;%%.

\bibitem{Kim:2016foj}
H.-C. Kim, S.~Kim, and J.~Park, ``{6d strings from new chiral gauge
  theories},''
\href{http://arxiv.org/abs/1608.03919}{{\ttfamily arXiv:1608.03919 [hep-th]}}.
%%CITATION = ARXIV:1608.03919;%%.

\bibitem{Shimizu:2016lbw}
H.~Shimizu and Y.~Tachikawa, ``{Anomaly of strings of 6d $\mathcal{N} = (1,0)$
  theories},''
\href{http://arxiv.org/abs/1608.05894}{{\ttfamily arXiv:1608.05894 [hep-th]}}.
%%CITATION = ARXIV:1608.05894;%%.

\bibitem{DelZotto:2016pvm}
M.~Del~Zotto and G.~Lockhart, ``{On Exceptional Instanton Strings},''
\href{http://arxiv.org/abs/1609.00310}{{\ttfamily arXiv:1609.00310 [hep-th]}}.
%%CITATION = ARXIV:1609.00310;%%.

\bibitem{Heckman:2016xdl}
J.~J. Heckman, P.~Jefferson, T.~Rudelius, and C.~Vafa, ``{Punctures for
  Theories of Class $\mathcal{S}_\Gamma$},''
  \href{http://dx.doi.org/10.1007/JHEP03(2017)171}{{\em JHEP} {\bfseries 03}
  (2017) 171},
\href{http://arxiv.org/abs/1609.01281}{{\ttfamily arXiv:1609.01281 [hep-th]}}.
%%CITATION = ARXIV:1609.01281;%%.

\bibitem{Apruzzi:2016nfr}
F.~Apruzzi, F.~Hassler, J.~J. Heckman, and I.~V. Melnikov, ``{From 6D SCFTs to
  Dynamic GLSMs},''
\href{http://arxiv.org/abs/1610.00718}{{\ttfamily arXiv:1610.00718 [hep-th]}}.
%%CITATION = ARXIV:1610.00718;%%.

\bibitem{Razamat:2016dpl}
S.~S. Razamat, C.~Vafa, and G.~Zafrir, ``{4d $\mathcal{N} = 1$ from 6d
  (1,0)},''
\href{http://arxiv.org/abs/1610.09178}{{\ttfamily arXiv:1610.09178 [hep-th]}}.
%%CITATION = ARXIV:1610.09178;%%.

\bibitem{Cordova:2016emh}
C.~Cordova, T.~T. Dumitrescu, and K.~Intriligator, ``{Multiplets of
  Superconformal Symmetry in Diverse Dimensions},''
\href{http://arxiv.org/abs/1612.00809}{{\ttfamily arXiv:1612.00809 [hep-th]}}.
%%CITATION = ARXIV:1612.00809;%%.

\bibitem{DelZotto:2017pti}
M.~Del~Zotto, J.~J. Heckman, and D.~R. Morrison, ``{6D SCFTs and Phases of 5D
  Theories},''
\href{http://arxiv.org/abs/1703.02981}{{\ttfamily arXiv:1703.02981 [hep-th]}}.
%%CITATION = ARXIV:1703.02981;%%.

\bibitem{Haghighat:2017vch}
B.~Haghighat, W.~Yan, and S.-T. Yau, ``{ADE String Chains and Mirror
  Symmetry},''
\href{http://arxiv.org/abs/1705.05199}{{\ttfamily arXiv:1705.05199 [hep-th]}}.
%%CITATION = ARXIV:1705.05199;%%.

\bibitem{Chang:2017xmr}
C.-M. Chang and Y.-H. Lin, ``{Carving Out the End of the World or
  (Superconformal Bootstrap in Six Dimensions)},''
\href{http://arxiv.org/abs/1705.05392}{{\ttfamily arXiv:1705.05392 [hep-th]}}.
%%CITATION = ARXIV:1705.05392;%%.

\bibitem{Mekareeya:2017jgc}
N.~Mekareeya, K.~Ohmori, Y.~Tachikawa, and G.~Zafrir, ``{$E_8$ instantons on
  type-A ALE spaces and supersymmetric field theories},''
\href{http://arxiv.org/abs/1707.04370}{{\ttfamily arXiv:1707.04370 [hep-th]}}.
%%CITATION = ARXIV:1707.04370;%%.

\bibitem{Mekareeya:2017sqh}
N.~Mekareeya, K.~Ohmori, H.~Shimizu, and A.~Tomasiello, ``{Small instanton
  transitions for M5 fractions},''
\href{http://arxiv.org/abs/1707.05785}{{\ttfamily arXiv:1707.05785 [hep-th]}}.
%%CITATION = ARXIV:1707.05785;%%.

\bibitem{Seiberg:1997ax}
N.~Seiberg, ``{Notes on theories with 16 supercharges},''
  \href{http://dx.doi.org/10.1016/S0920-5632(98)00128-5}{{\em Nucl. Phys. Proc.
  Suppl.} {\bfseries 67} (1998) 158--171},
\href{http://arxiv.org/abs/hep-th/9705117}{{\ttfamily arXiv:hep-th/9705117}}.
%%CITATION = HEP-TH/9705117;%%.

\bibitem{Ganor:1996pc}
O.~J. Ganor, D.~R. Morrison, and N.~Seiberg, ``{Branes, Calabi-Yau spaces, and
  toroidal compactification of the $\mathcal{N} = 1$ six-dimensional $E_8$
  theory},'' \href{http://dx.doi.org/10.1016/S0550-3213(96)00690-6}{{\em Nucl.
  Phys.} {\bfseries B487} (1997) 93--127},
\href{http://arxiv.org/abs/hep-th/9610251}{{\ttfamily arXiv:hep-th/9610251}}.
%%CITATION = HEP-TH/9610251;%%.

\bibitem{Ganor:1996xg}
O.~J. Ganor, ``{Compactification of Tensionless String Theories},''
\href{http://arxiv.org/abs/hep-th/9607092}{{\ttfamily arXiv:hep-th/9607092}}.
%%CITATION = HEP-TH/9607092;%%.

\bibitem{Green:1984bx}
M.~B. Green, J.~H. Schwarz, and P.~C. West, ``{Anomaly Free Chiral Theories in
  Six-Dimensions},''
\href{http://dx.doi.org/10.1016/0550-3213(85)90222-6}{{\em Nucl. Phys.}
  {\bfseries B254} (1985) 327--348}.
%%CITATION = NUPHA,B254,327;%%.

\bibitem{Sagnotti:1992qw}
A.~Sagnotti, ``{A Note on the Green-Schwarz mechanism in open string
  theories},'' \href{http://dx.doi.org/10.1016/0370-2693(92)90682-T}{{\em Phys.
  Lett.} {\bfseries B294} (1992) 196--203},
\href{http://arxiv.org/abs/hep-th/9210127}{{\ttfamily arXiv:hep-th/9210127}}.
%%CITATION = HEP-TH/9210127;%%.

\bibitem{Sadov:1996zm}
V.~Sadov, ``{Generalized Green-Schwarz mechanism in F theory},''
  \href{http://dx.doi.org/10.1016/0370-2693(96)01134-3}{{\em Phys. Lett.}
  {\bfseries B388} (1996) 45--50},
\href{http://arxiv.org/abs/hep-th/9606008}{{\ttfamily arXiv:hep-th/9606008
  [hep-th]}}.
%%CITATION = HEP-TH/9606008;%%.

\bibitem{Morrison:2012np}
D.~R. Morrison and W.~Taylor, ``{Classifying bases for 6D F-theory models},''
  \href{http://dx.doi.org/10.2478/s11534-012-0065-4}{{\em Central Eur. J.
  Phys.} {\bfseries 10} (2012) 1072--1088},
\href{http://arxiv.org/abs/1201.1943}{{\ttfamily arXiv:1201.1943 [hep-th]}}.
%%CITATION = ARXIV:1201.1943;%%.

\bibitem{Hanany:1996ie}
A.~Hanany and E.~Witten, ``{Type IIB superstrings, BPS monopoles, and
  three-dimensional gauge dynamics},''
  \href{http://dx.doi.org/10.1016/S0550-3213(97)00157-0,
  10.1016/S0550-3213(97)80030-2}{{\em Nucl. Phys.} {\bfseries B492} (1997)
  152--190},
\href{http://arxiv.org/abs/hep-th/9611230}{{\ttfamily arXiv:hep-th/9611230}}.
%%CITATION = HEP-TH/9611230;%%.

\bibitem{Cachazo:2001sg}
F.~Cachazo, B.~Fiol, K.~A. Intriligator, S.~Katz, and C.~Vafa, ``{A Geometric
  unification of dualities},''
  \href{http://dx.doi.org/10.1016/S0550-3213(02)00078-0}{{\em Nucl. Phys.}
  {\bfseries B628} (2002) 3--78},
\href{http://arxiv.org/abs/hep-th/0110028}{{\ttfamily arXiv:hep-th/0110028}}.
%%CITATION = HEP-TH/0110028;%%.

\bibitem{Tachikawa:2009rb}
Y.~Tachikawa, ``{Six-dimensional $D_N$ theory and four-dimensional SO-USp
  quivers},'' \href{http://dx.doi.org/10.1088/1126-6708/2009/07/067}{{\em JHEP}
  {\bfseries 07} (2009) 067},
\href{http://arxiv.org/abs/0905.4074}{{\ttfamily arXiv:0905.4074 [hep-th]}}.
%%CITATION = ARXIV:0905.4074;%%.

\bibitem{Drukker:2010jp}
N.~Drukker, D.~Gaiotto, and J.~Gomis, ``{The Virtue of Defects in 4D Gauge
  Theories and 2D CFTs},''
  \href{http://dx.doi.org/10.1007/JHEP06(2011)025}{{\em JHEP} {\bfseries 06}
  (2011) 025},
\href{http://arxiv.org/abs/1003.1112}{{\ttfamily arXiv:1003.1112 [hep-th]}}.
%%CITATION = ARXIV:1003.1112;%%.

\bibitem{Tachikawa:2010vg}
Y.~Tachikawa, ``{$\mathcal{N} = 2$ S-duality via Outer-automorphism Twists},''
  \href{http://dx.doi.org/10.1088/1751-8113/44/18/182001}{{\em J. Phys.}
  {\bfseries A44} (2011) 182001},
\href{http://arxiv.org/abs/1009.0339}{{\ttfamily arXiv:1009.0339 [hep-th]}}.
%%CITATION = ARXIV:1009.0339;%%.

\bibitem{Witten:1997bs}
E.~Witten, ``{Toroidal Compactification Without Vector Structure},'' {\em JHEP}
  {\bfseries 02} (1998) 006,
\href{http://arxiv.org/abs/hep-th/9712028}{{\ttfamily arXiv:hep-th/9712028}}.
%%CITATION = HEP-TH/9712028;%%.

\bibitem{deBoer:2001px}
J.~de~Boer, R.~Dijkgraaf, K.~Hori, A.~Keurentjes, J.~Morgan, D.~R. Morrison,
  and S.~Sethi, ``{Triples, Fluxes, and Strings},'' {\em Adv. Theor. Math.
  Phys.} {\bfseries 4} (2002) 995--1186,
\href{http://arxiv.org/abs/hep-th/0103170}{{\ttfamily arXiv:hep-th/0103170}}.
%%CITATION = HEP-TH/0103170;%%.

\bibitem{Tachikawa:2015wka}
Y.~Tachikawa, ``{Frozen singularities in M and F theory},''
  \href{http://dx.doi.org/10.1007/JHEP06(2016)128}{{\em JHEP} {\bfseries 06}
  (2016) 128},
\href{http://arxiv.org/abs/1508.06679}{{\ttfamily arXiv:1508.06679 [hep-th]}}.
%%CITATION = ARXIV:1508.06679;%%.

\bibitem{Bhardwaj:2015oru}
L.~Bhardwaj, M.~Del~Zotto, J.~J. Heckman, D.~R. Morrison, T.~Rudelius, and
  C.~Vafa, ``{F-theory and the Classification of Little Strings},''
  \href{http://dx.doi.org/10.1103/PhysRevD.93.086002}{{\em Phys. Rev.}
  {\bfseries D93} no.~8, (2016) 086002},
\href{http://arxiv.org/abs/1511.05565}{{\ttfamily arXiv:1511.05565 [hep-th]}}.
%%CITATION = ARXIV:1511.05565;%%.

\bibitem{Cordova:2015vwa}
C.~Cordova, T.~T. Dumitrescu, and X.~Yin, ``{Higher Derivative Terms, Toroidal
  Compactification, and Weyl Anomalies in Six-Dimensional (2,0) Theories},''
\href{http://arxiv.org/abs/1505.03850}{{\ttfamily arXiv:1505.03850 [hep-th]}}.
%%CITATION = ARXIV:1505.03850;%%.

\bibitem{Heckman:2015ola}
J.~J. Heckman, D.~R. Morrison, T.~Rudelius, and C.~Vafa, ``{Geometry of 6D RG
  Flows},'' \href{http://dx.doi.org/10.1007/JHEP09(2015)052}{{\em JHEP}
  {\bfseries 09} (2015) 052},
\href{http://arxiv.org/abs/1505.00009}{{\ttfamily arXiv:1505.00009 [hep-th]}}.
%%CITATION = ARXIV:1505.00009;%%.

\bibitem{Heckman:2016ssk}
J.~J. Heckman, T.~Rudelius, and A.~Tomasiello, ``{6D RG Flows and Nilpotent
  Hierarchies},'' \href{http://dx.doi.org/10.1007/JHEP07(2016)082}{{\em JHEP}
  {\bfseries 07} (2016) 082},
\href{http://arxiv.org/abs/1601.04078}{{\ttfamily arXiv:1601.04078 [hep-th]}}.
%%CITATION = ARXIV:1601.04078;%%.

\bibitem{Mekareeya:2016yal}
N.~Mekareeya, T.~Rudelius, and A.~Tomasiello, ``{T-branes, Anomalies and Moduli
  Spaces in 6D SCFTs},''
\href{http://arxiv.org/abs/1612.06399}{{\ttfamily arXiv:1612.06399 [hep-th]}}.
%%CITATION = ARXIV:1612.06399;%%.

\bibitem{Gaiotto:2014kfa}
D.~Gaiotto, A.~Kapustin, N.~Seiberg, and B.~Willett, ``{Generalized Global
  Symmetries},'' \href{http://dx.doi.org/10.1007/JHEP02(2015)172}{{\em JHEP}
  {\bfseries 02} (2015) 172},
\href{http://arxiv.org/abs/1412.5148}{{\ttfamily arXiv:1412.5148 [hep-th]}}.
%%CITATION = ARXIV:1412.5148;%%.

\bibitem{DelZotto:2014fia}
M.~Del~Zotto, J.~J. Heckman, D.~R. Morrison, and D.~S. Park, ``{6D SCFTs and
  Gravity},'' \href{http://dx.doi.org/10.1007/JHEP06(2015)158}{{\em JHEP}
  {\bfseries 06} (2015) 158},
\href{http://arxiv.org/abs/1412.6526}{{\ttfamily arXiv:1412.6526 [hep-th]}}.
%%CITATION = ARXIV:1412.6526;%%.

\bibitem{Heckman:2015axa}
J.~J. Heckman and T.~Rudelius, ``{Evidence for C-theorems in 6D SCFTs},''
  \href{http://dx.doi.org/10.1007/JHEP09(2015)218}{{\em JHEP} {\bfseries 09}
  (2015) 218},
\href{http://arxiv.org/abs/1506.06753}{{\ttfamily arXiv:1506.06753 [hep-th]}}.
%%CITATION = ARXIV:1506.06753;%%.

\bibitem{Kumar:2010ru}
V.~Kumar, D.~R. Morrison, and W.~Taylor, ``{Global aspects of the space of 6D N
  = 1 supergravities},'' \href{http://dx.doi.org/10.1007/JHEP11(2010)118}{{\em
  JHEP} {\bfseries 11} (2010) 118},
\href{http://arxiv.org/abs/1008.1062}{{\ttfamily arXiv:1008.1062 [hep-th]}}.
%%CITATION = ARXIV:1008.1062;%%.

\bibitem{Klebanov:2000hb}
I.~R. Klebanov and M.~J. Strassler, ``{Supergravity and a confining gauge
  theory: Duality cascades and $\chi$SB resolution of naked singularities},''
  \href{http://dx.doi.org/10.1088/1126-6708/2000/08/052}{{\em JHEP} {\bfseries
  08} (2000) 052},
\href{http://arxiv.org/abs/hep-th/0007191}{{\ttfamily arXiv:hep-th/0007191}}.
%%CITATION = HEP-TH/0007191;%%.

\bibitem{Seiberg:1994pq}
N.~Seiberg, ``{Electric - Magnetic Duality in Supersymmetric Non-Abelian Gauge
  Theories},'' \href{http://dx.doi.org/10.1016/0550-3213(94)00023-8}{{\em Nucl.
  Phys.} {\bfseries B435} (1995) 129--146},
\href{http://arxiv.org/abs/hep-th/9411149}{{\ttfamily arXiv:hep-th/9411149}}.
%%CITATION = HEP-TH/9411149;%%.

\bibitem{Schafer-Nameki:2016cfr}
S.~Schafer-Nameki and T.~Weigand, ``{F-theory and 2d $(0, 2)$ theories},''
  \href{http://dx.doi.org/10.1007/JHEP05(2016)059}{{\em JHEP} {\bfseries 05}
  (2016) 059},
\href{http://arxiv.org/abs/1601.02015}{{\ttfamily arXiv:1601.02015 [hep-th]}}.
%%CITATION = ARXIV:1601.02015;%%.

\bibitem{Apruzzi:2016iac}
F.~Apruzzi, F.~Hassler, J.~J. Heckman, and I.~V. Melnikov, ``{UV Completions
  for Non-Critical Strings},''
  \href{http://dx.doi.org/10.1007/JHEP07(2016)045}{{\em JHEP} {\bfseries 07}
  (2016) 045},
\href{http://arxiv.org/abs/1602.04221}{{\ttfamily arXiv:1602.04221 [hep-th]}}.
%%CITATION = ARXIV:1602.04221;%%.

\bibitem{Gadde:2013lxa}
A.~Gadde, S.~Gukov, and P.~Putrov, ``{(0, 2) trialities},''
  \href{http://dx.doi.org/10.1007/JHEP03(2014)076}{{\em JHEP} {\bfseries 03}
  (2014) 076},
\href{http://arxiv.org/abs/1310.0818}{{\ttfamily arXiv:1310.0818 [hep-th]}}.
%%CITATION = ARXIV:1310.0818;%%.

\bibitem{Franco:2016nwv}
S.~Franco, S.~Lee, and R.-K. Seong, ``{Brane Brick Models and 2d (0,2)
  Triality},''
\href{http://arxiv.org/abs/1602.01834}{{\ttfamily arXiv:1602.01834 [hep-th]}}.
%%CITATION = ARXIV:1602.01834;%%.

\bibitem{Franco:2016qxh}
S.~Franco, S.~Lee, R.-K. Seong, and C.~Vafa, ``{Brane Brick Models in the
  Mirror},'' \href{http://dx.doi.org/10.1007/JHEP02(2017)106}{{\em JHEP}
  {\bfseries 02} (2017) 106},
\href{http://arxiv.org/abs/1609.01723}{{\ttfamily arXiv:1609.01723 [hep-th]}}.
%%CITATION = ARXIV:1609.01723;%%.

\bibitem{Hohenegger:2015btj}
S.~Hohenegger, A.~Iqbal, and S.-J. Rey, ``{Instanton-Monopole Correspondence
  from M-Branes on $\mathbb{S}^1$ and Little String Theory},''
\href{http://arxiv.org/abs/1511.02787}{{\ttfamily arXiv:1511.02787 [hep-th]}}.
%%CITATION = ARXIV:1511.02787;%%.

\bibitem{Hanany:1999sj}
A.~Hanany and A.~Zaffaroni, ``{Issues on orientifolds: On the brane
  construction of gauge theories with SO(2n) global symmetry},''
  \href{http://dx.doi.org/10.1088/1126-6708/1999/07/009}{{\em JHEP} {\bfseries
  07} (1999) 009},
\href{http://arxiv.org/abs/hep-th/9903242}{{\ttfamily arXiv:hep-th/9903242
  [hep-th]}}.
%%CITATION = HEP-TH/9903242;%%.

\bibitem{Bergman:2012kr}
O.~Bergman and D.~Rodriguez-Gomez, ``{5d quivers and their AdS(6) duals},''
  \href{http://dx.doi.org/10.1007/JHEP07(2012)171}{{\em JHEP} {\bfseries 07}
  (2012) 171},
\href{http://arxiv.org/abs/1206.3503}{{\ttfamily arXiv:1206.3503 [hep-th]}}.
%%CITATION = ARXIV:1206.3503;%%.

\bibitem{Bah:2017gph}
I.~Bah, A.~Hanany, K.~Maruyoshi, S.~S. Razamat, Y.~Tachikawa, and G.~Zafrir,
  ``{4d $\mathcal{N} = 1$ from 6d $\mathcal{N} = (1,0)$ on a torus with
  fluxes},''
\href{http://arxiv.org/abs/1702.04740}{{\ttfamily arXiv:1702.04740 [hep-th]}}.
%%CITATION = ARXIV:1702.04740;%%.

\bibitem{Douglas:1996sw}
M.~R. Douglas and G.~W. Moore, ``{D-branes, quivers, and ALE instantons},''
\href{http://arxiv.org/abs/hep-th/9603167}{{\ttfamily arXiv:hep-th/9603167}}.
%%CITATION = HEP-TH/9603167;%%.

\bibitem{Johnson:1996py}
C.~V. Johnson and R.~C. Myers, ``{Aspects of type IIB theory on ALE spaces},''
  \href{http://dx.doi.org/10.1103/PhysRevD.55.6382}{{\em Phys. Rev.} {\bfseries
  D55} (1997) 6382--6393},
\href{http://arxiv.org/abs/hep-th/9610140}{{\ttfamily arXiv:hep-th/9610140
  [hep-th]}}.
%%CITATION = HEP-TH/9610140;%%.

\bibitem{Lawrence:1998ja}
A.~E. Lawrence, N.~Nekrasov, and C.~Vafa, ``{On Conformal Field Theories in
  Four Dimensions},''
  \href{http://dx.doi.org/10.1016/S0550-3213(98)00495-7}{{\em Nucl. Phys.}
  {\bfseries B533} (1998) 199--209},
\href{http://arxiv.org/abs/hep-th/9803015}{{\ttfamily arXiv:hep-th/9803015
  [hep-th]}}.
%%CITATION = HEP-TH/9803015;%%.

\bibitem{Aharony:2007dj}
O.~Aharony and Y.~Tachikawa, ``{A Holographic Computation of the Central
  Charges of $d=4$, $\mathcal{N} = 2$ SCFTs},''
  \href{http://dx.doi.org/10.1088/1126-6708/2008/01/037}{{\em JHEP} {\bfseries
  01} (2008) 037},
\href{http://arxiv.org/abs/0711.4532}{{\ttfamily arXiv:0711.4532 [hep-th]}}.
%%CITATION = ARXIV:0711.4532;%%.

\bibitem{Hyakutake:2000mr}
Y.~Hyakutake, Y.~Imamura, and S.~Sugimoto, ``{Orientifold planes, type I Wilson
  lines and nonBPS D-branes},''
  \href{http://dx.doi.org/10.1088/1126-6708/2000/08/043}{{\em JHEP} {\bfseries
  08} (2000) 043},
\href{http://arxiv.org/abs/hep-th/0007012}{{\ttfamily arXiv:hep-th/0007012
  [hep-th]}}.
%%CITATION = HEP-TH/0007012;%%.

\bibitem{Hanany:2000fq}
A.~Hanany and B.~Kol, ``{On orientifolds, discrete torsion, branes and M
  theory},'' \href{http://dx.doi.org/10.1088/1126-6708/2000/06/013}{{\em JHEP}
  {\bfseries 06} (2000) 013},
\href{http://arxiv.org/abs/hep-th/0003025}{{\ttfamily arXiv:hep-th/0003025
  [hep-th]}}.
%%CITATION = HEP-TH/0003025;%%.

\bibitem{Bergman:2001rp}
O.~Bergman, E.~G. Gimon, and S.~Sugimoto, ``{Orientifolds, RR torsion, and K
  theory},'' \href{http://dx.doi.org/10.1088/1126-6708/2001/05/047}{{\em JHEP}
  {\bfseries 05} (2001) 047},
\href{http://arxiv.org/abs/hep-th/0103183}{{\ttfamily arXiv:hep-th/0103183
  [hep-th]}}.
%%CITATION = HEP-TH/0103183;%%.

\bibitem{Morrison:1996xf}
D.~R. Morrison and N.~Seiberg, ``{Extremal transitions and five-dimensional
  supersymmetric field theories},''
  \href{http://dx.doi.org/10.1016/S0550-3213(96)00592-5}{{\em Nucl. Phys.}
  {\bfseries B483} (1997) 229--247},
\href{http://arxiv.org/abs/hep-th/9609070}{{\ttfamily arXiv:hep-th/9609070}}.
%%CITATION = HEP-TH/9609070;%%.

\bibitem{Minahan:1996cj}
J.~A. Minahan and D.~Nemeschansky, ``{Superconformal Fixed Points with $E_n$
  Global Symmetry},''
  \href{http://dx.doi.org/10.1016/S0550-3213(97)00039-4}{{\em Nucl. Phys.}
  {\bfseries B489} (1997) 24--46},
\href{http://arxiv.org/abs/hep-th/9610076}{{\ttfamily arXiv:hep-th/9610076
  [hep-th]}}.
%%CITATION = HEP-TH/9610076;%%.

\bibitem{Hayashi:2017jze}
H.~Hayashi and K.~Ohmori, ``{5d/6d DE instantons from trivalent gluing of web
  diagrams},''
\href{http://arxiv.org/abs/1702.07263}{{\ttfamily arXiv:1702.07263 [hep-th]}}.
%%CITATION = ARXIV:1702.07263;%%.

\bibitem{Chacaltana:2012ch}
O.~Chacaltana, J.~Distler, and Y.~Tachikawa, ``{Gaiotto duality for the twisted
  $A_{2N−1}$ series},'' \href{http://dx.doi.org/10.1007/JHEP05(2015)075}{{\em
  JHEP} {\bfseries 05} (2015) 075},
\href{http://arxiv.org/abs/1212.3952}{{\ttfamily arXiv:1212.3952 [hep-th]}}.
%%CITATION = ARXIV:1212.3952;%%.

\bibitem{Bershadsky:1997sb}
M.~Bershadsky and C.~Vafa, ``{Global anomalies and geometric engineering of
  critical theories in six-dimensions},''
\href{http://arxiv.org/abs/hep-th/9703167}{{\ttfamily arXiv:hep-th/9703167
  [hep-th]}}.
%%CITATION = HEP-TH/9703167;%%.

\end{thebibliography}\endgroup

\end{document}